\documentclass[12pt]{article}

\usepackage{amsmath,amssymb,graphicx} %use drftcite in draftmode
\usepackage{epsf}
\usepackage{pstricks}
\usepackage{cite}
\newcommand{\beq}{\begin{eqnarray}}% can be used as {equation} or  {eqnarray}
\newcommand{\eeq}{\end{eqnarray}}
\newcommand{\fpos}{n_{e^+}}
\newcommand{\fpbar}{n_{\bar p}}
\newcommand{\vpos}{v_{e^+}}
\newcommand{\fluxpos}{\Phi_{e^+}}
\newcommand{\fluxpho}{\Phi_{\gamma}}
\newcommand{\fluxap}{\Phi_{\bar p}}
\newcommand{\fluxp}{\Phi_{p}}
\newcommand{\fluxep}{\Phi_{e^++e^-}}
\newcommand{\rsun}{r_\odot}
\newcommand{\zsun}{z_\odot}
\newcommand{\thetasun}{\theta_\odot}
\newcommand{\rhosun}{\rho_\odot}
\newcommand{\R}{{\cal R}}
\newcommand{\mdm}{{m_{\chi}}}
\newcommand{\Qpbar}{Q_{\bar p}}

\newcommand{\rsm}{R_{\rm SM}}
\newcommand{\adm}{\alpha_{\rm DM}}

\newcommand{\centeron}[2]{{\setbox0=\hbox{#1}\setbox1=\hbox{#2}\ifdim

\wd1>\wd0\kern.5\wd1\kern-.5\wd0\fi
\copy0

\kern-.5\wd0\kern-.5\wd1\copy1\ifdim\wd0>\wd1
                                       \kern.5\wd0\kern-.5\wd1\fi}}
\newcommand{\ltap}{\>\centeron{\raise.35ex\hbox{$<$}}
                               {\lower.65ex\hbox{$\sim$}}\>}
\newcommand{\gtap}{\>\centeron{\raise.35ex\hbox{$>$}}
                               {\lower.65ex\hbox{$\sim$}}\>}

\newcommand\ZZ{\hbox{\zfont Z\kern-.4emZ}}
\font\zfont = cmss10 %scaled \magstep1

\def\gappeq{\mathrel{ \rlap{\raise.5ex\hbox{$>$}}
                      {\lower.5ex\hbox{$\sim$}}  } }
\def\lappeq{\mathrel{ \rlap{\raise.5ex\hbox{$<$}}
                      {\lower.5ex\hbox{$\sim$}}  } }
\begin{document}
\begin{titlepage}
\begin{flushright}
\end{flushright}

\vskip.5cm
\begin{center}
{\LARGE Dark Matter Sees The Light}
\vspace{.2cm}
 
\vskip.2cm
\end{center}
\vskip0.2cm

\begin{center}
{\bf Patrick Meade, Michele Papucci and Tomer Volansky}
\end{center}
\vskip 8pt

\begin{center}
{\it Institute for Advanced Study\\
Princeton, NJ 08540} \\
\vspace*{0.3cm}
\end{center}

\vglue 0.3truecm

\begin{abstract}
  \vskip 3pt \noindent We construct a Dark Matter (DM) annihilation
  module that can encompass the predictions from a wide array of
  models built to explain the recently reported PAMELA and
  ATIC/PPB-BETS excesses.  We present a detailed analysis of the
  injection spectrums for DM annihilation and quantitatively
  demonstrate effects that have previously not been included from the
  particle physics perspective. With this module we demonstrate the
  parameter space that can account for the aforementioned excesses and
  be compatible with existing high energy gamma ray and neutrino
  experiments.  However, we find that it is relatively generic to have
  some tension between the results of the HESS experiment and the
  ATIC/PPB-BETS experiments within the context of annihilating DM.  We
  discuss ways to alleviate this tension and how upcoming
  experiments will be able to differentiate amongst the various
  possible explanations of the purported excesses.
\end{abstract}

\end{titlepage}

\newpage

\renewcommand{\thefootnote}{(\arabic{footnote})}

%%%%%%%%%%%%%%%%%%%%%%%%%%%%%%%%%%%%%%%%%%%%%%%%%%%%%%
%%%%%%%%%%%%%%%%%%%%%%%%%%%%%%%%%%%%%%%%%%%%%%%%%%%%%%
\section{Introduction}\label{sec:intro}
\setcounter{equation}{0}
\setcounter{footnote}{0}
%%%%%%%%%%%%%%%%%%%%%%%%%%%%%%%%%%%%%%%%%%%%%%%%%%%%%%
%%%%%%%%%%%%%%%%%%%%%%%%%%%%%%%%%%%%%%%%%%%%%%%%%%%%%%
Recently there has been a series of experimental results suggesting
that we may have indirectly detected dark matter (DM) within our
Galaxy.  The combination of the positron fraction measured by the
PAMELA experiment~\cite{pamelapos} and the ATIC/PPB-BETS
experiments~\cite{ppbbets,atic}, have led to a compelling picture of
DM being responsible for a new population of positrons at high
energies.  These excesses, if confirmed, could in principle have
alternative explanations through either refining our understanding of
charged particle propagation within our Galaxy, or by identifying  new
astrophysical sources of positrons coming, for instance, from
pulsars~\cite{pulsars}.  It is intriguing therefore, that new
experimental data expected in the near future could not only confirm
or contradict those results, but also allow us to possibly determine
the physics behind these excesses.  Furthermore, in the case of DM, such
experiments could strongly constrain the various DM models.

Broadly speaking, the plethora of DM models bifurcate into either
annihilating~\cite{udm,annihilating} or decaying~\cite{decaying} DM.  In
this paper we choose to focus on the former possibility as being the
source of the electronic excesses.  The above experiments then place
strong restrictions on the models, so we adopt the
following phenomenological inputs as constraints:
\begin{itemize}
\item There is an excess in the flux ratio $\fluxpos/\fluxep$ observed
  by the PAMELA experiment extending to {\em at least} 100
  GeV~\cite{pamelapos}.
\item There is an excess in the ATIC/PPB-BETS experiments for the flux
  of charged electrons and positrons, $\fluxep$, extending to energies of
  $\sim700$ GeV~\cite{atic,ppbbets}.
\item There is no excess observed by the PAMELA
  experiment in the antiproton flux~\cite{pamelaantip}. 
\item In the absence of large local overdensity in the DM
  distribution (boost factor), the annihilation cross section in
  our galaxy needs to be $\mathcal{O}(100)$ times larger than a
  standard thermal WIMP.
\end{itemize}
These facts are not easily reconciled. The last of these assumptions
follows from the large measured rates combined with the higher mass
scale indicated by the ATIC/PPB-BETS anomaly.  For the case of a WIMP DM,
a large enhancement of the cross section
is needed~\cite{sommerfeld,breitwigner,wimponium}.  Alternatively, a large boost factor (BF) is
required.  However, such a possibility seems unlikely in light of the
results from N-body simulations\cite{nbodyboost}.  Furthermore, a model must
prefer annihilation into leptonic final states so that the antiproton
fraction is not overpopulated.  

There has been a recent explosion in model building that attempts to
incorporate the necessary ingredients to explain these excesses.
Typically, these models explain the electronic activity by either assuming a
symmetry that forbids hadronic production, or otherwise
postulating an intermediate light state that can only decay into light
leptons due to kinematics.  Most of these studies
have either stopped at the heuristic level of explanation, or
attempted quantitatively only to postdict certain experiments.  It is
therefore desirable to consider a larger set of experimental data in
order to better establish the correct model-building direction.

\noindent We attempt to address the following questions:
\begin{itemize}
\item Given a model that can explain the PAMELA and ATIC/PPB-BETS
  data, what are the experimental bounds arising from other
  experiments?
\item What are the viable classes of models?
\item For these models, what are the implications
  for upcoming experiments?
\end{itemize}

The most logical additional signature which has not been entirely
explored is the one coming from photons.  Whenever there are charged particles
in the final state there will be additional photons radiated, leading to a model independent signature~\cite{maxim}.  Additional sources of photons may contribute depending on the specific details of the model.  Recently there have been a few papers~\cite{bellgamma,strumiagamma,bergstrom} that have studied
the bounds from high energy photons in models that explain the
excesses.  The authors of ~\cite{strumiagamma,bergstrom} reached the conclusion that for most dark
matter density profiles, experimental results rule out the possibility
of annihilating DM as an explanation of the excesses.  In~\cite{strumiagamma,bellgamma} the case where DM directly annihilates into a pair of SM leptons was studied.  We reach a similar conclusion to~\cite{strumiagamma}, that
such models disagree with the experimental data collected for high
energy photons.  In~\cite{bergstrom} models where DM annihilates
through a light state and then into leptons was studied.  In our paper
we focus on these models and reach a different conclusion than the
authors of~\cite{bergstrom}.  While we find there exists some tension
between models that explain ATIC/PPB-BETS and high energy photons,
they are not ruled out by an order of magnitude.  We also
include several effects that have not yet been studied in the
literature, that can ameliorate this tension, such as dark sector radiation.

To study the implications of the present experimental data, we construct a
module that incorporates many of the required features necessary for a
model to explain the excesses.  This module has several parameters
that allow us to interpolate between different classes of models.

Our main results are summarized as follows. For annihilating DM
scenarios that explain both PAMELA and ATIC/PPB-BETS:
\begin{itemize}
\item Rather generically, such models are in tension with
  constraints from high-energy photons.  
\item Photons are more constraining than the antiprotons measurements.
  In particular, models that produce antiprotons and still fit the
  PAMELA data in many cases produce too many photons to be consistent
  with the measurements.
\item The tension is not sufficient to exclude all models.  However, it
  requires a factor of order ${\cal O}(2-5)$ that may arise from various sources, e.g. a local boost factor or a less cuspy DM
  profile.
\end{itemize}
We find a number of interesting implications based on these results.
Upcoming experiments have the power to exclude the full region of
parameter space, for models that explain the ATIC/PPB-BETS excesses
with annihilating DM.   From this point of view decaying DM models are attractive
because they generate fewer photons at the center of the
galaxy.  Models of annihilating DM that do not seek to explain the
ATIC/PPB-BETS are also viable, and can be tested with several ongoing
experiments. In particular, we stress the importance of the currently
running experiment, FERMI \cite{fermi}, for helping determine the
underlying nature of these excesses.

This paper is organized as follows.  In Section~\ref{sec:udm} we
construct a DM module that enables us to investigate the parameter
space of models.  We then calculate the particle physics input for the
relevant experiments, namely the injection spectrums for $e^+$,
$\bar{p}$, $\gamma$ and $\nu$.  In Section~\ref{sec:ai} we discuss the
astrophysical inputs for our study.  We review the methods used in
this study for propagating the various particles from their source to
Earth.  Additionally, we further discuss the experimental inputs that
we use, and comment on many of the uncertainties associated in
calculating the fluxes for them.  In Section~\ref{sec:pb} we present
the resulting astrophysical fluxes calculated from our
particle physics module.  We demonstrate
how the various particle physics and astrophysics parameters affect
the predicted fluxes for $e^+$, $\bar{p}$, $\gamma$ and $\nu$.
Finally, in Section~\ref{sec:imp} we discuss the implications and
interpretations of the regions of parameter space that we find to be
consistent with the experiments studied.
In Appendix~\ref{app:box} we discuss the so called ``leaky box"
approximation which we use to estimate the positron background flux. 

%%%%%%%%%%%%%%%%%%%%%%%%%%%%%%%%%%%%%%%%%%%%%%%%%%%%%%
%%%%%%%%%%%%%%%%%%%%%%%%%%%%%%%%%%%%%%%%%%%%%%%%%%%%%%
\section{Unified Dark Matter Module}\label{sec:udm}
\setcounter{equation}{0} \setcounter{footnote}{0}
%%%%%%%%%%%%%%%%%%%%%%%%%%%%%%%%%%%%%%%%%%%%%%%%%%%%%%
%%%%%%%%%%%%%%%%%%%%%%%%%%%%%%%%%%%%%%%%%%%%%%%%%%%%%%

We are interested in understanding the bounds and predictions for
characteristic DM models that could be used to explain the excesses
observed.  In this work we do not focus on the bounds for a particular
model nor are we completely model independent.  Instead, we construct
a module that contains the most important components that we identify
from the particle physics perspective.  This module can then be
appropriately recast to reflect the predictions from a wide array of
models that have been, and inevitably will be, built.

We construct the dark matter module loosely in accord with the Unified
Dark Matter model of Arkani-Hamed {\it et al.}\cite{udm}.  This model
offers the intriguing possibility to describe not only the PAMELA and
ATIC/PPB-BETS excesses, but also the INTEGRAL~\cite{integral} and DAMA~\cite{dama} excesses
simultaneously.  We treat~\cite{udm} as representative of a class of
ideas, and choose to incorporate those features of~\cite{udm} that are
relevant to studying the experimental consequences for indirect
searches in the high energy $e^+$, $\bar{p}$ and $\gamma$ channels.

\begin{figure}[htbp] %  figure placement: here, top, bottom, or page
   \centering
   \includegraphics[width=3in]{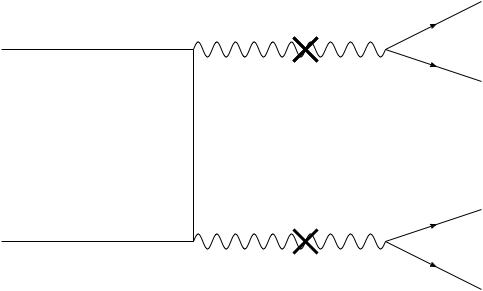} 
   \caption{An example of $\chi$ annihilation,
     $\chi\chi\rightarrow\phi\phi\rightarrow e^+e^-e^+e^-$ via the
     mixing of $\phi$ and $\gamma$.}
   \label{fig:ann}
\end{figure}

The components we choose to include in our module for explaining the
excesses are the following.  We assume there are heavy DM particle(s),
$\chi$(s), that are charged under some ``dark" gauge group, and possibly
the SM electroweak (EW) gauge groups as well.  Additionally, the dark gauge
group is broken and therefore the sector consists of light gauge bosons
which we collectively refer to as $\phi$.  The light gauge bosons are
required for two reasons: On the one hand, they allow for a
kinematical explanation for the electron but no antiproton excess
measured by the above experiments.
On the other hand, they play part in the Sommerfeld mechanism that
can enhance the usual thermal WIMP cross-section to the required rate
today. With this in mind,  we allow the $\chi$'s to annihilate either through
the SM EW gauge bosons, $V=(W,Z)$, or the $\phi$'s:
\begin{equation}\label{eq:ann}
\chi\chi\rightarrow V V \;\;\mathrm{or}\;\; \chi\chi \rightarrow \phi\phi.
\end{equation}
If $\chi$ annihilates through SM gauge bosons, then the final states
are clear.  The annihilation into $\phi$ needs further explanation.
If one assumes that there are measurable consequences in experiments
then $\phi$ ultimately needs to decay into SM final states.  There are
two possibilities that allow $\phi$ to decay.  Either the SM matter
fields are charged under the dark gauge group, in which case $\phi$
can decay into SM fields based on their charge assignments.
Otherwise, the $\phi$ gauge bosons can mix with the SM gauge bosons
and thereby decay through this mixing.  While either scenario is
possible in principle, we will choose the latter and couple $\phi$ to
the SM matter through gauge boson mixing.  On general grounds the
vector $\phi$ will decay back to SM states by mixing through the
photon or the Z boson. However, given the lightness of $\phi$, the
decays going through Z mixing will be further suppressed by at least
$m_{\phi}^{2}/m_{Z}^{2}$. Unless the $\gamma$-$\phi$ mixing is much
smaller than the Z-$\phi$ mixing, one can then assume that $\phi$ couples
to SM particles proportionally to their electric charge. Hence the
$\phi$ decay branching fractions are completely determined by its
mass.  An example annihilation is shown in Figure~\ref{fig:ann}.

With these considerations we have come up with a minimal module that
has five particle physics parameters:
\begin{itemize}
\item $m_\chi$ -  sets the mass scale for the DM annihilation.
\item $m_\phi$ - determines what states $\phi$ can decay into and
  the kinematics of the decay products.
\item $\adm$ - the strength of the gauge coupling in the dark
  sector.
\item $\langle \sigma v \rangle\equiv \langle \sigma
  v\rangle_{VV}+\langle \sigma v\rangle_{\phi\phi}$  - a free parameter
  for the overall cross section.
\item $R_{SM} = \frac{\langle \sigma v\rangle_{VV}}{\langle \sigma v
    \rangle}$ -  a free parameter for the relative contributions of the
  annihilations into the SM vs dark gauge bosons.
\end{itemize}
While some of these parameters may seem redundant, they allow us to
cover the parameter space of a large number of models without having
to calculate within each model separately.  In particular, the inclusion of
$\adm$ as a separate parameter is noticeable as we keep $\langle
\sigma v \rangle$ and $R_{SM}$ free.

Once we stipulate that $\phi$ is a dark gauge boson, we need to allow
for the possibility that the dark gauge group is nonabelian.  In fact
this is exactly what is desired in~\cite{udm} to explain DAMA and
INTEGRAL anomalies.  If $\phi$ represents collectively the gauge bosons of a
nonabelian group, then there are additional processes for annihilation
into the SM compared to those shown in Figure~\ref{fig:ann}.
\begin{figure}[htbp] %  figure placement: here, top, bottom, or page
   \centering
   \includegraphics[width=3in]{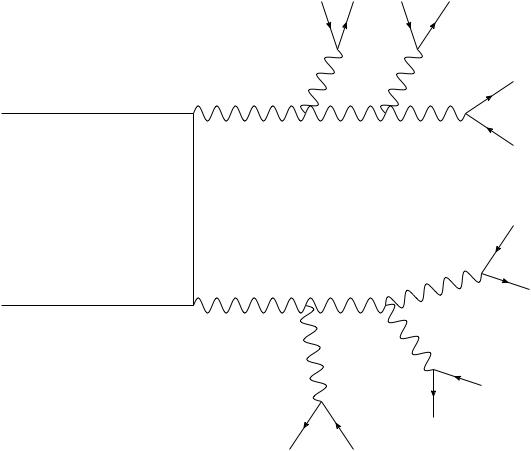} 
   \caption{An example of $\chi\chi$ annihilation, with the inclusion
     of the dark sector shower.}
   \label{fig:sho}
\end{figure}
Just like in QCD once the gauge bosons are produced, they can shower
and split into new gauge bosons.  We give an example of this process
in Figure~\ref{fig:sho}, where after showering the dark gauge boson subsequently decays into SM final states through photon mixing.
This process has not been quantitatively explored before in this
context and we demonstrate the effect in later sections.  As we shall see, while
naively there is no large enhancement as the $\phi$'s are 
massive, a significant change in the resulting energy spectrum arises since we assume
$m_\chi \gg m_\phi$.

There additionally could be another annihilation channel for $\chi$, i.e. 
$\chi\chi\rightarrow Z\phi$.  This is strongly dependent on the model,
and could in principle be used as a separate parameter in our module.
For instance this mode will not occur with any appreciable rate in
models where $\chi$ is a single Majorana particle. We choose not to include
this as a separate parameter, and instead one can infer bounds on this
mode from our $R_{SM}$ appropriately rescaled.

%%%%%%%%%%%%%%%%%%%%%%%%%%%%%%%%%%%%%%%%%%%%%%%%%%%%%%
%%%%%%%%%%%%%%%%%%%%%%%%%%%%%%%%%%%%%%%%%%%%%%%%%%%%%%
\subsection{Calculation of Injection Spectrum}\label{sec:udmdndx}
%%%%%%%%%%%%%%%%%%%%%%%%%%%%%%%%%%%%%%%%%%%%%%%%%%%%%%
%%%%%%%%%%%%%%%%%%%%%%%%%%%%%%%%%%%%%%%%%%%%%%%%%%%%%%

In this section we calculate the particle physics input for all dark
matter indirect experiments.  For the experiments that we are
interested in, we simply need the injection spectrum of $e^+,\bar{p}$, $\gamma$, and $\nu$ coming from the annihilation of the $\chi$'s.  Therefore we
need to calculate the inclusive annihilation of $\chi\chi$ at
threshold and then extract the energy distribution, $dN/dE$, for each particle we are interested
in.

To calculate $dN/dE$ we implement our module in several MC programs
and scan over the particle physics parameters $m_\chi$,$m_\phi$ and
$\adm$ that define it.  We do {\em not} calculate $\langle
\sigma v\rangle$ and $R_{SM}$ from first principles, instead we will
{\em fit} the experimental data for PAMELA and ATIC/PPB-BETS to fix
these parameters.  The result of this fit will then dictate what boost
factor is needed compared to a standard thermal WIMP annihilation
cross section.  If one calculated the Sommerfeld effect within a model,
this would constrain the values of the masses and gauge couplings
given the necessary cross section.  However, since we do not wish to
focus on one particular model alone, we do not require this consistency
check.  Additionally once the mass and $\adm$ are given, a model could
predict the ratio $R_{SM}$.  However, $R_{SM}$ can also be an independent
of $\adm$, for instance if the DM fields $\chi$ are singlets under the
SM and mixed through a Yukawa coupling.

To implement the module in event generators we make the specific
choice of $SU(2)$ for the dark gauge group, and we define $\chi$ to be
a bi-doublet under the SM $SU(2)$ and the hidden $SU(2)$.  This choice
{\em only} affects the dark gauge boson parton shower directly, since
$\langle \sigma v\rangle$ and $R_{SM}$ are fit to data.  However,
since we will scan over $\adm$, a different choice of coupling can still be
be used to approximately interpolate amongst different Casimirs and extrapolate the results to
different gauge groups.  

The parameter space we choose to cover is
shown in Table~\ref{tab:params}.  Our choice of whether to implement a
parton shower in the dark sector is a binary one.  This allows us to
cover the case where the dark gauge group is $U(1)$ and the gauge
bosons don't shower.
\begin{table}[htdp]
\begin{center}
\begin{tabular}{|c|c|c|c|c|}
  \hline
  No dark sector parton shower &\\
  \hline
  $m_\chi$ & 200 GeV - 3.5 TeV\\
  $m_\phi$ & 200 MeV, 500 MeV, 1.2 GeV, 8 GeV, 15 GeV\\
  $\adm$ & arbitrary\\
  \hline
  Dark sector parton shower & \\
  \hline
  $m_\chi$ & 200 GeV - 3.5 TeV (discrete grid)\\
  $m_\phi$ & 200 MeV, 500 MeV, 1.2 GeV, 8 GeV, 15 GeV\\
  $\adm$ & $10^{-3},10^{-2},4\cdot10^{-2},10^{-1}$\\
  \hline
\end{tabular}
\end{center}
\caption{Parameters scanned in the DM module.}
\label{tab:params}
\end{table}

To implement the decays of the $\phi$ particle we use different
effective field theories depending on the mass of $\phi$ that we are
interested in.  When $m_{\phi}\gg \Lambda_{QCD}$, we coupled
$\phi$ directly to the quarks at parton level, which can then be
showered and hadronized. On the other hand if \mbox{$m_{\phi}
  \lesssim$~a few GeV},  this procedure is a bad
approximation. In this case we couple $\phi$ directly to mesons.  We
take the corresponding cross sections from experimental data of
$e^{+}e^{-}\rightarrow hadrons$ exclusive processes
\cite{expeehad}. In particular for the 500~MeV case, we assume $\phi$
decays to $e^{+}e^{-},\,\mu^{+}\mu^{-},\pi^{+}\pi^{-}$ and the ratio
is approximately 2:2:1 as shown in Table~\ref{tab:bf}. For the 1.2~GeV
case we include other mesonic 2-body modes as well as 3- and 4-body
decays that are non-negligible.  We did not implement any other
intermediate mass between 1.2~GeV and 8~GeV because multi-body final
states become increasingly important and there is not enough experimental
information to completely reconstruct the exclusive final states. The
8~GeV and 15~GeV were chosen to have $m_{\phi}$ separated enough from
the quarkonia resonance region (where the hadronization model we use
from Pythia~\cite{pythia} suffers
from large uncertainties) and be above and below the $b\bar b$
threshold.  We catalogue the decay modes implemented and branching
fractions of $\phi$ in Table~\ref{tab:bf}.

\begin{table}[htdp]
\begin{center}
$\begin{array}{cc}
%\begin{array}{c}
\begin{tabular}{|c|c|c|c|c|}
  \hline
  $m_\phi$ & Mode & BF \\
  \hline
  200 MeV & $\phi\rightarrow e^+e^-$ & 1 \\
  \hline
  500 MeV &  $\phi\rightarrow e^+e^-$ & $4\cdot 10^{-1}$ \\
  &  $\phi\rightarrow \mu^+\mu^-$ & $4\cdot 10^{-1}$ \\
  &  $\phi\rightarrow \pi^+\pi^-$ & $2\cdot 10^{-1}$ \\
  \hline
  1.2 GeV &  $\phi\rightarrow e^+e^-$ & $3.4\cdot 10^{-1}$ \\
  &  $\phi\rightarrow \mu^+\mu^-$ & $3.3\cdot 10^{-1}$ \\
  &  $\phi\rightarrow \omega\pi^0$ & $7.9\cdot 10^{-2}$ \\
  &  $\phi\rightarrow \pi^+\pi^-\pi^0\pi^0$ & $7.5\cdot 10^{-2}$ \\
  &  $\phi\rightarrow \pi^+\pi^-$ & $6.4\cdot 10^{-2}$ \\
  &  $\phi\rightarrow K^+K^-$ & $4.5\cdot 10^{-2}$ \\ 
  &  $\phi\rightarrow \pi^+\pi^+\pi^-\pi^-$ & $4.1\cdot 10^{-2}$\\
  &  $\phi\rightarrow \pi^+\pi^-\pi^0$ & $2.4\cdot 10^{-2}$\\
  &  $\phi\rightarrow K^0\bar{K}^0$ & $5\cdot 10^{-3}$ \\
  \hline
  \end{tabular}
%  \end{array}
&
\begin{tabular}{|c|c|c|c|c|}
  \hline
  $m_\phi$ & Mode & BF \\
  \hline
  8 GeV &  $\phi^+\rightarrow e^+e^-$ & $1.6\cdot 10^{-1}$\\
  &  $\phi\rightarrow \mu^+\mu^-$ & $1.6\cdot 10^{-1}$ \\
  &  $\phi\rightarrow \tau^+\tau^-$ & $1.6\cdot 10^{-1}$ \\
  &  $\phi\rightarrow u\bar{u}$ & $2.1\cdot 10^{-1}$\\
  &  $\phi\rightarrow d\bar{d}$ & $5.2\cdot 10^{-2}$\\
  &  $\phi\rightarrow c\bar{c}$ & $2.1\cdot 10^{-1}$\\
  &  $\phi\rightarrow s\bar{s}$   & $5.2\cdot 10^{-2}$\\
  \hline
  15 GeV &  $\phi^+\rightarrow e^+e^-$ & $1.5\cdot 10^{-1}$\\
  &  $\phi\rightarrow \mu^+\mu^-$  & $1.5\cdot 10^{-1}$ \\
  &  $\phi\rightarrow \tau^+\tau^-$   & $1.5\cdot 10^{-1}$ \\
  &  $\phi\rightarrow u\bar{u}$  & $2.0\cdot 10^{-1}$\\
  &  $\phi\rightarrow d\bar{d}$  & $5.0\cdot 10^{-2}$\\
  &  $\phi\rightarrow c\bar{c}$ & $2.0\cdot 10^{-1}$\\
  &  $\phi\rightarrow s\bar{s}$ & $5.0\cdot 10^{-2}$\\
  &  $\phi\rightarrow b\bar{b}$  & $4.8\cdot 10^{-2}$\\
  \hline
\end{tabular}
\end{array}
$
\end{center}
\caption{Branching Fractions for $\phi$.  The values for $0.5$ and $1.2$
  GeV are extracted from experimental data 
  for exclusive $e^++e^-\rightarrow hadrons$ processes \cite{expeehad}, while for 8 and 15 GeV are computed using BRIDGE.}
\label{tab:bf}
\end{table}

To calculate the injection spectrums using existing Monte Carlo (MC) tools
is quite difficult.  The kinematic regime we study is based on very
heavy particles annihilating through very light particles, that
subsequently decay.  In this
regime most MC generators that we have used have difficulties.  To
generate our injection spectrums we were forced to use a variety of generators linked
together depending on the task: MadGraph/MadEvent~\cite{madgraph},
BRIDGE~\cite{bridge}, SHERPA~\cite{sherpa}, and Pythia~\cite{pythia}.

For the SM annihilations $\chi\chi\rightarrow VV\rightarrow \mathrm{fermions}$, we generate parton
level events keeping spin correlations.  We then shower and hadronize,
including the effects of photons that are showered from the $W$ gauge
bosons~\cite{wpho}.  To generate
$\chi\chi\rightarrow\phi\phi\rightarrow SM$ without the dark sector
parton shower we again generate parton level events including spin
correlations. We then shower and hadronize either the fundamental
particles, or just shower the charged mesons and their decay products.
Unlike~\cite{bergstrom}, when applicable, we include the
$\mathcal{O}(1)$ effects of calculating photons showered from muon decays.

For the case when we include the dark sector parton shower, we first
need to calculate the massive vector boson splitting function for the
$SU(2)$ case that we have implemented.  Given the kinematics that we
scan over, $m_\chi \gg m_\phi$, the effects of the massive splitting
function are minor and it is sufficient to use the massless splitting
functions,
\begin{equation}\label{masslesssplit}
P_{\phi\rightarrow\phi\phi}(z)=\frac{\adm}{2\pi}\frac{2 (1-z(1-z))^2}{z (1-z)}.
\end{equation} 
Nevertheless in our computation we include the complete massive splitting
function which only differs by the inclusion of
another term that is subdominant over most of our kinematic range.  To
generate events including the dark sector parton shower, we compute
the $2\rightarrow 2$ matrix element.  We
then shower in the dark sector and decay the $\phi$ keeping some of
the spin correlations.  Finally, we shower and hadronize the SM particles
that come from the $\phi$ decays, including the effects listed previously.

\begin{figure}[htbp] %  figure placement: here, top, bottom, or page
   $\begin{array}{cc}
   \includegraphics[width=3in]{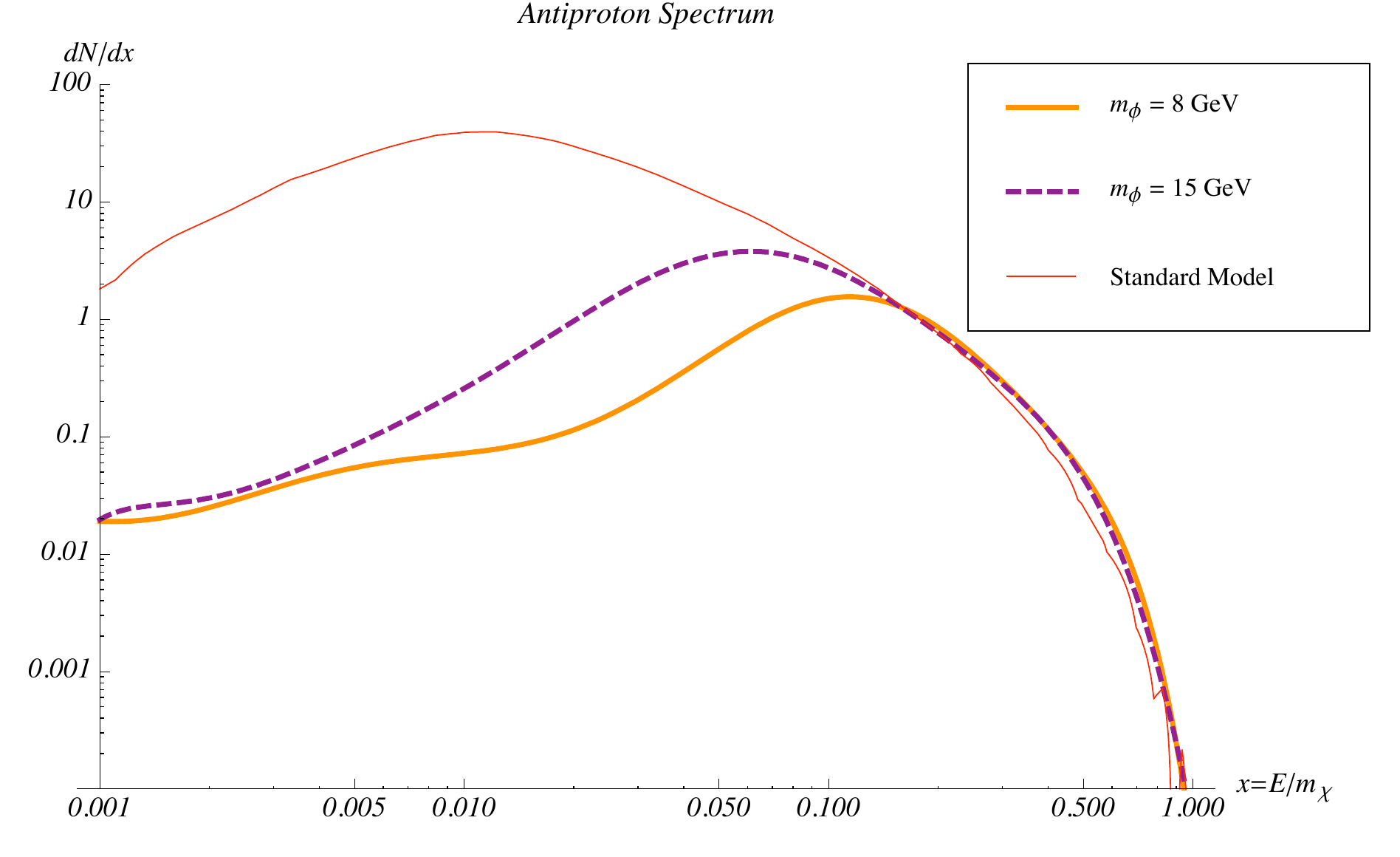}&\includegraphics[width=3in]{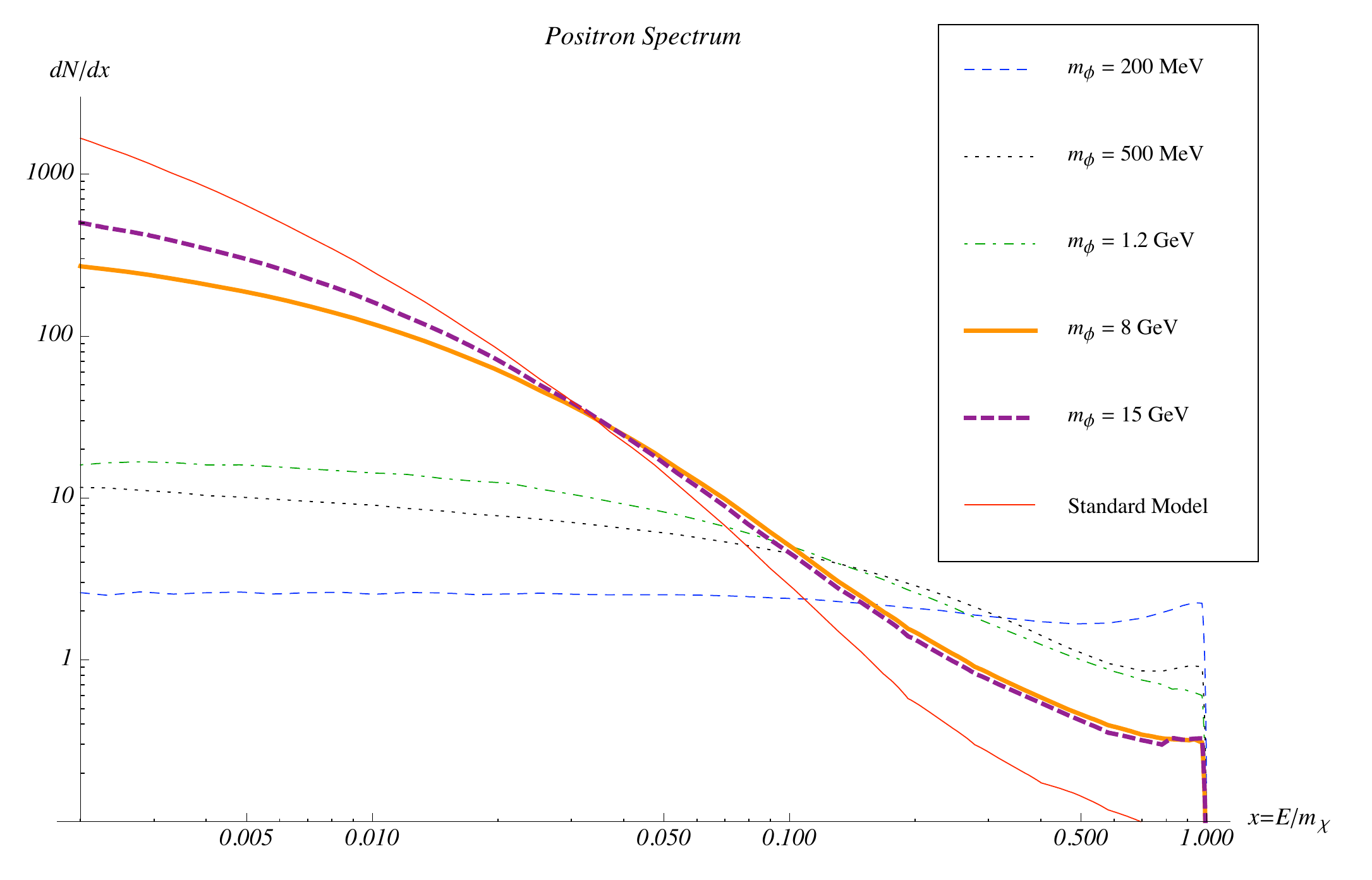}\\
   \includegraphics[width=3in]{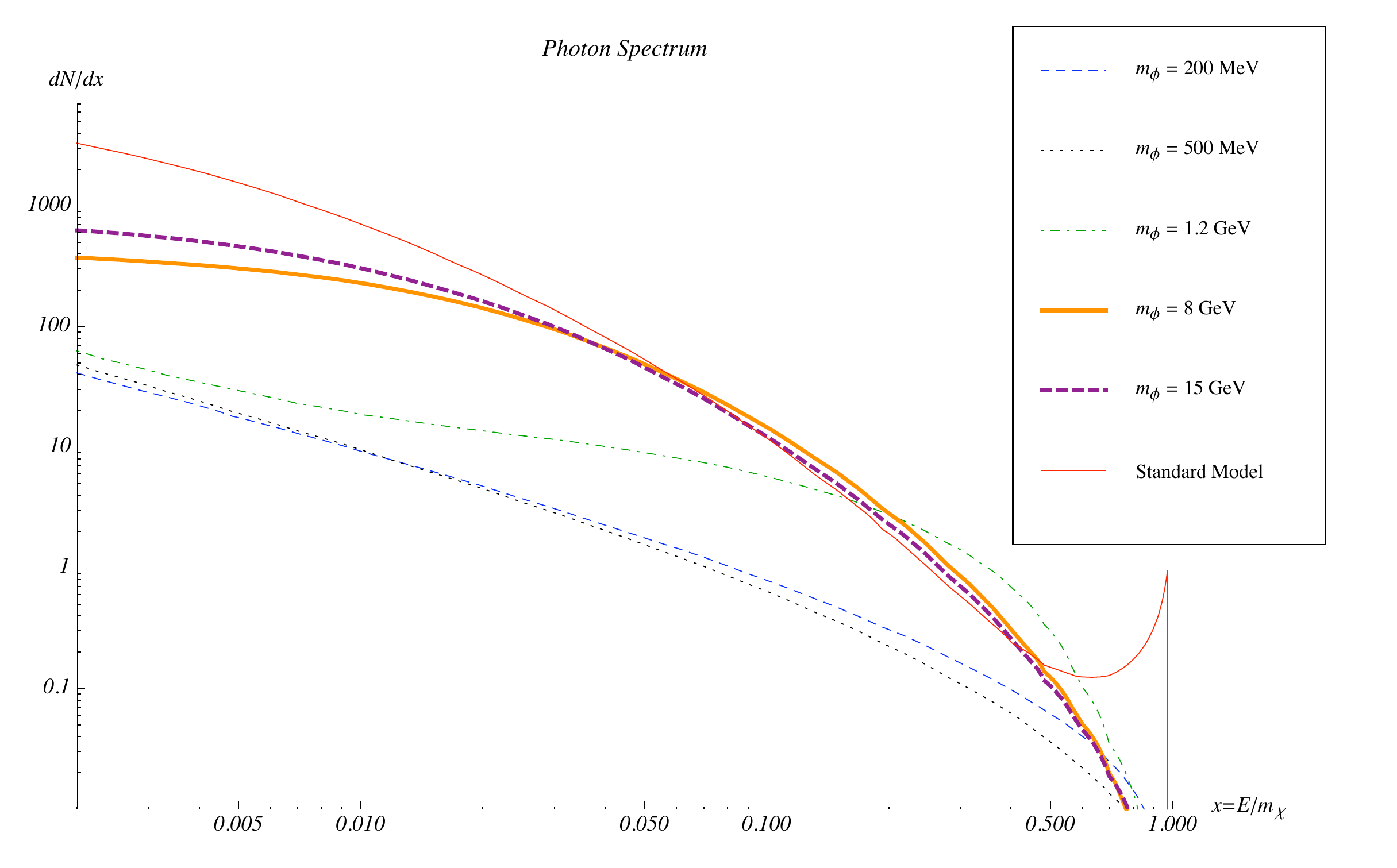}&\includegraphics[width=3in]{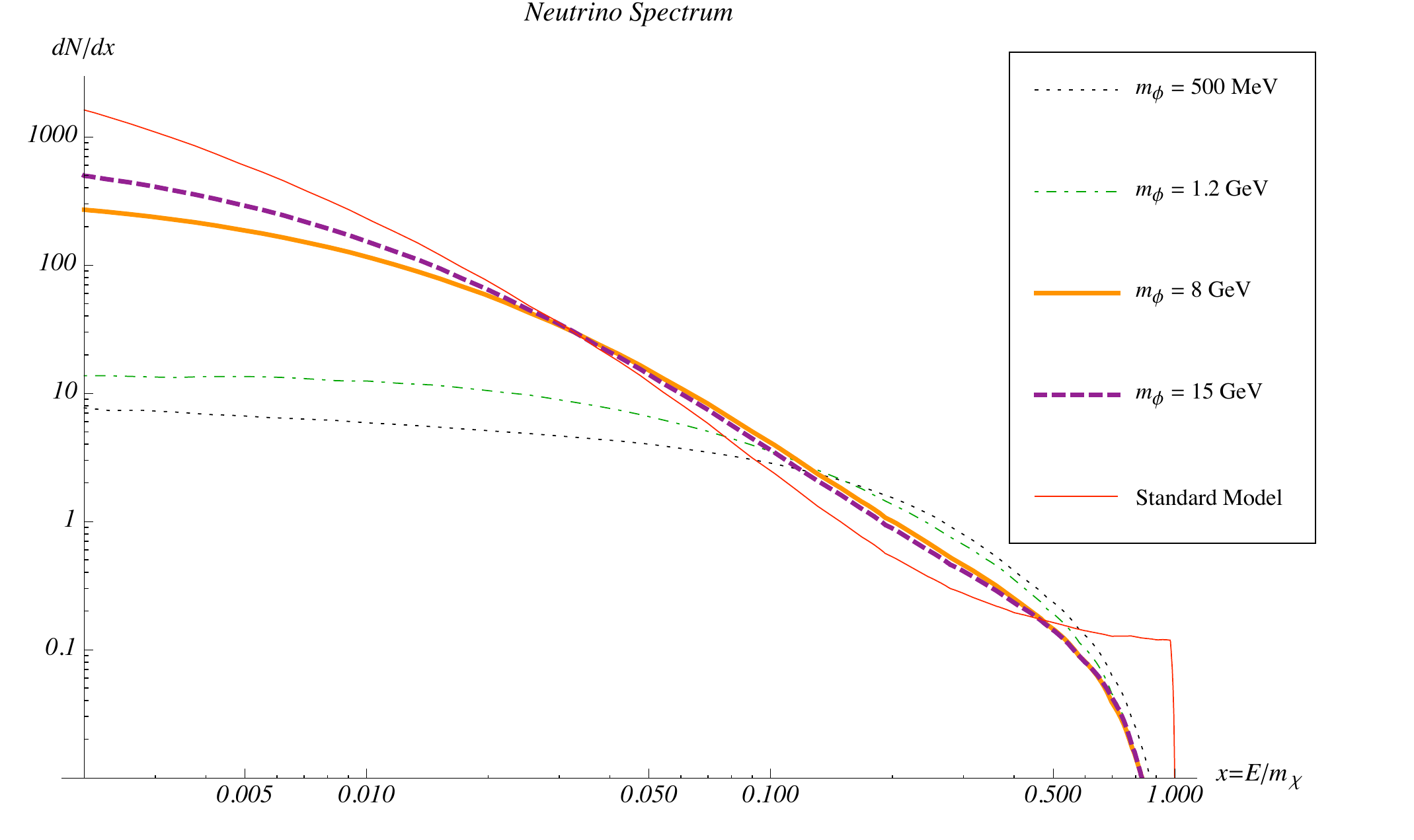}\\
   \end{array}$
   \caption{$dN/dx$ for $\bar{p},e^+,\gamma,$ and $\nu$ for the
     various $m_\phi$ used.  The SM contribution from
     $\chi\chi\rightarrow VV$ is also shown.  In the range of $x$
     plotted we never reach $\Lambda_{QCD}/m_\chi$ thus the shapes are
     universal and independent of $m_\chi$, even for the antiprotons.}
   \label{fig:dndx}
\end{figure}

\begin{figure}[t] 
  \centering
  $\begin{array}{cc}
    \includegraphics[width=3in]{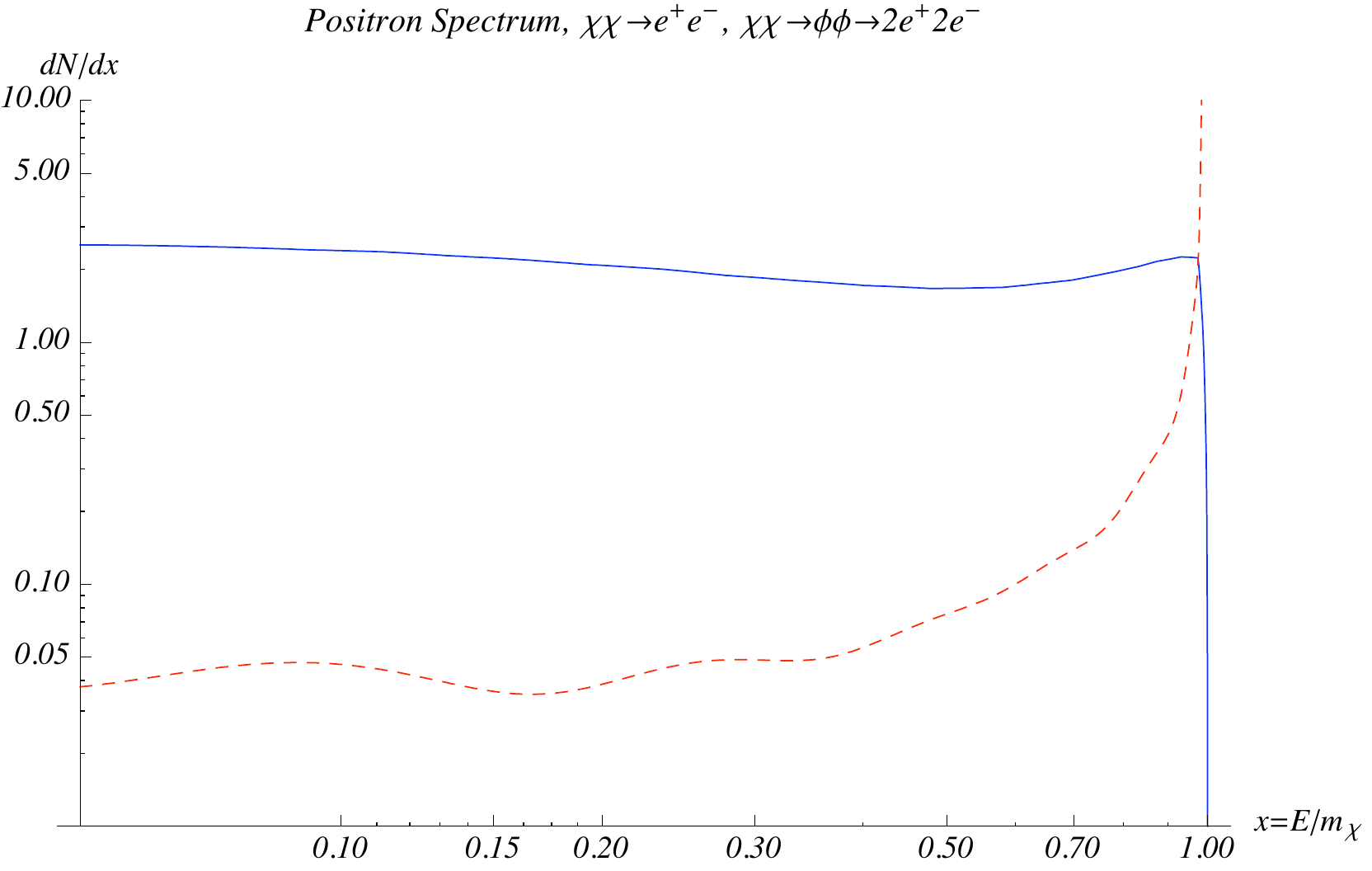}&\includegraphics[width=3in]{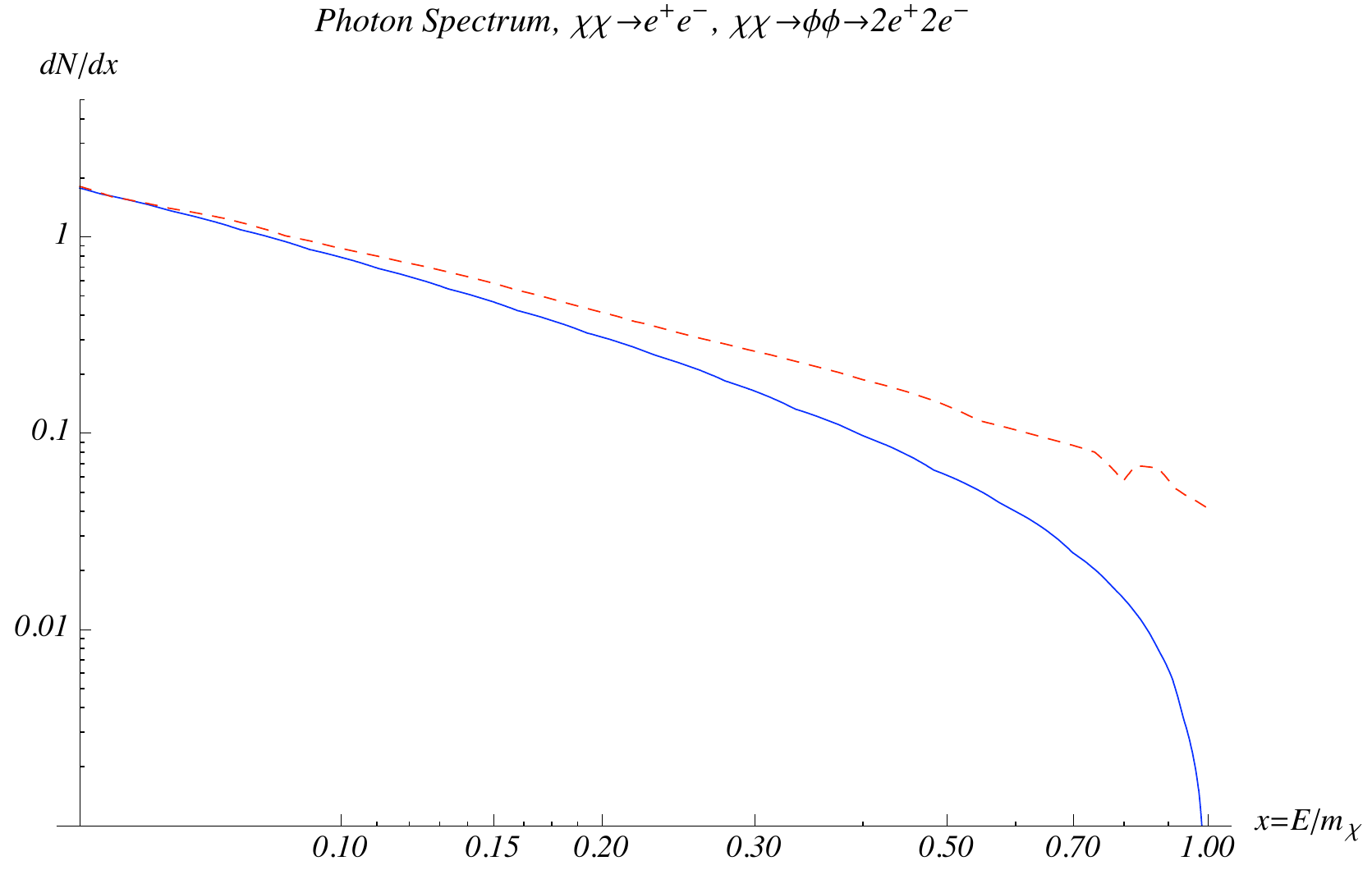}
  \end{array}$
  \caption{Photon and positron spectrum for the two body decay
    $\chi\chi\rightarrow e^+e^-$ (dashed red line).  For comparison, we also plot the
    four body decay $\chi\chi\rightarrow2\phi\rightarrow 2e^+2e^-$
    (solid blue line).}
  \label{fig:EE} 
\end{figure}

In Figure~\ref{fig:dndx} we plot $dN/dx$, where $x=E/m_\chi$, for
$\bar{p},e^+,\gamma,$ and $\nu$ for different $m_\phi$ with no parton
shower included.  We additionally include the SM annihilation channel
$\chi\chi\rightarrow VV$.  For later reference, we also plot in
Figure~\ref{fig:EE} the photon and positron spectrum for the two body decay $\chi\chi\rightarrow
e^+e^-$.  Both the SM annihilation channels will be used  below for comparison with different models
and to study the constraints on annihilating though the SM gauge
bosons.  As one can see, the larger the number of open decay channels for
$\phi$, the larger the population of low energy particles, namely the
softer the spectrum.  One can
see that in the $8$ GeV, $15$ GeV and SM gauge boson cases, showering
and hadronization effects from QCD  induce several orders of
magnitude increase in $dN/dx$.  Moreover, showering effects also
softens the spectrum of the direct decay into $e^+e^-$, as is clear
from Figure~\ref{fig:EE}.  

\begin{figure}[ht] %  figure placement: here, top, bottom, or page
  $\begin{array}{cc}
    \includegraphics[width=3in]{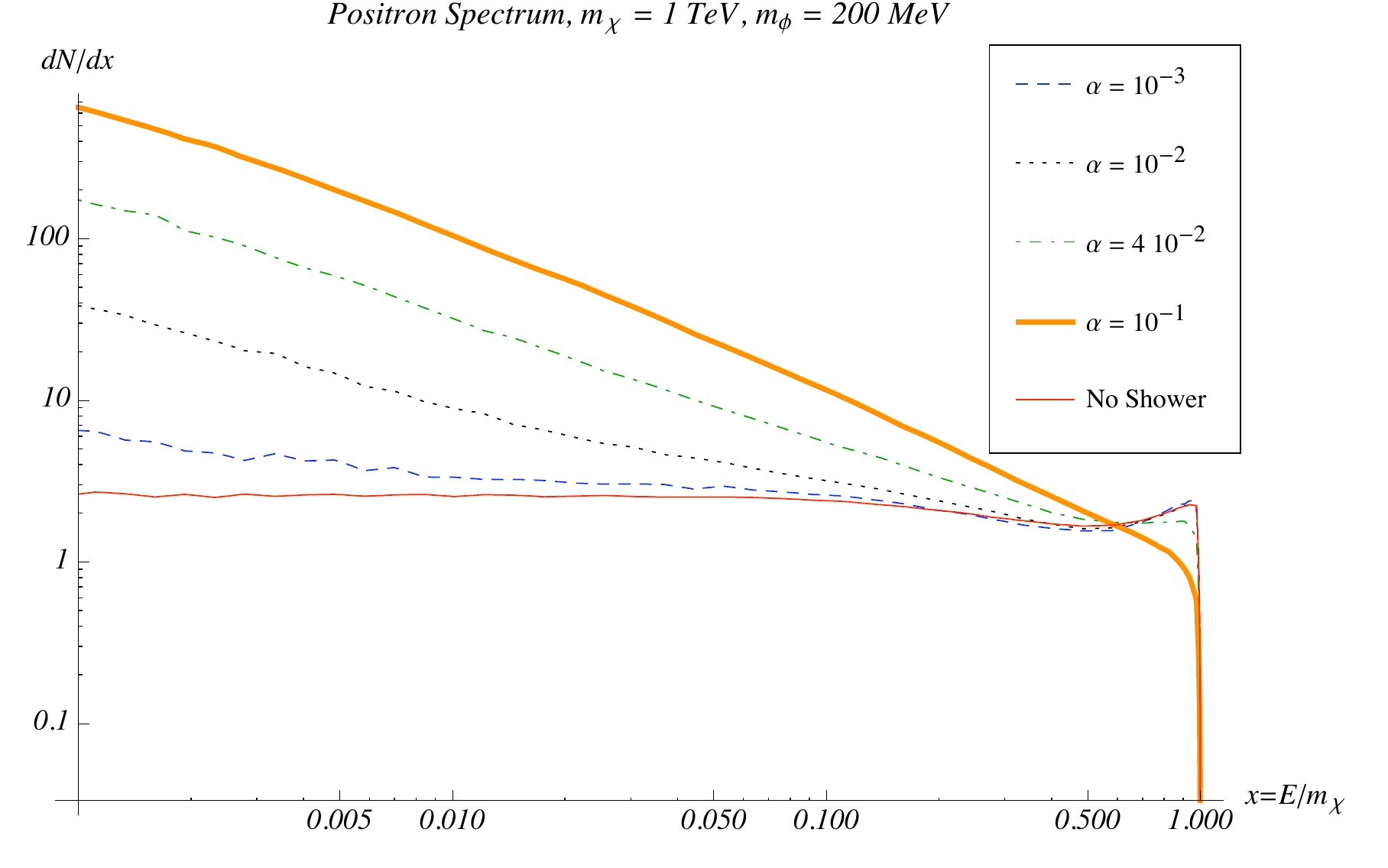}&\includegraphics[width=3in]{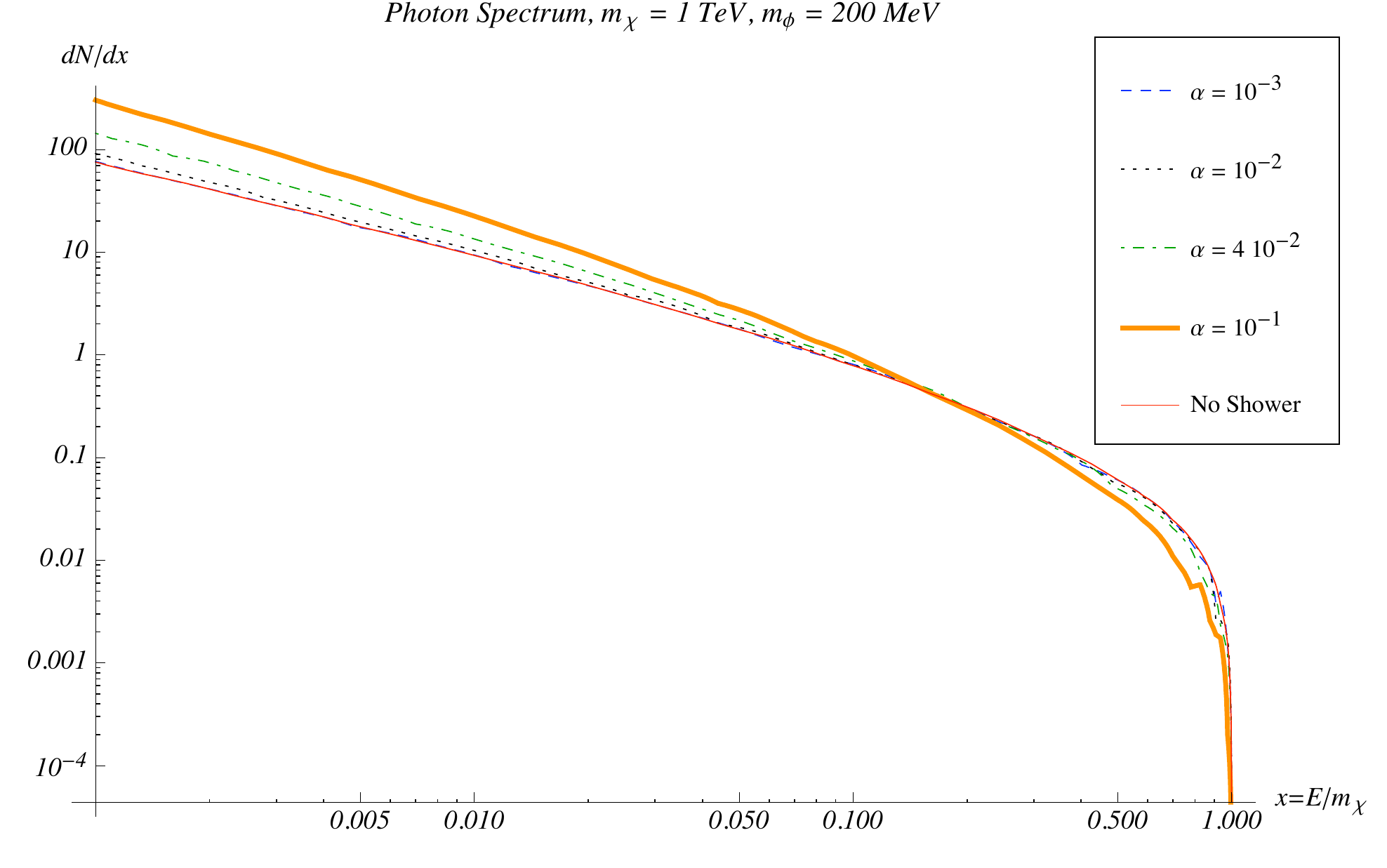}
\\ 
    \includegraphics[width=3in]{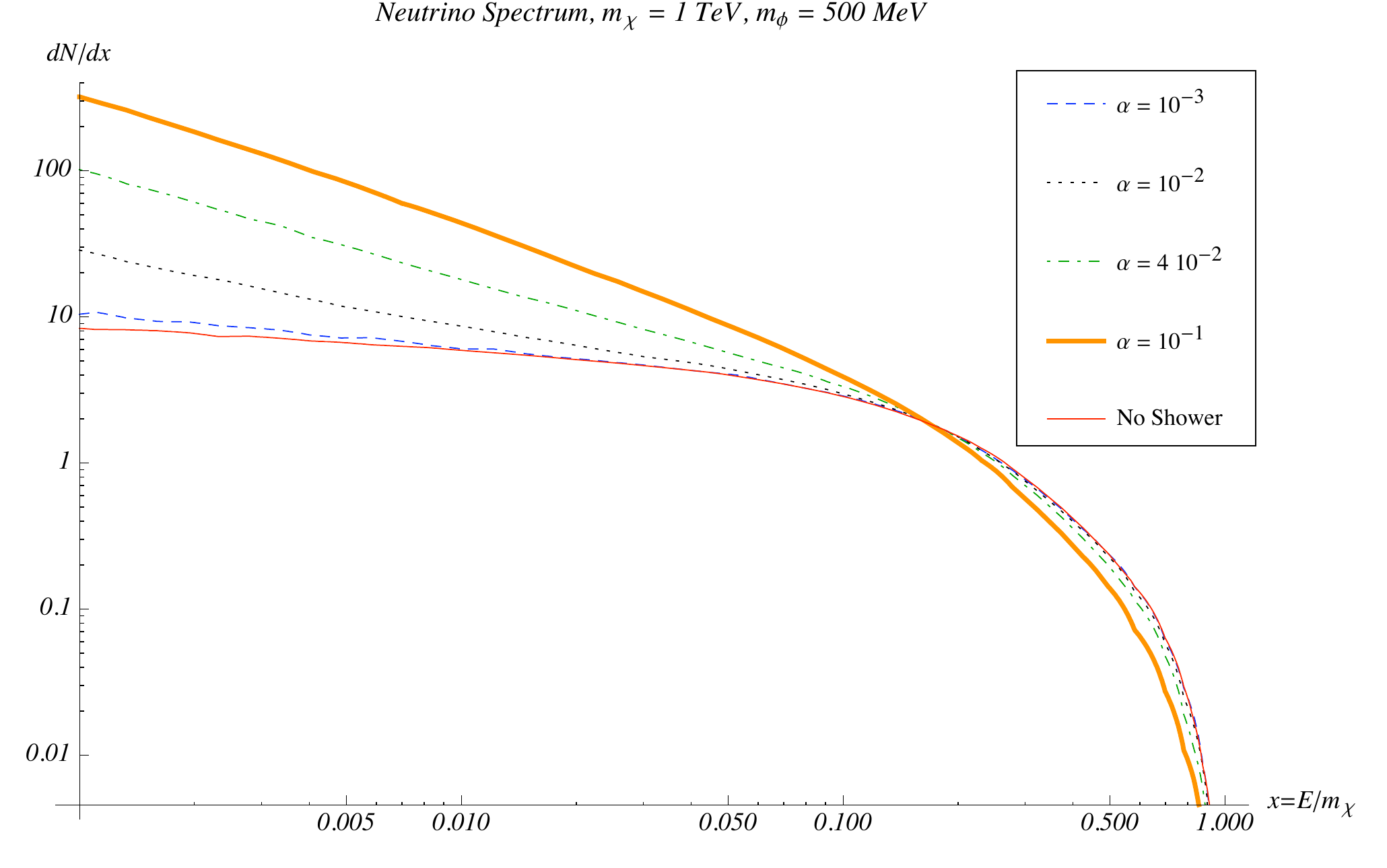}&\includegraphics[width=3in]{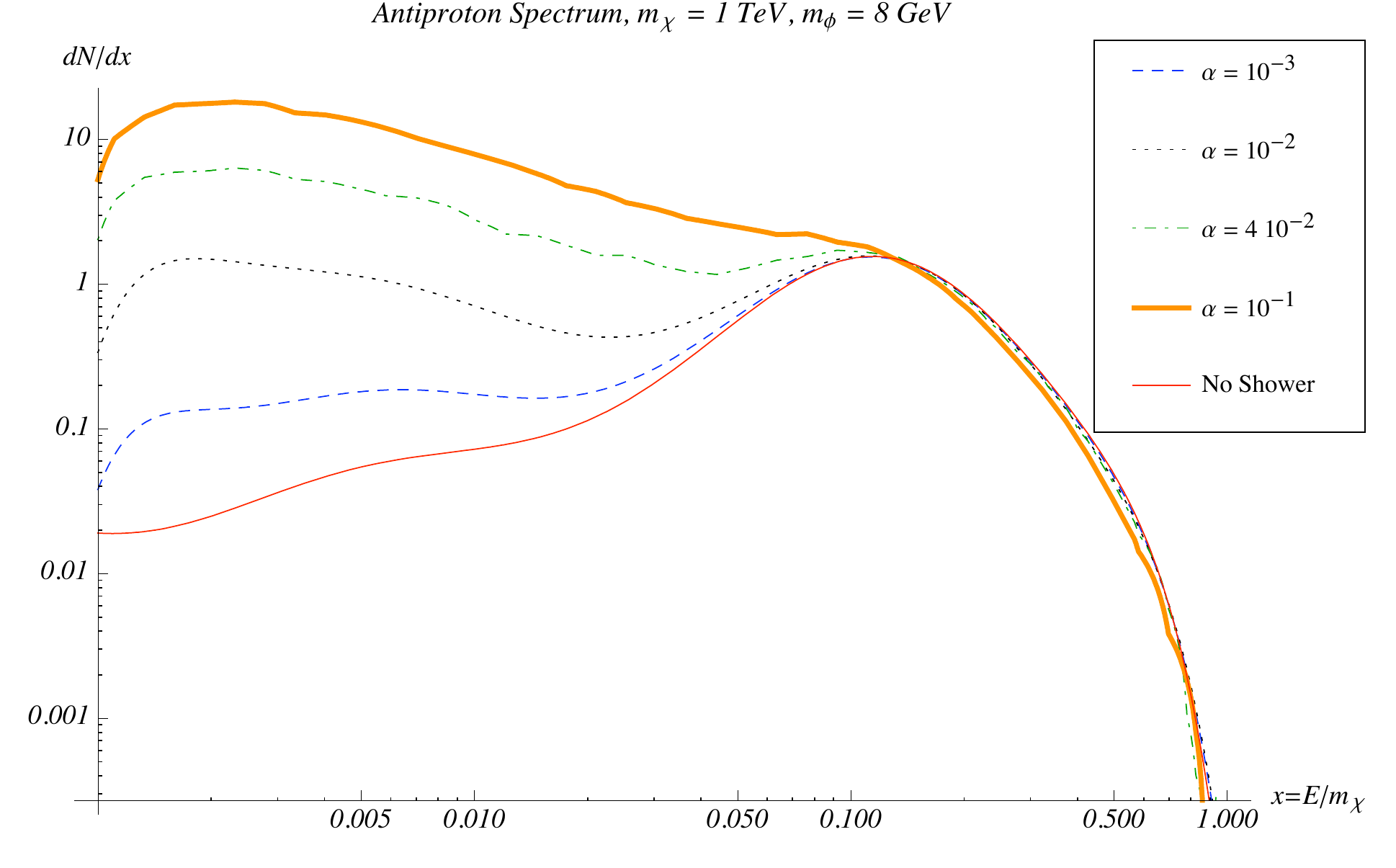}
\\
  \end{array}$
  \caption{The effect of varying $\adm$ over the range given
    in Table~\ref{tab:params}.  We plot $dN/dx$ for a given
    particle type with a fixed $m_\chi=1$ TeV and various $m_\phi$.  }
  \label{fig:dndxsho}
\end{figure}

In Figure~\ref{fig:dndxsho} we plot $dN/dx$, for $\bar{p},e^+,\gamma,$
and $\nu$ for different values of $\adm$ to illustrate the effects of
the dark sector parton shower.  Without showering, the spectrum, $dN/dx$, is
universal and independent of $m_\chi$, this is no longer true when
showering is taken into account.  The reason for this is that the
number of energy decades for showering depends on the ratio $m_\chi/m_\phi$.  Hence
in order to demonstrate the $\adm$ dependence, we fix the value of
$m_\chi$ and $m_\phi$.  One should note that showering can have a
large effect on the $dN/dx$.  Hidden sector showering not only softens the spectrum via radiation, but also through additional $\phi$ decays.

\subsubsection{Particle Physics Uncertainties}\label{sec:punc}

In this section we review some of the uncertainties in calculating the
$dN/dE$ for $e^+$, $\bar{p}$, $\nu$ and $\gamma$.  Most channels are
rather clean from the particle physics perspective, but there are
several issues that can have a significant effect on the $dN/dx$ that
we compute.  These issues are hadronic uncertainties in the
calculation of the  $dN/dx$ for $\bar p$, showering uncertainties which
change the photon spectrum and the effects associated with the possibility of having different matrix
elements which can affect all particles.

Hadronic uncertainties are due to the fact that once we have SM quarks
 we need to turn these partons into mesons and baryons
through hadronization/fragmentation.  While QCD can correctly describe
the parton showering evolution, the process of hadronization relies
upon phenomenological models that are tuned to data.  While the
effects of hadronization are important in some specific collider searches, when
discussing high energy jets at colliders, one rarely talks about
exclusive channels that label specific numbers of mesons and baryons.
On the other hand,  in the case of DM we look at the fully exclusive $\bar{p}$
channel. To hadronize we use the Pythia program.  Pythia employs
several possible fragmentation models for creating baryons such as
diquark, popcorn and advanced popcorn.  The Pythia manual~\cite{pythia},
states that the tuning of their baryon production models are done with
a global fit to the data and that the resulting fragmentation functions for individual baryons can be lower or higher
than the actual data.  To quantify this effect we used the Pythia
default popcorn algorithm, and compare it with the advanced popcorn
algorithm and SHERPA's AHADIC++ algorithm.  We find that there is an
$\mathcal{O}(1)$ effect in using different hadronization/fragmentation
algorithms for the $\bar{p}$ injection spectrum.

For showering, there are various levels of sophistication
that one may try to employ.  For instance in comparison
with~\cite{bergstrom} we include the effects of showering off the
decay products of the muon while they only include the shower from the
muon itself.  We find that this is an
$\mathcal{O}(1)$ effect over the entire range of energies that we
examine.  Additionally there are further uncertainties depending on
how the parton shower is implemented.  For instance, when SHERPA is used to compute the radiative decays of the vector $\phi$, taking into account the full matrix element structure for  $l^+l^- \gamma$, it 
tends to underpopulate the high energy gammas compared to the parton shower approximation.
Since the above effect is model dependent, we do not consider it here and instead, rely solely on the parton shower.
Nevertheless, one should keep in mind that this discrepancy induces another $\mathcal{O}(1)$ effect that may reduce or enhance our results.

As in the case of $\phi$ decays, differences between matrix elements can introduce a nontrivial $dN/dx$ shape
dependence which translates into different fluxes
observed in experiments.  This has been studied in the literature
e.g.~\cite{mdm}.  Including non trivial effects of spin
correlation, typically do not introduce large changes in the positron
spectrum~\cite{mdm}, since usually the spectrum changes only
at very high energy.  However, depending on the form of the positron
injection spectrum, this can lead to non trivial changes in the
region of interest for gamma ray searches.  While we only take the
matrix element for $2\rightarrow4$ assuming there is an intermediate
light gauge boson, one should keep in mind that there could be a factor of
a few uncertainty if the underlying model was different.

%%%%%%%%%%%%%%%%%%%%%%%%%%%%%%%%%%%%%%%%%%%%%%%%%%%%%% ''
%%%%%%%%%%%%%%%%%%%%%%%%%%%%%%%%%%%%%%%%%%%%%%%%%%%%%%
\section{Astrophysics Inputs}\label{sec:ai}
\setcounter{equation}{0} \setcounter{footnote}{0}
%%%%%%%%%%%%%%%%%%%%%%%%%%%%%%%%%%%%%%%%%%%%%%%%%%%%%%
%%%%%%%%%%%%%%%%%%%%%%%%%%%%%%%%%%%%%%%%%%%%%%%%%%%%%%

In this section we review the astrophysical propagation of $e^+$,
$\bar{p}$, $\nu$ and $\gamma$.  Due to a lack of theoretical
understanding and experimental data, propagation models suffer
from large uncertainties and there are various 
models that exist in the literature.  The choice of a model is not only
important for estimating the DM signal, but also for correctly
evaluating the backgrounds.  Using different models may result in
significantly different  predictions.

As an example, we could, in principle,  use a
program such as GALPROP~\cite{galprop} and calculate both the
background and signal propagation by using a best fit propagation
model.  However, we aim to be as data driven as possible and
introduce the minimal amount of theory necessary to estimate both the signal
and background.  We therefore use  semi-analytical propagation
method which we review below.  This approach has been
studied extensively in the literature \cite{Delahaye:2008ua,mmm}
and it therefore allows for a simple estimation of the underlying
uncertainties.  In later sections we will briefly revisit some of these
uncertainties. 

We can also employ a data-driven analytic approach to estimate the
backgrounds.  Whenever we can, we use known measurements or otherwise
conservative models for the evaluation of the backgrounds.  Below we
discuss the propagation model, the experimental data and the background
estimation for each of the relevant channels.  We discuss in each case
the uncertainties involved and explain how we take those into account
when fitting the predictions to the data.

%%%%%%%%%%%%%%%%%%%%%%%%%%%%%%%%%%%%%%%%%%%%%%%%%%%%%%
%%%%%%%%%%%%%%%%%%%%%%%%%%%%%%%%%%%%%%%%%%%%%%%%%%%%%%
\subsection{DM Halos}\label{sec:dm-halos}
%%%%%%%%%%%%%%%%%%%%%%%%%%%%%%%%%%%%%%%%%%%%%%%%%%%%%%
%%%%%%%%%%%%%%%%%%%%%%%%%%%%%%%%%%%%%%%%%%%%%%%%%%%%%% 

In this section we review the DM density profiles that we use as
inputs when calculating the fluxes from DM annihilation. 
Most of the dark matter profiles that we consider here are inferred
from N-body simulations.  Starting in the mid-nineties, a paradigm emerged~\cite{nfw}
where by examining the results of dark matter N-body simulations for
many different galaxies, a type of universality for the
density profiles appeared.  This led to the famous NFW profile~\cite{nfw} for
dark matter that is in common use today.  Since then other groups have
examined this universality and found similar results.  Nevertheless, there is
some disagreement between groups concerning how cuspy the dark matter
profile is at the center of the galaxy.  Most standard dark matter
profiles can be parameterized using the $(\alpha,\beta,\gamma)$
parametrization~\cite{abg}
\begin{equation}
\rho (r)=\rho_{\odot}\left[\frac{r_\odot}{r}\right]^\gamma
\left[\frac{1+(r_\odot/r_s)^\alpha}{1+(r/r_s)^\alpha}\right]^{(\beta-\gamma)/\alpha} .
\end{equation}
The two most commonly used profiles of this type are NFW~\cite{nfw}
and Moore~\cite{moore}\footnote{The Moore profile has been
  attributed to several values of $\alpha,\beta,\gamma$.  The
  original profile, often referred to as M99, is  very cuspy and has
  $\gamma=1.5$.  Sometimes this is used in dark matter calculations and it artificially
  inflates the rate of annihilations in the center of the Galaxy.
  While there is no longer evidence for this cuspy of profile, a
  more modern study by Moore {\em et al.}~\cite{moore2} found a best
  fit with $\gamma=1.2$ which is usually denoted as Moore2004.}.  Additionally the Isothermal
profile~\cite{isothermal} also fits into this parameterization, but it
is based on older approximations for dark matter halos and there is no
evidence for it in N-body simulations.  While we do not consider the Isothermal profile viable, we show it below for the sake of comparison.  
In
(\ref{abgs}) we list the values of $(\alpha,\beta,\gamma,r_s)$ that
correspond to the commonly used profiles.
\begin{equation}\label{abgs}
\begin{array}{|c|c|c|c|c|}
\hline
\mathrm{Profile Name} & \alpha & \beta & \gamma & r_s (\mathrm{kpc}) \\
\hline
 \mathrm{NFW} & 1 & 3 & 1 & 20 \\
 \mathrm{Moore} & 1 & 3 & 1.16 & 30\\
 \mathrm{Isothermal} & 2 & 2 & 0 & 5\\
 \hline
 \end{array}
 \end{equation}
 Recently N-body simulations have been able to increase their
 resolution by using $\mathcal{O}(10^9)$ particle simulations.  From
 this, one is able to test the validity of the commonly used profiles
 such as NFW.  It turns out that, as the resolution has increased and
 one has been able to probe further and further in towards the center
 of the Galaxy, the cuspiness does not persist.  This has
 led~\cite{einasto0} to propose that, instead of the commonly used
 profiles in~(\ref{abgs}), the best fit profile is the so called Einasto
 profile~\cite{einasto}\footnote{This is sometimes referred to as a
   Merritt profile in the literature~\cite{merritt}.}:
 \begin{equation}
 \rho(r)=
 \rho_\odot\exp\left[\frac{-2}{\alpha}\left(\left(\frac{r}{r_s}\right)^\alpha-1\right)\right]. 
 \end{equation}
 What determines the deviation from the power law behavior is the
 Einasto $\alpha$ parameter which was found in~\cite{einasto0} to have
 a mean value of $0.172$ when fitting to simulations based on a few
 hundred million particles carried out by the Virgo consortium.
 Recently billion particle simulations have been carried out.  The
 Aquarius~\cite{einastoevidence}, Via Lactea~\cite{vialactea}, and
 GHALO~\cite{darkheart} simulations all fit the Einasto
 profile and $\alpha$ has been found to lie in the range
 $[0.142,0.177]$.  To be conservative, we choose a larger range of $\alpha$ consistent with~\cite{einasto0}.  Below, we will examine $\alpha=0.12,0.17,0.20$.

 The cuspiness of the profile is not the only important feature, the
 local density is obviously crucial as well.  Unfortunately our local
 density can not be pinned down very well from N-body
 considerations alone since the simulations do not include baryons.
 The value of the local density used in most dark matter detection
 studies is $\rho_{\odot} \simeq 0.3
 \mathrm{\ GeV}/\mathrm{cm}^3$~\cite{localdensity,pdg}, however this is
 based upon an older understanding of potential DM density profiles.
 A more recent study that has attempted to fix this value more
 carefully and construct a probability distribution for $\rho_\odot$,
 finds a canonical value of $\rho_{\odot} \simeq 0.4
 \mathrm{\ GeV}/\mathrm{cm}^3$ that can vary by up to a factor of
 $\sim10$, but is likely to be less not more than a factor of
 2~\cite{localdensityrecent}.  While the N-body simulations can not
 constrain this value very well, the recent study based on the {\em
   Via Lactea II} billion particle simulation finds a value consistent
 with this, albeit with a possible large variation due to local
 clumping of dark matter in their simulation~\cite{grainy}.  For our
 purposes we will use a value of $\rho_{\odot} \simeq 0.3
 \mathrm{\  GeV}/\mathrm{cm}^3$ so our results can be easily compared
 with those of other groups.  Nonetheless,  but one should keep in mind this
 potential uncertainty when examining our results.

Finally, the cuspy profiles which diverge as $r\rightarrow0$, should be regularized.  Similar to \cite{Barrau:2005au},
we regularize the NFW \cite{nfw} and Moore \cite{moore} profiles, by assuming at small $r < r_c = 100$ pc, the DM density takes the form
\begin{eqnarray}
  \label{eq:dmreg}
  \frac{\rho(r<r_c)}{\rho(r_c)} =  A_1 + A_2\ {\rm sinc}\left(\frac{\pi r}{r_c}\right) + A_3\
  {\rm sinc}^2\left(\frac{\pi r}{r_c}\right).
\end{eqnarray}
The coefficients $A_i$ are determined by demanding that $\rho(r)$ and
it's first derivative are continuous and that the regularization does
not alter the overall DM mass in the Galaxy.   As we show later, the DM signal is rather sensitive to $r_c$.

\begin{figure}[t] 
   \centering
   \includegraphics[width=3in]{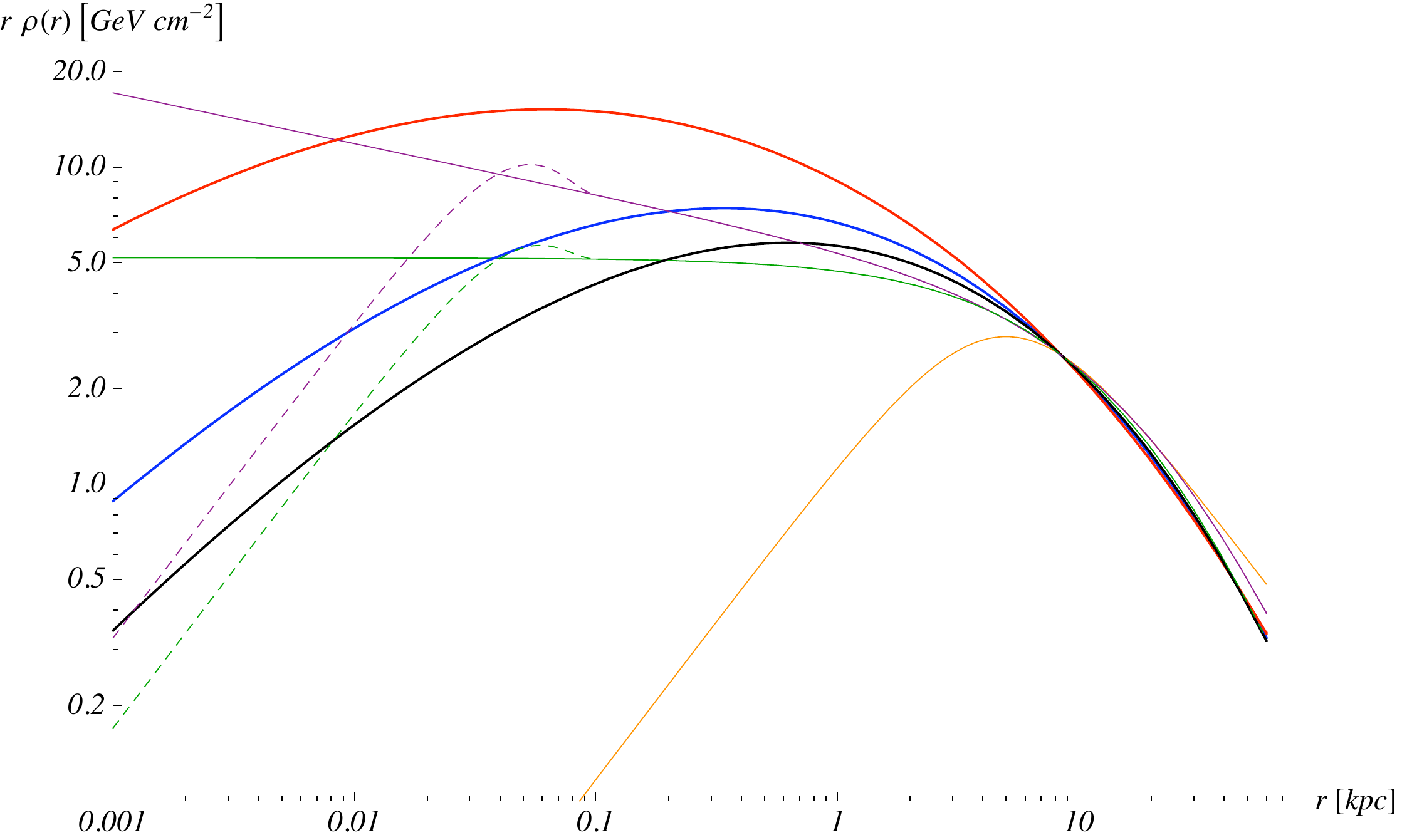}
   \caption{$r \rho(r)$ for the different DM profiles.  When
     appropriate, we also plot the regularized profile, with $r_c=100$
     pc.  Thick lines (from the bottom up): Einasto profile with
     $\alpha=0.2,0.17,0.12$. Thin lines: isothermal, NFW, Moore
     (2004). The dashed lines represent the effect of the
     regularization described in the text.}
   \label{fig:profiles}
\end{figure}

In Figure~\ref{fig:profiles} we plot the various DM profiles as a function of the radial
distance from the Galactic center (GC).  When appropriate, we also
show the regularized profile.

%%%%%%%%%%%%%%%%%%%%%%%%%%%%%%
\subsection{Positrons}\label{sec:posin}
%%%%%%%%%%%%%%%%%%%%%%%%%%%%%%
 
\subsubsection{Propagation}
\label{sec:propagation}

To calculate the positron flux at the Earth one needs to understand
how positrons propagate through our Galaxy.  Due to our limited
understanding of the latter, we wish to use a simplified model which employs a minimal
set of assumptions.  Perhaps the simplest model is the so called leaky
box model which assumes a free homogeneous diffusion of charged
particles within the Galactic disk.  Since in its simplest form the
model does not take particle cooling into account and since DM is not
homogeneously distributed in the Galaxy, it is insufficient for calculating the DM signal.   However, it is useful for calculating backgrounds and we will return to this model in Section~\ref{sec:experiments}
and the Appendix.

Below we adopt a widely studied diffusion model which does take
cooling into account.  Ignoring other effects in the propagation (such
as convection and re-acceleration) is consistent for
positrons at energies above $\sim10$ GeV \cite{Delahaye:2008ua}.
Moreover, the popularity of this model among the DM community allows
one to compare our results with those of others, thereby emphasizing
the new particle-physics features of our study.  Where appropriate, we
will remark on the uncertainties both in our theoretical
understanding of the physical mechanisms involved and in the
experimental tuning of the model \cite{mmm}.

To this end, the diffusion-loss equation in the steady state regime takes the form
\cite{Ginzburg:1976dj,Strong:2007nh}
\begin{equation}
\label{eq:10c}
   - K(E)\nabla^2 \fpos
  (E,x) - \frac{\partial}{\partial E} \left(b(E) \fpos
    (E,x)\right) = Q_{e^+}(E,x).
\end{equation}
Here $\fpos(E,x)$ is the positron number density per unit energy
which is related to the positron flux through $\fluxpos =
\vpos\fpos/4\pi$.  $Q_{e^+}(E,x)$ is the source term for the
positrons which for the background will be discussed in
section~\ref{sec:experiments} while for the DM source is given by
\begin{eqnarray}
  \label{eq:3}
  Q_{e^+}(E,r) = \frac{1}{2}\frac{\rho(r)^2}{M^2_{\rm DM}}\sum_k
  \langle \sigma v\rangle_k \frac{dN_{e^+}^k}{dE}. 
\end{eqnarray}
Finally  
\begin{equation}
\label{eq:9b}
b(E)=\frac{E^2}{\mathrm{GeV}\ \tau_E},
\end{equation}
is the energy loss coefficient due to Inverse Compton Scattering (ICS) and Synchrotron Radiation, with $\tau_E = 10^{16}$ s.  Here and
below, $E$ denotes the kinetic energy of the corresponding particle.
The diffusive region is assumed to be a flat cylinder parameterized by
$(r,z,\theta)$, with $z\in [-L,L]$ and $r\in [0,R]$, $R = 20$ kpc.
The positron number density is taken to vanish at the boundaries and
particles may propagate freely outside of it.  The solar system is
located at $(\rsun,\zsun,\thetasun)=(8.5$ kpc, $0,0)$ .

Diffusion arises due to the interactions of charged particles with the
galactic magnetic field inhomogeneities.  Such interactions produce
stable and unstable spallation products that may be used to extract
the height of the diffusion region, $L$, and the diffusion coefficient
$K(E)$ which is usually taken to be of the form
\begin{equation}
\label{eq:10}
K(E)=K_0 \beta (\R /\mathrm{GV})^\delta.
\end{equation}
Here $\beta=v/c$ and $\R=pc/eZ$ is the rigidity of the particle. For electrons
and positrons, $e{\cal R} \simeq E$.  The most stringent constraint on
the above parameters comes from the B/C measurements of the HEAO-3
experiment \cite{Engelmann:1990} and other balloon experiments, the
most recent being ATIC-2 \cite{Panov:2007fe}.  The range for $K_0$,
$\delta$ and $L$ has been studied e.g. in
\cite{Maurin:2001sj,galprop}.  We adopt the parameters
of~\cite{Donato:2003xg}, shown in Table~\ref{tab:pbar}. The (MIN, MAX)
are taken to extremize the antiproton flux, while the (M1,
M2) extremize the positron flux.  However, as it will be shown below, the differences between MIN and M2 are
small for the positrons, while MIN encompasses a larger spread for the
antiprotons.
\begin{table}[h]
\vskip 0.5cm
\begin{center}
{\begin{tabular}{|c||c|c|c|c|}
\hline
Model  & $\delta$ & $K_0$ [kpc$^2$/Myr] & $L$ [kpc] & $V_C$ [km/s]\\
\hline \hline
MIN  & 0.85 &  0.0016 & 1  & 13.5\\
MED  & 0.70 &  0.0112 & 4  &  12\\
MAX (M1)  & 0.46 &  0.0765 & 15 & 5\\
M2 & 0.55 & 0.00595 & 1 & 8\\
\hline
\end{tabular}}
\end{center}
\caption{Parameters for the cosmic ray propagation model, compatible
  with measurements of B/C.  The names correspond to minimum, medium
  or maximum primary antiproton and positron fluxes. The convective
  wind parameter $V_C$ is only relevant to the propagation of
  antiprotons, as discussed 
  in section \ref{sec:antin}.}
\label{tab:pbar}
\end{table}
These sets of parameters are used as representatives of the
uncertainties in the diffusion model.  It is important however to note
that these parameters are model dependent and their spread may be
larger at high energies.  For example, $\delta$ is expected to
change with energy,  because an extrapolation of $\delta \sim 0.6$ to
higher energies predicts large anisotropies already at $10^{15}$ eV,
contradicting observations\footnote{It is possible that such a change
  is indicated in the ATIC-2 measurements \cite{Panov:2007fe} (which
  are not taken into account in \cite{Maurin:2001sj}), although the
  large statistical uncertainties and possible normalization problems
  do not allow for a decisive conclusion.}\cite{Hillas:2005cs}.

\begin{figure}[t] 
   \centering
   \includegraphics[width=3in]{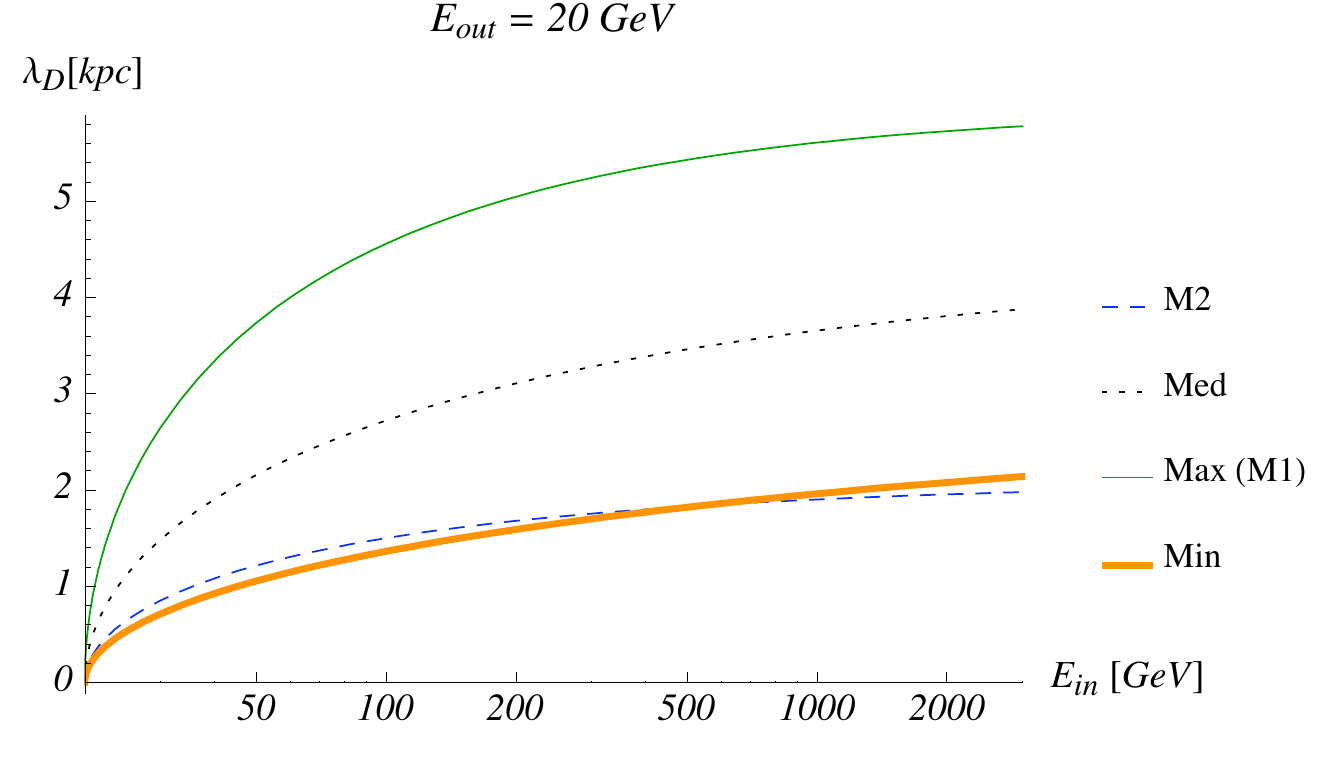}\includegraphics[width=3in]{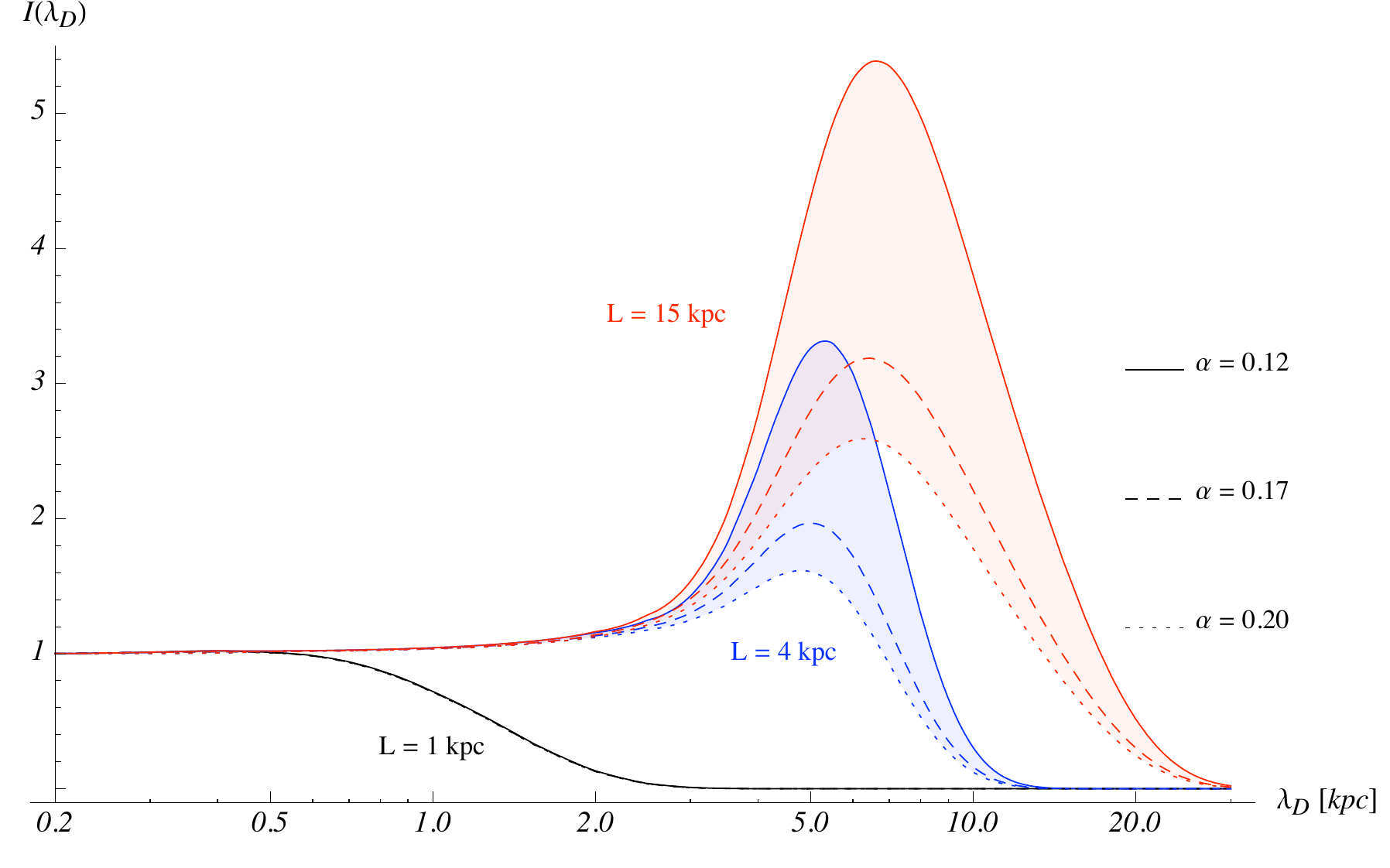}
   \caption{ {\it Left:} The effective diffusion length,
     $\lambda_D$, for the propagations models we study as a function of the injection spectrum for a $20$
     GeV electron detected at the Earth. {\it Right:} $I(\lambda_D)$
     for $L=1,4,15$ kpc and the Einasto profile
     with $\alpha=(0.12,0.17,0.20)$.}
   \label{fig:lambdaD}
\end{figure}

Eq.~\eqref{eq:10c} sets a natural energy-dependent diffusion length scale\cite{Baltz:1998xv,mmm}
\begin{eqnarray}
  \label{eq:1}
  \lambda_D(E,E_S) = \sqrt{4K_0(t_E-t_{E_S})},
\end{eqnarray}
where $t_{E}\equiv \tau_E (1-\delta)^{-1}(E/{\rm GeV})^{\delta-1} $ is
the energy-dependent pseudo time.  Using the models given in
Table~\ref{tab:pbar}, we plot $\lambda_D$ for detected electron with
energy of $20$ GeV, as a function of the injection energy in
Fig.~\ref{fig:lambdaD}.  A fundamental difference exists between the
background and the DM contributions to the spectrum: while the injection
spectrum of the former peaks at low energies (see eq.~\eqref{eq:8}),
the one of the DM does not, as can be seen in Figs. \ref{fig:dndx},\ref{fig:dndxsho}.  It then
follows that while background electrons travel short distances, of
order $\lambda_D\lesssim 2-3$ kpc, the bulk of the electrons
originating from DM annihilations travel larger distances and may be
created closer to the center of the Galaxy where the DM density is
larger.  We will return to this point shortly.

A semi-analytic solution to the diffusion equation above is found to
take the form \cite{mmm}
\begin{eqnarray}
  \label{eq:2}
  \fluxpos(E,\rsun) = \frac{\vpos}{4\pi b(E)} \int_E^\mdm dE'
  Q_{e^+}(E',\rsun)I(\lambda_D(E,E')). 
\end{eqnarray}
where
\begin{eqnarray}
\label{eq:20}
I(\lambda_D) = \sum_{n,m=1}^\infty J_0(\zeta_n \rsun/R) \sin(m\pi/2)
{\rm exp}\left[ -\left(\left(\frac{m\pi}{2L}\right)^2 +
    \left(\frac{\zeta_n}{R}\right)^2\right)\frac{\lambda_D^2}{4}\right]R_{n,m}
\end{eqnarray}
and
\begin{eqnarray}
\label{eq:20b}
R_{n,m} = \frac{2}{J_1(\zeta_n)^2L R^2}\int_0^R r dr \int_{-L}^L dz
J_0(\zeta_n r/R)\sin\left(m\pi
(z+L)/2L\right)\left(\frac{\rho(r,z)}{\rhosun}\right)^2,  
\end{eqnarray}
Here $J_i$ are the i-th order Bessel functions of the first kind and
$\zeta_n$ is the n-th zero of $J_0$.  We plot $I(\lambda_D)$ for the
Einasto profiles with $\alpha = (0.12,0.17,0.20)$ and in Figure~\ref{fig:lambdaD} for the three
different values of $L$ considered here.  As can be seen
in the plot, $I(\lambda_D)$ may provide an enhancement to the flux if
$\lambda_D$ is sufficiently large.  Since $\lambda_D$ is largest for
positrons arriving to the Earth with low energy, the spectrum tends to be softer in those cases
where the positrons propagate for significant distances.  This, in turn,
weakens any feature that may exists at the high end of the spectrum.
As suggested in Fig.~\ref{fig:lambdaD}, the effect is most significant
for the MAX model.  Consequently, as we shall see in
Sect.~\ref{sec:pb}, the MAX model fits less well to the ATIC data, but
on the other hand requires a smaller cross-section to fit the PAMELA data, therefore
predicting less photons from the GC.  Conversely, the
MIN model fits better to the ATIC data but predicts a larger
cross-section and therefore a larger number of photons.  This tension,
in part, will be responsible for excluding significant parts of the
parameter space in these DM annihilation models that try to fit PAMELA
and ATIC simultaneously.

\subsubsection{Experiments}
\label{sec:experiments}

The three relevant experiments for the positron study are
PAMELA~\cite{pamelapos}, ATIC-2~\cite{atic} and PPB-BETS\cite{ppbbets}.
The former recently provided the positron-to-electron flux ratio up to
$100$ GeV while the latter two experiments measure the positron plus
electron flux up to energies of a few TeV.

For the purpose of testing the predictions of the theory, we need an
estimation of the backgrounds.  To this end, we will attempt to be as
data-driven as possible.  The two relevant backgrounds needed are the
positron flux in the energy range between $10$ to $100$ GeV and the
$e^++e^-$ flux at energies up to $1$ TeV.  Starting with the latter,
the spectrum has been measured rather accurately.  In the absence of
the ATIC/PPB-BETS bump, which is apparent above $\sim100$ GeV, the data fits a
power-law spectrum
\begin{eqnarray}
  \label{eq:14}
  \fluxep(E) = A_1 \left(\frac{E}{{\rm GeV}}\right)^{\gamma_1} \ {\rm \ cm^{-2}\ s^{-1}\ sr^{-1}\
  GeV^{-1}} . 
\end{eqnarray}
For finding a best fit to $A_{1}$ and $\gamma_{1}$ we used the
measurements of ATIC-2 and HEAT \cite{Barwick:1997kh} in the energy
range $20-100$ GeV.  The lower cutoff ensures that uncertainties from
solar modulation are insignificant while the upper prevents probing
the bump.  We find $A_1 = 0.046 \pm 0.015 $ and
$\gamma_1=-3.30\pm0.08$.  This result is in agreement with the
preliminary analysis presented by the PAMELA collaboration
\cite{Boezio}.  Taking the best fit at face value for our background
is incorrect because it is impossible to disentangle the
signal from background even at low energies.  Therefore, when fitting
together with the DM signal, we allow both $A_1$ and $\gamma_1$ to
float around their central values given above.

No precise measurements of the positron flux exist at energies above
$10$ GeV.  To calculate the background, we therefore use the leaky box
approximation which is well suited for energies $\lesssim 100$ GeV
 and above $\gtrsim10$ GeV\footnote{We thank Eli Waxman and Boaz Katz for drawing our attention
  to this point.}.  We derive the prediction in
Appendix \ref{app:box} and simply quote the result here
\begin{eqnarray}
  \label{eq:8}
  \fluxpos(E) \simeq  4\times 10^{-3} \ \left(\frac{E}{\rm
      GeV}\right)^{-2.84-\delta\pm 0.02} \ {\rm \ cm^{-2}\ s^{-1}\
    sr^{-1}\ 
  GeV^{-1}}.
\end{eqnarray}
which is in agreement with \cite{Delahaye:2008ua}.  The parameter $\delta$
is typically of order $\delta \sim 0.6$ but for the propagation models
considered above, we vary it according to the values of Table
\ref{tab:pbar}.  As described in the Appendix, the normalization of
the flux suffers from large uncertainties which arise from the
assumptions in the theory and uncertainties in the measurements of the nuclear spallation
cross-sections.  Thus when fitting the data together with the DM
signal, we allow the normalization to float around the above value.
The spectral index is instead kept fixed.

%%%%%%%%%%%%%%%%%%%%%%%%%
\subsection{Antiprotons}\label{sec:antin}
%%%%%%%%%%%%%%%%%%%%%%%%%

\subsubsection{Propagation}
\label{sec:propagation-1}

We
describe the propagation of $\bar p$ at equilibrium with a diffusion equation similar to that of
the positrons \cite{Hisano:2005ec}
\begin{eqnarray}
\label{eq:8a}
 - K(E)\nabla^2 \fpbar(E,x)
 + \frac{\partial}{\partial z} \left( V_C(z)\fpbar(E,x) \right)
+ 2 h \delta(z)\Gamma_{\rm ann} \fpbar(E,x)=  \Qpbar (E,x) 
\end{eqnarray}
where $\fpbar(T,\vec{r})$ is the number density of antiprotons per
unit energy.  As in the case of positrons, $K(E)$ is the diffusion
coefficient given in Eq.~\eqref{eq:10}.  Energy loss, however, is
negligible for antiprotons due to their mass and we therefore do not
consider it.  The last two terms on the \mbox{l.h.s.} of
eq. \eqref{eq:8a} correspond to convective wind and interaction with
the Interstellar Medium (ISM) in the galactic plane respectively.  The
convective wind is assumed to be of the form $V_C(z) =sign(z) V_C$ and
is directed outwards from the galactic plane.  The annihilation width
between antiprotons and protons is given by
\begin{eqnarray}
  \label{eq:11}
  \Gamma_{\rm ann} = (n_{\rm H}+4^{2/3}n_{\rm He})\sigma_{\bar p 
    p}^{\rm ann}v_{\bar p},
\end{eqnarray}
where $n_{\rm H,He}\simeq 1 {\rm\ cm}^{-3}$ are the hydrogen and
helium number density and $v_{\bar p}$ is the velocity
of the antiprotons.  A parameterization for $\sigma_{p \bar p}$ can be
found in \cite{Tan:1984ha,Hisano:2005ec}.  Both the annihilation and
convective terms can be neglected at energies above $\gtrsim 10$ GeV.
Nevertheless we keep these terms for completeness.

\begin{figure}[t] 
   \centering
   \includegraphics[width=3in]{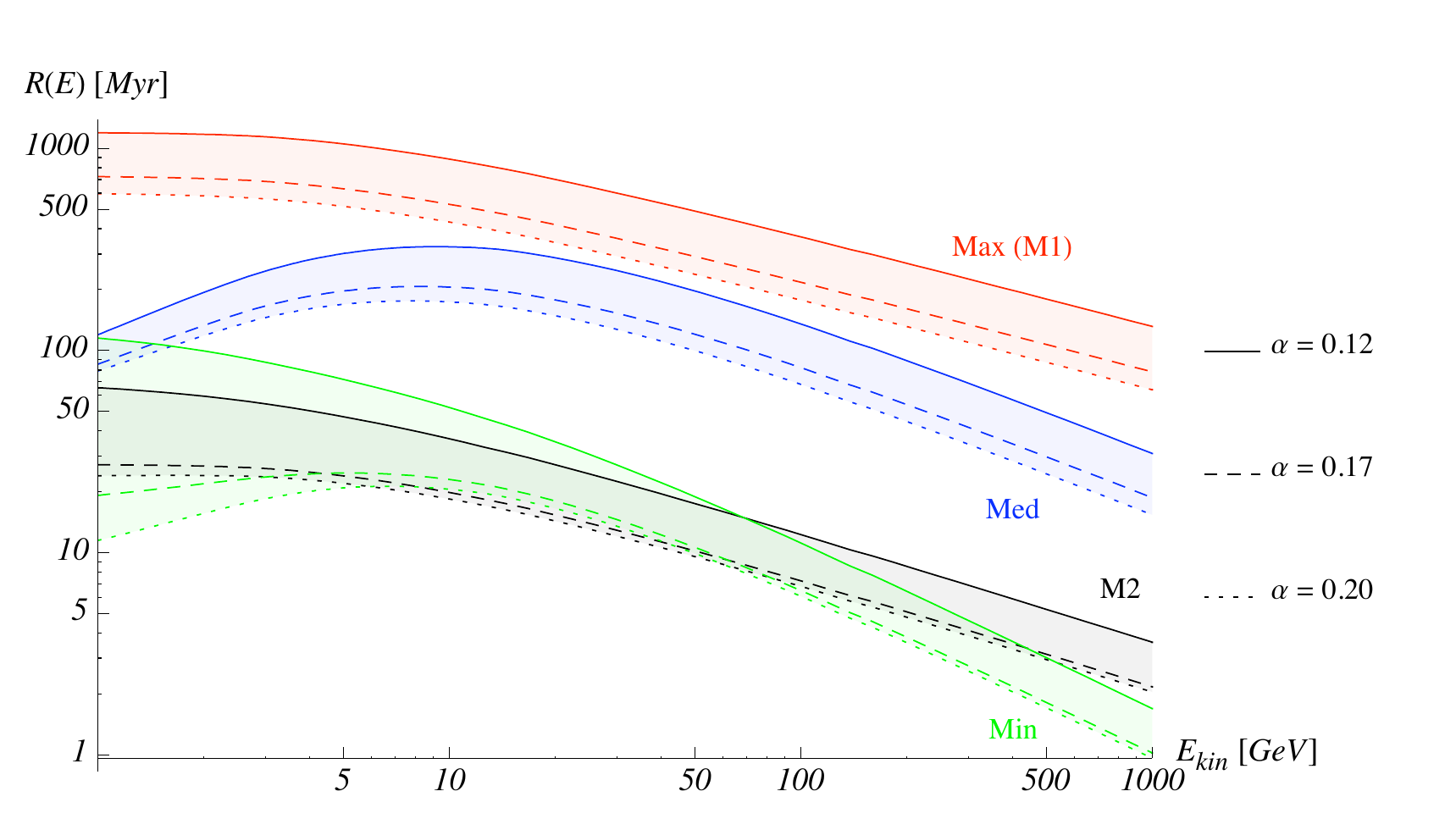}
   \caption{$R(E)$
     for the propagation models we consider and the Einasto profile
     with $\alpha=(0.12,0.17,0.20)$.}
   \label{fig:RE}
\end{figure}

As in the positron case, propagation takes place within the disk of
half-height $L$ and radius $R=20$ kpc.  In order to be able to fit the
data both for positrons and antiprotons and since we would like to
quantify the uncertainties, we use the same models described in Table
\ref{tab:pbar}.  An semi-analytical solution to the flux $\fluxap =
v_{\bar p} \fpbar /4\pi$ is given by \cite{Donato:2001ms}
\begin{eqnarray}
  \label{eq:12}
  \fluxap(E, \rsun) = \frac{c}{4\pi}\ Q_{\bar p} \ R(E)
\end{eqnarray}
where as before, 
\begin{eqnarray}
  \label{eq:6}
  Q_{\bar p} = \frac{1}{2}\frac{\rho(r)^2}{\mdm^2} \sum_k
  \langle\sigma v\rangle_k \frac{dN_{\bar p}^k}{dE} .
\end{eqnarray}
$R(E)$ encodes the astrophysical propagation information and was found
to take the form \cite{Donato:2003xg}, 
\begin{eqnarray}
\label{eq:21}
R(E) = \sum_{n=1}^\infty J_0(\zeta_n \rsun/R) {\rm
  exp}\left[-\frac{V_C L}{2K(E)}\right]\frac{y_n(L)}{A_n \sinh(S_n
  L/2)} 
\end{eqnarray}
with,
\begin{eqnarray}
\label{eq:21b}
y_{n}(L)& = &\frac{4}{J_1(\zeta_n)^2 R^2}
\\ \nonumber 
&& \times\int_0^R r dr \int_{0}^L dz\
J_0(\zeta_n r/R)\ \sinh(S_n(L-z)/2){\rm 
  exp}\left[\frac{V_C(L-z)}{2K(E)}\right]\
\left(\frac{\rho(r,z)}{\rhosun}\right)^2, 
\\ \nonumber \\ \nonumber
A_n &=& 2 h\Gamma_{\rm ann} + V_C + K(E)S_n\coth(S_nL/2), 
\\ \nonumber \\ \nonumber
S_n &=& \sqrt{(V_C/K(E))^2 + 4\zeta_n^2/R^2}.
\end{eqnarray}
In Figure~\ref{fig:RE} we plot this function for the the Einasto
profiles with $\alpha = (0.12,0.17,0.20)$ and for the propagation
models of Table~\ref{tab:pbar}.

\subsubsection{Experiments}
\label{fsec:experiments-1}

The relevant experimental data is the recent PAMELA measurement for
the antiproton-to-proton flux ratio \cite{pamelaantip} up to the
energy of $100 \rm GeV$.

The proton flux, has been well measured by the
BESS~\cite{besslow,besshigh} and AMS-01~\cite{ams} experiments and the
best fit to the data (for demodulated protons) is of the form
~\cite{donatoppbar}:
\begin{equation}
\Phi_{p_{\mathrm{bkgd}}}= A \beta^{P_1} {\cal R}^{-P_2}\,
\mathrm{cm}^{-2}\,\mathrm{s}^{-1}\,\mathrm{sr}^{-1}\,(\mathrm{GeV/n})^{-1} 
\end{equation}
where as before ${\cal R}$ is the rigidity and $\beta$ is the
velocity.  We use the following values for the coefficients
\begin{equation}
\begin{array}{c|c|c|c}
{\cal R}  & A  & P_1 & P_2 \\
  \hline
 <20\,\mathrm{GV} & 1.94\pm 0.13  & 0.7 \pm 0.52 & 2.76 \pm 0.03  \\
 >20\,\mathrm{GV} & 2.42\pm 0.18  &   0 & 2.84\pm 0.02 \\
\end{array}
\end{equation}
in agreement with \cite{donatoppbar}.

The antiproton flux is known to come primarily from cosmic ray protons
interacting with the ISM.  Unfortunately the flux generated by this
process does not have a characteristic power law shape (as many other
astrophysical processes) in the range of energy that we examine.  To
calculate the approximate shape for the background we need to take
into account the spallation cross section given
in~\cite{donatoppbar}, and then propagate the products.
Fortunately, as was found in \cite{donatoppbar}, changing the model of
propagation does not significantly change the shape of the antiproton
flux but does change the normalization.  We therefore use the results
of \cite{donatoppbar}, parametrized as follows \cite{mdm},
\begin{equation}
\label{eq:22}
\log_{10}\fluxp =  -1.356 + 0.114\tau -0.645\tau^2-0.614\tau^3 + 0.02\tau^4+0.168\tau^5-0.038\tau^6,
\end{equation}
where $\tau = \log_{10} (E/{\rm GeV})$.  To fit to the PAMELA data one needs to take into account solar modulation~\cite{solar}. We
modulate the background according to the PAMELA modulation potential
$\phi = 0.58$ GV and allow the normalization to float.

%%%%%%%%%%%%%%%%%%%%%%%%%%%%%%%%%%%%%%%%%%%%%%%%%%%%%%
%%%%%%%%%%%%%%%%%%%%%%%%%%%%%%%%%%%%%%%%%%%%%%%%%%%%%%
\subsection{Photons}\label{sec:gammain}
%%%%%%%%%%%%%%%%%%%%%%%%%%%%%%%%%%%%%%%%%%%%%%%%%%%%%%
%%%%%%%%%%%%%%%%%%%%%%%%%%%%%%%%%%%%%%%%%%%%%%%%%%%%%%
  \subsubsection{Propagation}
  Photons from dark matter are one of the cleanest channels possible
  to study from the point of view of propagation.  This is because
  once produced, photons freely travel to Earth and thus are
  insensitive to the propagation parameters that complicate studies of
  $e^+$ and $\bar{p}$. Since photons are not affected by propagation,
  the flux at the Earth depends only on the injection spectrum and on
  how many annihilations occur.  While this simplifies the calculation
  of the photon flux, it also introduces a large amount of
  uncertainty.  Unlike positrons and antiprotons which reach us
  typically from distances less than $\rsun$, photons from the
  galactic center, where the dark matter density is the highest, can
  always reach us. N-body simulations can not resolve distances less
  than approximately 100-200 pc (as discussed in~\ref{sec:dm-halos}),
  and thus there is a fair amount of uncertainty in the DM density
  profile at the galactic center.

The differential photon flux (in units of ${\rm cm^{-2}\ s^{-1}\ sr^{-1}\
  GeV^{-1}}$) is given by,
\begin{eqnarray}
  \label{eq:5}
  \fluxpho = \frac{\rsun}{4\pi} Q_\gamma\ \bar J {\Delta\Omega}   \ , 
\end{eqnarray}
As before, $Q_\gamma$ is the photon source encoding the particle
physics contribution to the flux,
\begin{eqnarray}
  \label{eq:6a}
  Q_\gamma = \frac{1}{2}\frac{\rho(r)^2}{\mdm^2}\sum_k\langle\sigma v\rangle_k \frac{dN_\gamma^k}{dE}.
\end{eqnarray}
Much like $I(E)$ and $R(E)$ in the positron and anti-proton case,
$\bar J$ encapsulates the astrophysics dependence and is given by,
\begin{equation}
\label{eq:J}
  \bar J=\frac{1}{\Delta\Omega}\int d\Omega\int_{\rm line-of-sight}
  \frac{dl}{\rsun}\left(\frac{\rho(r)}{\rhosun}\right)^2. 
\end{equation}
where $\Delta\Omega$ is the opening solid angle for a given experiment. The flux received in a given direction is proportional to the
integral over the line-of-sight of the DM density squared. By writing
the photon flux in this factorized form we simply need to compute the
$\bar{J}$ for a given experiment and profile.

In principle, for cuspy profiles the integral in Eq.~$\ref{eq:J}$ could
diverge. For the profiles that we consider $\bar{J}$ in fact does not
diverge.  However, as discussed in \ref{sec:dm-halos}, depending on
the assumed regulation of the DM density inside of $\mathcal{O}(100)$
pc, the actual value of $\bar{J}$ computed can vary significantly as a
function of $r_{c}$ in Eq.~\ref{eq:dmreg}.  We will quantify this
uncertainty further in Section~\ref{sec:astro} when we show the
dependence of photon flux on the choice of the dark matter density
profile.
 
 \subsubsection{Experiments}

 The experiments that we will be primarily concerned with are those
 that measure high energy gamma rays.  Specifically we will focus on
 the HESS experiment which is the most sensitive to high mass DM
 annihilation.  The FERMI-LAT experiment will also be of importance in
 the near future and we discuss it further in Section~\ref{sec:imp}.
 If one is looking for signals of DM annihilation into photons, the
 best place to look is in regions of potential high DM density with
 small astrophysical backgrounds.  For instance high mass/light ratio
 dwarf spheroidal galaxies~\cite{dwarfsearch} or subhalos of our
 galaxy~\cite{subhalos} are both prime candidates for a clean signal
 of DM annihilations.    However, for the purpose of constraining models of DM, it is more important to observe regions that have effectively a large $\bar J$ for sufficiently long time.  
  For the HESS experiment this is found in two regions, the Galactic
 Center (GC)~\cite{hessgc} and the Galactic Ridge (GR)~\cite{hessgr}.

 The GC as studied by HESS has a $\Delta \Omega=10^{-5}$.  To
 calculate $\bar{J}$ for the Galactic Center one can write
\begin{eqnarray}
\label{OmegaGC}
&\int d\Omega = \int_{\cos\psi}^1 2\pi d(\cos\psi^\prime)\ ,
\\
&\Delta\Omega = 2\pi(1-\cos\psi)\ , \qquad r =
\sqrt{\rsun^2 + l^2 - 2l\rsun \cos\psi^\prime}.
\end{eqnarray}
The GC was observed by HESS for 48.7 hours during
2004, and is cataloged as J1745-290.

The GR is a larger region that includes the GC.  It is defined in galactic latitude and longitude as the region $\vert
l\vert <0.8^\circ$ and $\vert b\vert<0.3^\circ$.  The GR has several
known suspected point sources including J1745-290 that the HESS
experiment modeled in order to subtract it in~\cite{hessgr}.  The
experiment fits a point source to two specific regions, J1745-290 and
another source, G0.9+1, that lie within the Ridge.  The best fit for
these sources was then subtracted and the differential flux $dN/dE$
was given for the remaining portion which defines the Ridge.
Unfortunately the $dN/dE$ for each galactic latitude and
longitude has not been published and therefore one can not mimic their subtraction scheme.
However, since the hypothesized point sources are the dominant sources
of photons we can excise these entirely without introducing a large
amount of uncertainty into the calculation of $\bar{J}$.
Additionally, beyond these partial subtractions, a further subtraction
to reduce systematics is also performed.  The HESS experiment observes
the regions $0.8^\circ <\vert b\vert<1.5^\circ$ and uniformly
subtracts this ``background" from the GR data.  We will perform this
same subtraction when plotting the observed photon flux.  To calculate
the value of $\bar{J}$ for the GR (which is a rectangle not a circle)
we use a slightly different form of $d\Omega$.  For a region defined
by longitudinal and latitudinal boundaries $|\theta| < \Delta l$,
$|\psi| < \Delta b$, we have,
 \begin{eqnarray}
\label{OmegaGCa}
&\int d\Omega = 4\int_{0}^{\Delta l} d\theta^\prime \int_{0}^{\sin\Delta b}d(\sin\psi^\prime)\ ,
\\
&\Delta\Omega = 4\Delta l \Delta b \ ,\qquad r=\sqrt{\rsun^2 + l^2 -
  2l\rsun \cos\psi^\prime\cos\theta^\prime}. 
\end{eqnarray}

To finish our discussion of the experimental inputs for photons we
will tabulate the $\bar{J}$ calculated for the various experiments and
profiles in Table~\ref{tab:J}.
\begin{table}[h]
\vskip 0.5cm
\begin{center}
  {\begin{tabular}{|l|c||c|}
      \hline
      Location & Profile  & $\bar{J}$\\
      \hline \hline
      & Cored isothermal & $13.6$
      \\
      Galactic Center  & NFW &  $4076 - 10170$
      \\
      $(\Delta\Omega = 2\cdot 10^{-5})$  & Moore & $13128 - 51388$
      \\
      & Einasto $\alpha=0.17$ & $6610$
      \\
      & Einasto $\alpha=0.12$ & $65500$
      \\
      & Einasto $\alpha=0.2$ & $2306$
      \\
      \hline
      \hline
      & Cored isothermal & $0.02$
      \\
      Galactic Ridge  & NFW &  $1295 - 1541$
      \\
      & Moore & $3836 - 5653$
      \\
      & Einasto $\alpha=0.17$ & $1614$
      \\
      & Einasto $\alpha=0.12$ & $10886$
      \\
      & Einasto $\alpha=0.2$ & $602$
      \\
      \hline
\end{tabular}}
\end{center}
\caption{The values for $\bar{J}$ for the various
 galactic regions we consider.  For
  the NFW and Moore profiles in the Milky way, we show the range for
  $J$ for $0 \leq   r_c \leq 0.1 \,\rm kpc$.}
\label{tab:J}
\end{table}

%%%%%%%%%%%%%%%%%%%%%%%%%%%%%%%%%%%%%%%%%%%%%%%%%%%%%%
%%%%%%%%%%%%%%%%%%%%%%%%%%%%%%%%%%%%%%%%%%%%%%%%%%%%%%
\subsection{Neutrinos}\label{sec:neutrinoin}
%%%%%%%%%%%%%%%%%%%%%%%%%%%%%%%%%%%%%%%%%%%%%%%%%%%%%%
%%%%%%%%%%%%%%%%%%%%%%%%%%%%%%%%%%%%%%%%%%%%%%%%%%%%%%
\subsubsection{Propagation}
Dark Matter annihilations can also produce neutrinos. After being
produced, neutrinos propagate till the Earth where they are
detected. The standard strategy to detect these neutrinos is to look
at those that convert in rock nearby the detectors and observe
the charged leptons, in particular muons. Therefore the quantity actually
measured is the muon flux, that can be related to the neutrino flux at
production by
\begin{equation}
  \frac{d \Phi_{\mu}}{d E_{\mu}} = R(E_{\mu})
  \int_{E_{\mu}}^{m_{\chi}}d E_{\nu} \sum_{i=e,\mu,\tau} \frac{d
    \Phi_{\nu_{i}}}{d E_{\nu}} P_{i\rightarrow\mu} \sum_{N=p,n}
  \rho_{N} \frac{d \sigma_{\nu N}(E_{\mu}/E_{\nu})}{d E_{\mu}} + (\nu
  \leftrightarrow \bar{\nu}) 
\end{equation}
where $d \Phi_{\nu_{i}}/d E$ is the flavor specific neutrino flux, the
sum is taken over all the leptonic flavors and the term
$P_{i\rightarrow\mu}$ comes from the fact that neutrinos oscillate
during their travel\footnote{$P_{i\rightarrow\mu}$ is determined only
  by the mixing angles since the baseline is very long. For simplicity
  we have taken $\theta_{13}=0$.}. The muon production cross section
in neutrino nucleon interactions is denoted by $\sigma_{\nu N}$
\cite{Ritz:1987mh,Jungman:1995df}, while $\rho_{p,n}$ are the proton
and neutron number densities (taken to be equal for standard
rock). Finally $R(E)$ is the muon range~\cite{pdg}, i.e. the distance
traveled by a muon of energy $E$ before stopping or before getting below the
detection energy threshold of the experiment.

\subsubsection{Experimental Status}
Good places to look for neutrinos produced in DM annihilations
are the GC (like in the case of photons) but also the Sun
and the Earth, where DM may be trapped in the gravitational field and
their density may grow large enough to allow for a sizable annihilation
rate. Another signal that one can look for is the total diffuse
neutrino flux \cite{Yuksel:2007ac}.

The flux coming from the Sun and the Earth has been recently
re-investigated in light of the PAMELA and ATIC~\cite{nupamela}. In particular the authors
of \cite{Delaunay:2008pc} have found that for DM annihilating mainly
into leptonic final states, the neutrino flux from the Sun and the
Earth is out of reach of IceCube unless the DM annihilates directly
into a pair of neutrinos. In our case IceCube can still be of
relevance if the $\phi$ mass is high enough in the multi-GeV range
where hadronic decays are also allowed, but in this case the bounds on
the photon flux will be also very strong as it will be shown in
Sect.~\ref{sec:pb}.  For this reason here we will only discuss the
neutrino flux coming from DM annihilations in the Galactic Center.

In this case the present bounds comes mostly from detectors located in
the northern hemisphere, since there the Galactic Center is below the
horizon most of the time. In particular the current best limit is from
SuperKamiokande \cite{Desai:2004pq}, which looks at the up-going muons
produced in the rocks below the detector. The collaboration reports an
upper bound on the total muon flux above a threshold of $1.6\,\rm
GeV$, as a function of the half-opening angle of a cone pointing
towards the center of the Galaxy.
As we show in the next Section, this bound is not powerful enough to give any appreciable constraints on the parameters of our module.

In the future this limit will be improved by Antares, a neutrino
telescope in deep Mediterranean waters, which started taking data in
2007, and on a longer timescale by Km3Net and Megaton-size neutrino
detectors like Hyper-Kamiokande.

%%%%%%%%%%%%%%%%%%%%%%%%%%%%%%%%%%%%%%%%%%%%%%%%%%%%%%
%%%%%%%%%%%%%%%%%%%%%%%%%%%%%%%%%%%%%%%%%%%%%%%%%%%%%%
\section{Results}\label{sec:pb}
\setcounter{equation}{0} \setcounter{footnote}{0}
%%%%%%%%%%%%%%%%%%%%%%%%%%%%%%%%%%%%%%%%%%%%%%%%%%%%%%
%%%%%%%%%%%%%%%%%%%%%%%%%%%%%%%%%%%%%%%%%%%%%%%%%%%%%%

In this section we combine the particle physics inputs that were
calculated in Section~\ref{sec:udm} and the propagation methods that were
discussed in Section~\ref{sec:ai} to calculate the fluxes for various
experiments.  We are primarily interested in answering the following questions:
\begin{enumerate}
\item Given our particle physics framework, what is the preferred
  parameter space that can explain PAMELA and ATIC/PPB-BETS?
\item Given this parameter space, what are the regions that are
  compatible with existing searches for $\gamma$'s and $\nu$'s?
\end{enumerate}

To answer these questions we need to define what we mean by the
preferred parameter space and how we bound it.  We implement a $\chi^2$ function in order to fit a given data set with our particle physics and background parameters  (the relevant set is shown in Table~\ref{tab:variables}).  To understand the allowed regions and how the parameters affect the fluxes we perform the following procedure.  We pick a parameter of interest and we marginalize over all the other parameters.  We then extract the fluxes as a function of the chosen parameter.  In this section we will only marginalize a $\chi^2$ function for the  PAMELA and ATIC/PPB-BETS data.  We then plot the resulting fluxes as a function of this parameter both for PAMELA, ATIC and PPB-BETS, and for HESS and SuperK data, where applicable.   The allowed parameter regions are those points with good fits that do not conflict with the HESS and SuperK data.

Below, we separate the parameters that we are interested in into two categories: particle physics discussed in Section~\ref{sec:particle-physics}, and astrophysics discussed in Section~\ref{sec:astro}.  By examining these parameters independently, we study their influence on the fluxes, and the uncertainties in our predictions.  We postpone the fits to all data and their implications to Section~\ref{sec:imp}.

\begin{table}[h]
\vskip 0.5cm
\begin{center}
\begin{tabular}{|c|c|c|c|c|c|c|c|}
\hline
Background Parameters & $N_{e^+}$, $N_{e^++e^-}$ , $N_{\bar{p}}$ ,
$\gamma_{e^++e^-}$, $N_\gamma$ ,$\gamma_\gamma$\\ 
\hline
Particle Physics Parameters & $m_\chi$, $m_\phi$ , $\langle \sigma v
\rangle$ , $R_{\mathrm{SM}}$, $\alpha_{DM}$\\ 
\hline
\end{tabular}
\end{center}
\caption{A summary of the parameters used for fitting the data.}
\label{tab:variables}
\end{table}

%%%%%%%%%%%%%%%%%%%%%%%%%%%%%%%%%%%%%%%%%%%%%%%%%%%%%%
%%%%%%%%%%%%%%%%%%%%%%%%%%%%%%%%%%%%%%%%%%%%%%%%%%%%%%
\subsection{Particle Physics}
\label{sec:particle-physics}
%%%%%%%%%%%%%%%%%%%%%%%%%%%%%%%%%%%%%%%%%%%%%%%%%%%%%%
%%%%%%%%%%%%%%%%%%%%%%%%%%%%%%%%%%%%%%%%%%%%%%%%%%%%%%

%%%%%%%%%%%%%%%%%%%%%%%%%%%%%%%%%%%%%%%%%%%%%%%%%%%%%%
%%%%%%%%%%%%%%%%%%%%%%%%%%%%%%%%%%%%%%%%%%%%%%%%%%%%%%
\subsubsection{Light Gauge Boson Mass}
\label{sec:light-gauge-boson}
%%%%%%%%%%%%%%%%%%%%%%%%%%%%%%%%%%%%%%%%%%%%%%%%%%%%%%
%%%%%%%%%%%%%%%%%%%%%%%%%%%%%%%%%%%%%%%%%%%%%%%%%%%%%%

In this section we isolate the effects of varying the light
gauge boson mass, $m_\phi$, on the fluxes.  To understand how the particle physics
parameter $m_\phi$ influences the results, we fix both the dark matter
profile and propagation model and marginalize over the rest of the parameters.  We choose the Einasto profile with
$\alpha=0.17$ and the MED propagation model.

\begin{figure}[hp]
\begin{center}
$\begin{array}{lr}
\includegraphics[width=3.1in]{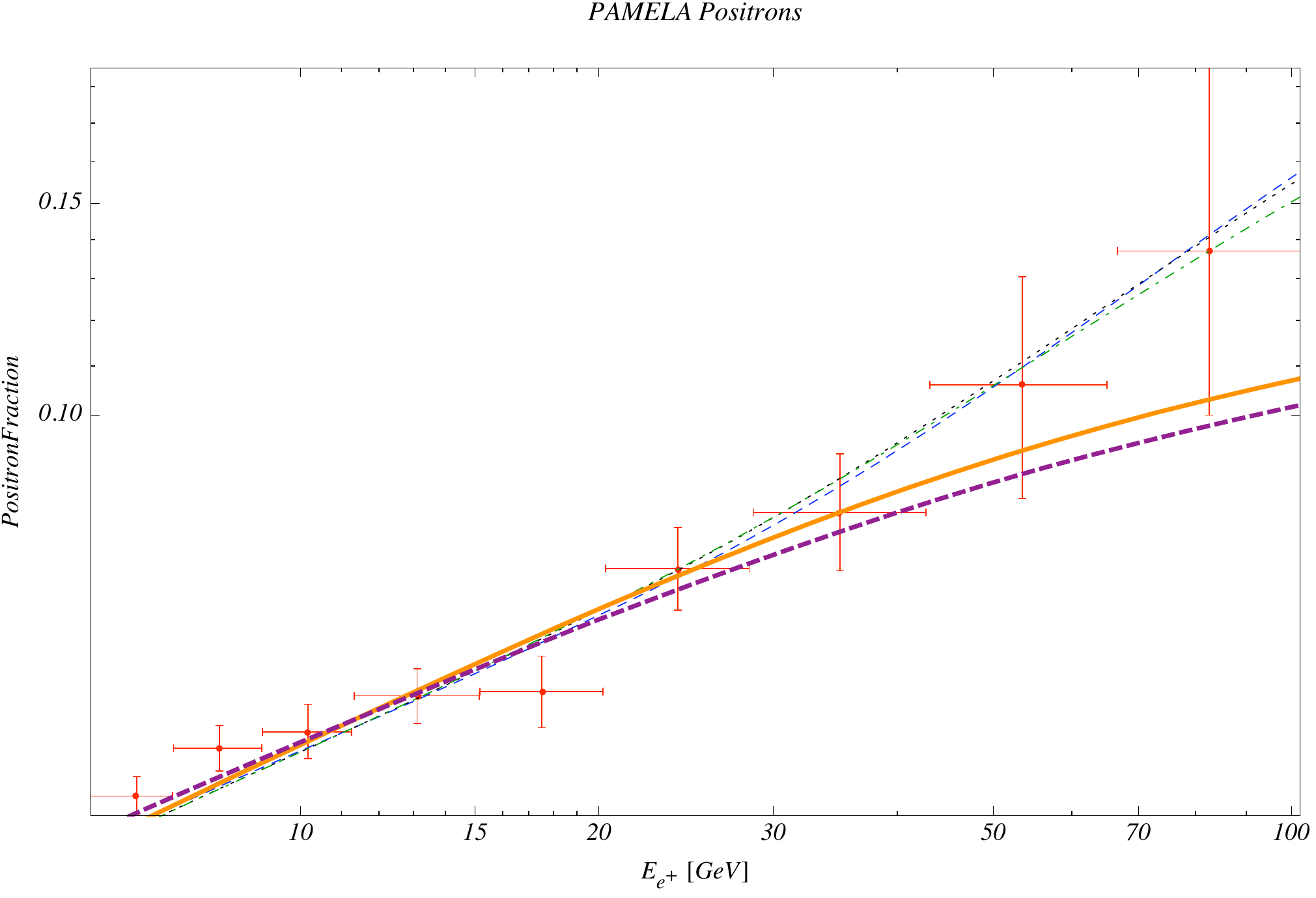}
& \includegraphics[width=3.1in]{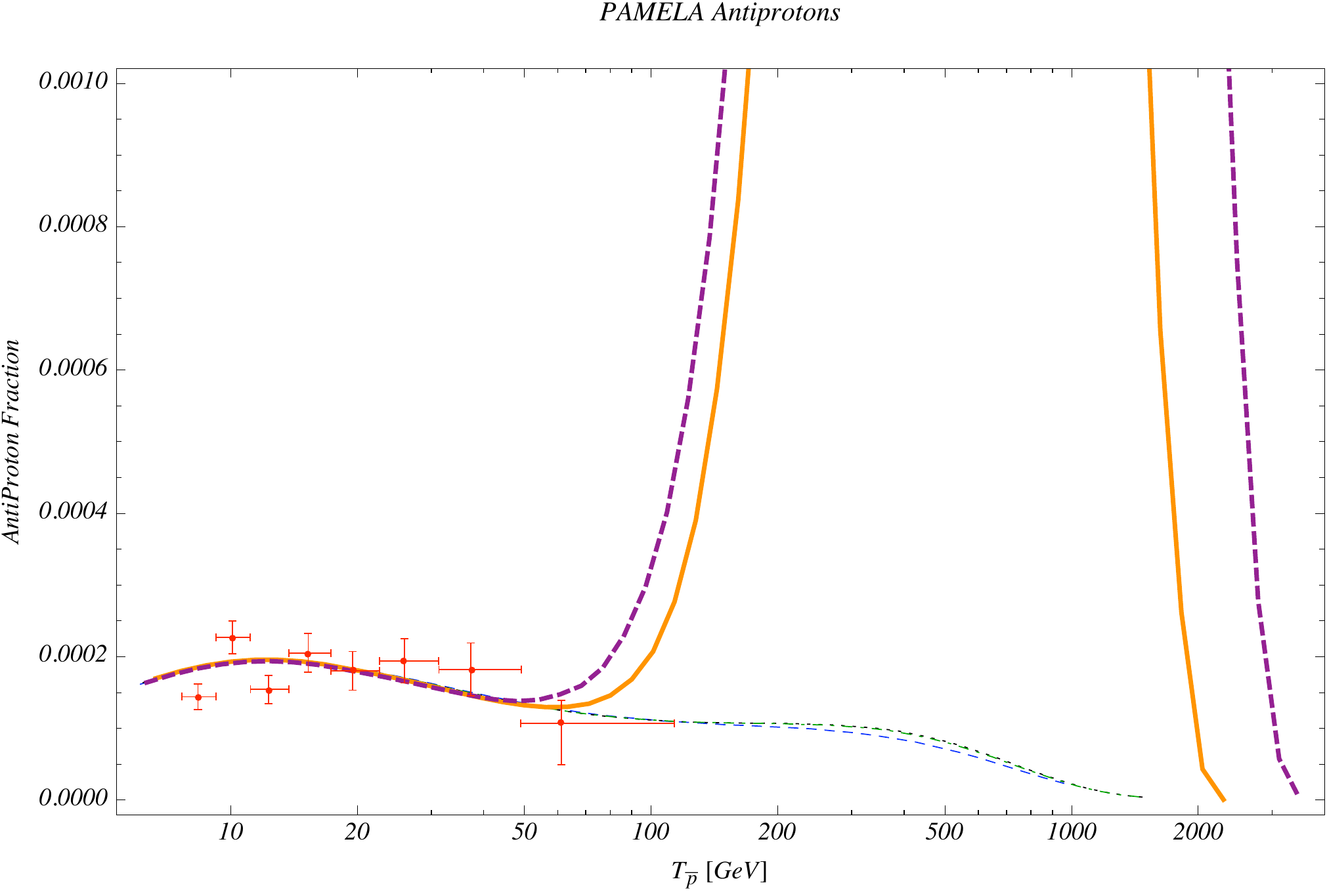} 
\\
\includegraphics[width=3.35in]{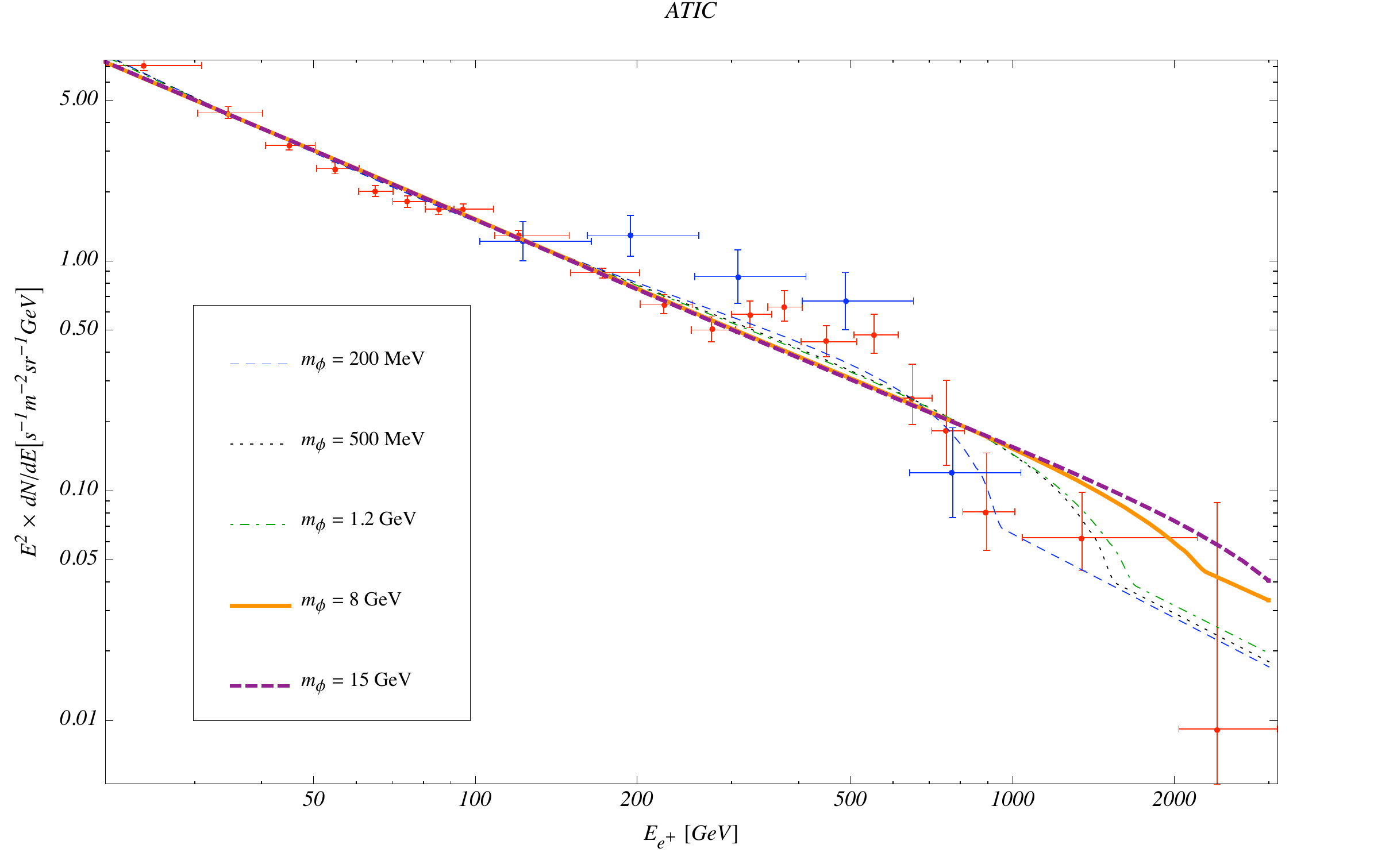} &
\includegraphics[width=3.28in]{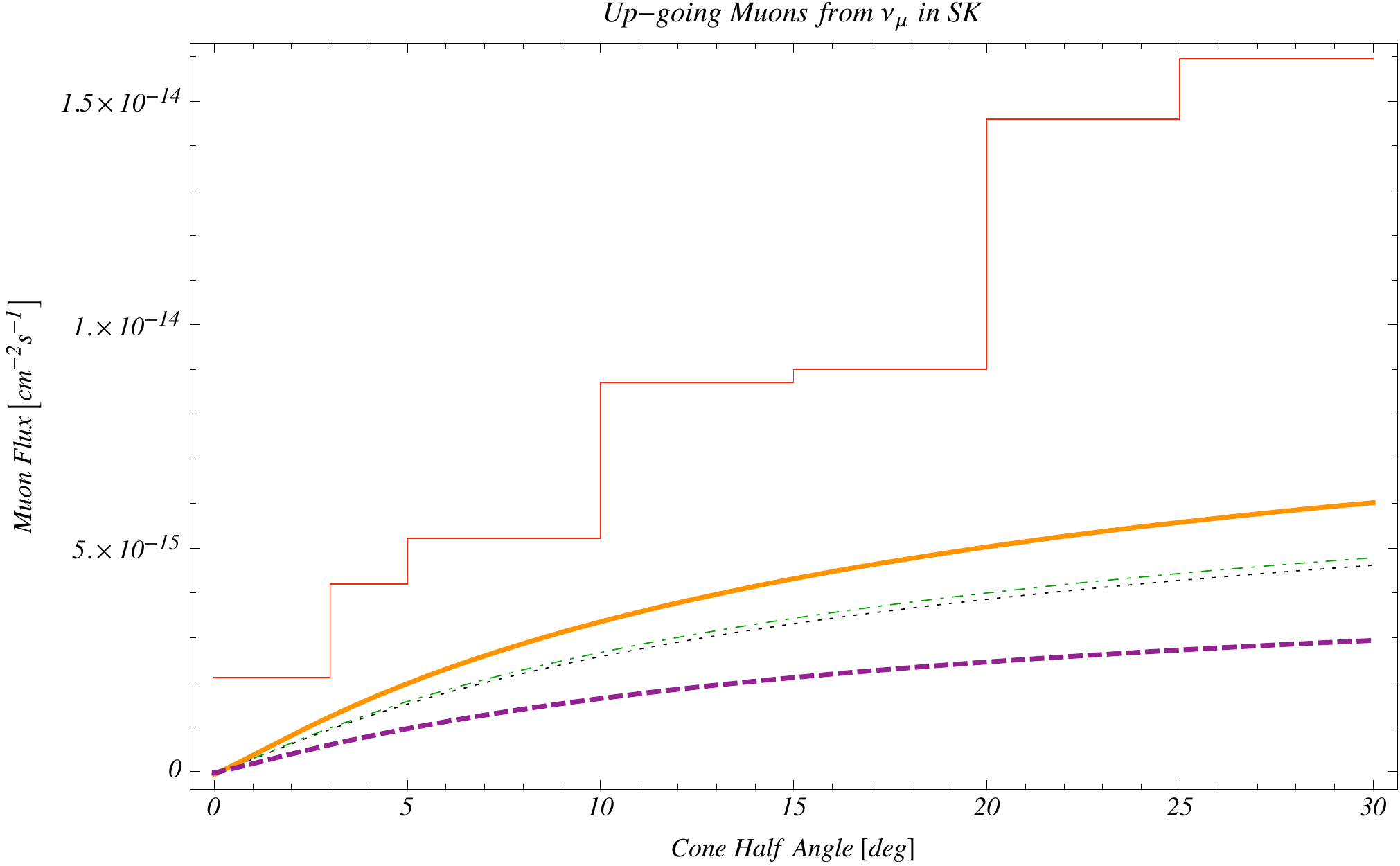}   
\\ 
\hspace{-0.3cm}\includegraphics[width=3.3in]{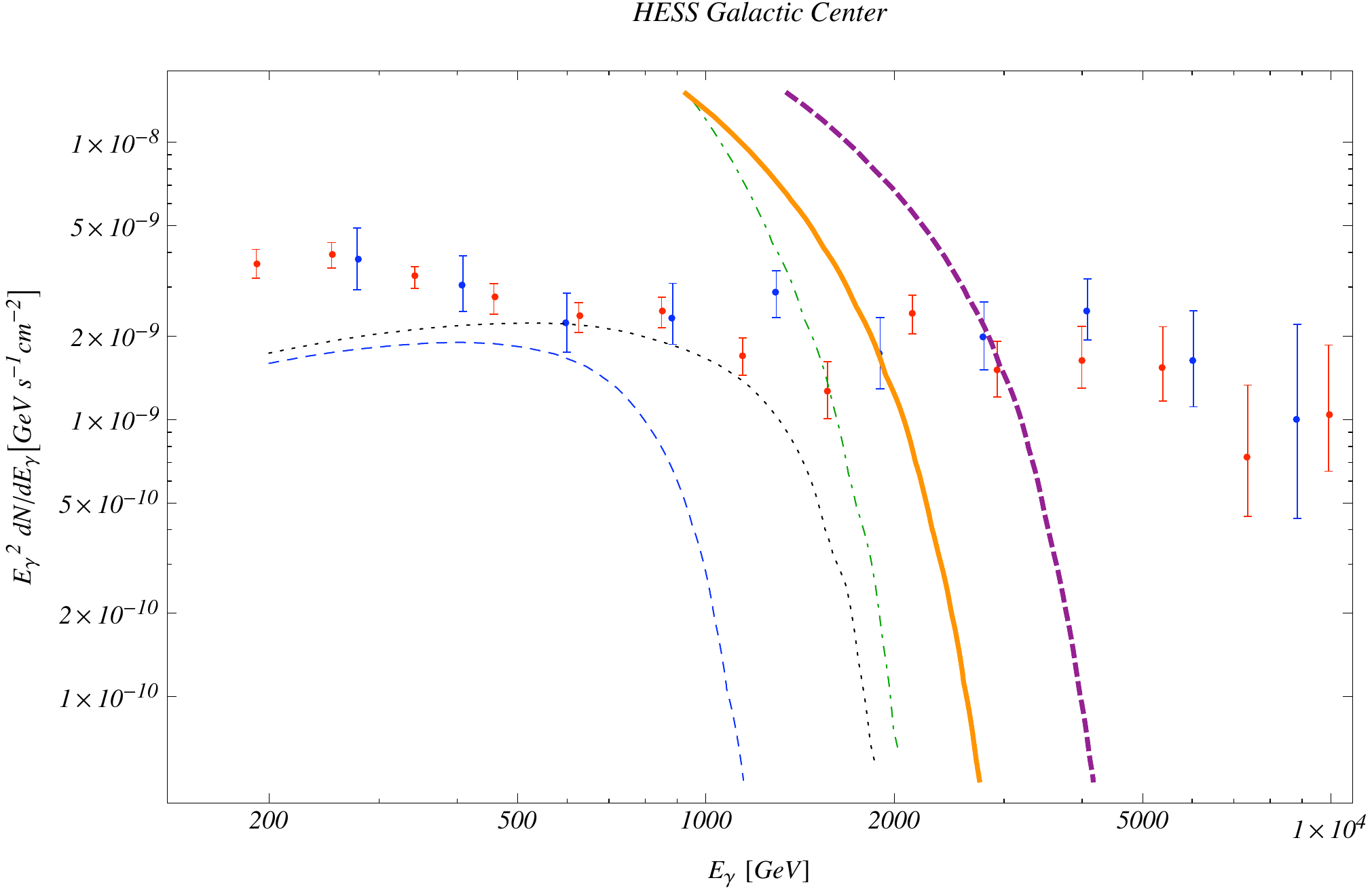}  &  
\includegraphics[width=3.2in]{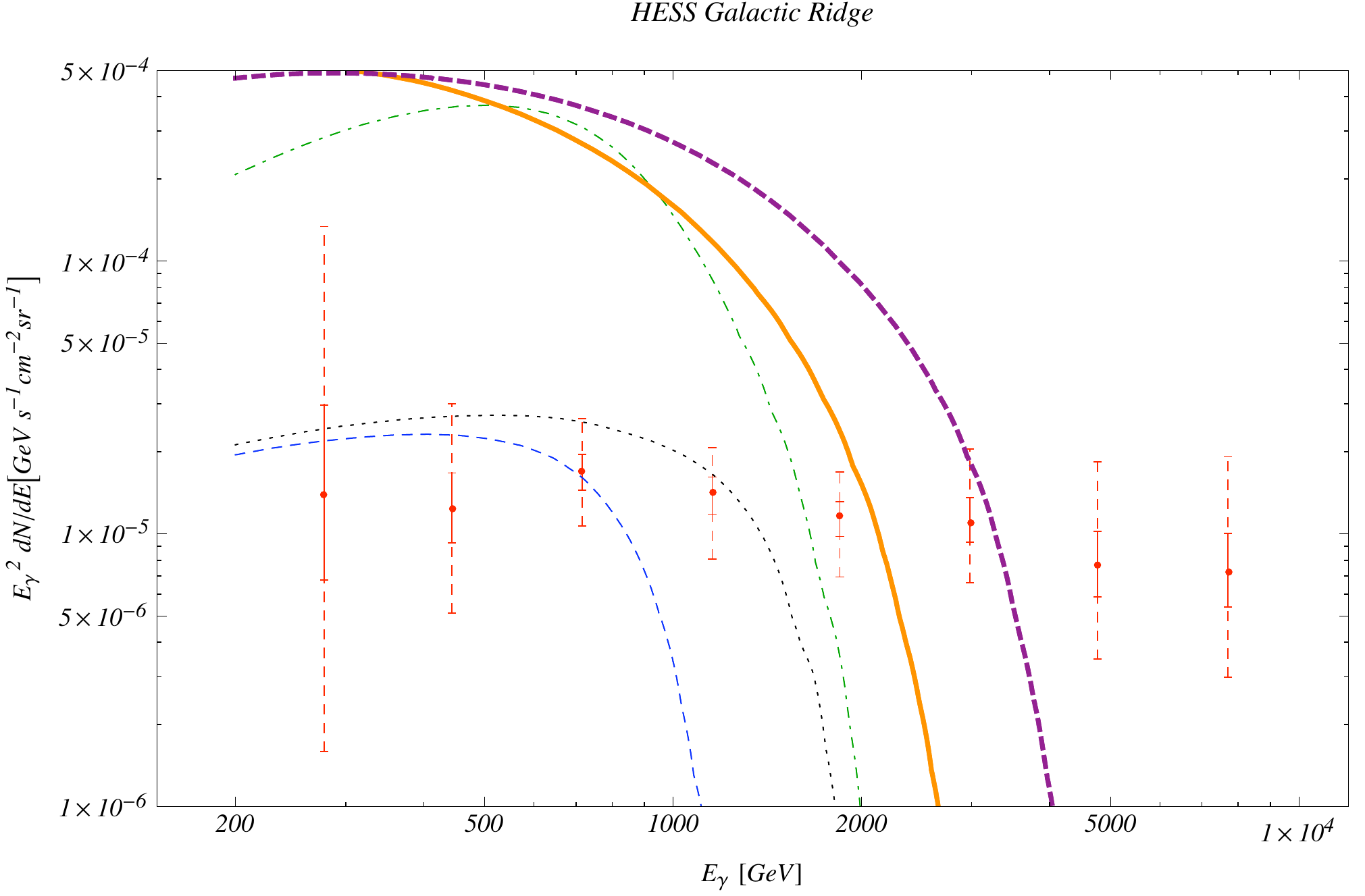}
\end{array}$
\end{center}\
\caption{Best fit to the PAMELA, ATIC and PPB-BETS experiments for
  Einasto profile with $\alpha=0.17$, propagation model MED fixed and
  $m_\phi=0.2,0.5,1.2,8,15$ GeV.  The predictions for neutrinos and
  photons are shown in the last three plots overlayed with the SuperK
  and HESS measurements respectively. Around each bin of the HESS
  data, we indicate the $1\sigma$ (solid) error bar.  For the GR we
  also indicate the $3\sigma$ (dasshed) error bar.  The best fit
  parameters are shown in Table~\ref{tab:iVfit}.}
\label{fig:lightmass}
\end{figure}

In Fig.~\ref{fig:lightmass} we plot
the best fits for PAMELA, ATIC/PPB-BETS as well as the bounds from
neutrinos and HESS experimental data for the
GC and GR in Figure~\ref{fig:lightmass}.  In Table~\ref{tab:iVfit} we give the best fit values
for these plots together with the $\chi^2/{\rm dof}$.  The number of dof's that go into our fit is 33, from 
assuming $40$ independent measurement points (coming from
PAMELA, ATIC and PPB-BETS) and $7$ variables.  As one can see, the
$\chi^2/{\rm dof}$ is typically of order $2$ which for $33$ dof's is a rather poor fit.  We
trace this fact to the two features in the ATIC data: The plateau
above $100$ GeV and a bump at around $500$ GeV which cannot be
explained together with only one DM state.  It remains to be seen if
these two features in the spectrum survive future measurements.  In
the mean time, we view the values of $\chi^2$ as no more than an
indicator for comparing different models.

\begin{table}[t]
\vskip 0.5cm
\begin{center}
{\begin{tabular}{|c|c|c|c|c||c|}
\hline
$m_\phi$ (GeV) & $\mdm$ (TeV) & $10^{23}\times \langle\sigma v\rangle
(\mathrm{cm^3 s^{-1}})$ & $(N,\gamma)_{e^++e^-}$ & $N_{e+}$ & $\chi^2/ {\rm dof}$\\
\hline \hline
0.2 & 0.96 & 0.94 & (0.6,-3.21) & 0.7& 1.9
\\
0.5 & 1.54 & 2.70 & (0.6,-3.20) & 0.6 &2.1
\\
1.2 & 1.46 & 2.51  & (0.6,-3.19) & 0.6&2.1
\\
8 & 2.3 & 5.41 & (0.3,-3.06) & 0.2&2.7
\\
15 & 3.5 & 9.17 & (0.3,-3.06) & 0.3& 2.9
\\
\hline
\end{tabular}}
\end{center}
\caption{The best fit values for the plots in
  Fig.~\ref{fig:lightmass}.  $\gamma_{e^++e^-}$ is the best fit value
  for the spectral index of the background electron plus positron
  flux.  $N_i$ is the fraction of the normalization found for the best
  fit background without a DM signal.  These normalizations are found
  in Section~\ref{sec:ai}.  Finally, the SM ratios, $\rsm$ are all
  smaller than $1\%$. and so are not shown. }
\label{tab:iVfit}
\end{table}

The values plotted for $m_\phi$ are discussed in
Section~\ref{sec:udm}.  From Table~\ref{tab:iVfit} we see that small
$m_\phi$ fit the data better.  This is because as $m_\phi$ increases the positron spectrum becomes softer due to  new kinematically accessible decay channels.  As $m_\phi$ increases $dN/dx$ increasingly deviates from the relatively flat spectrum when $\phi$ only annihilates into $e^++e^-$.  Compared to the flat $dN/dx$, to fit PAMELA and ATIC/PPB-BETS for higher $m_\phi$, one needs a combination of boost factor and increase in mass to attempt to make up for the softer spectrum.

For the case of $8$ and $15$ GeV, the large DM mass (and
therefore the poor fits) is driven by the more constraining antiproton flux.
Indeed, for $m_\phi$ above
twice the proton mass, the PAMELA antiproton-to-proton spectrum
becomes an even  stronger constraint.  In particular, for $m_\phi=8$ and $15$
GeV, the DM mass must be sufficiently large in order to avoid generating
too many antiprotons at energies within the PAMELA reach.  This is indeed
apparent in Fig.~\ref{fig:lightmass} which shows a large bump in the
antiproton spectrum just above the PAMELA $60$ GeV bin.  Whenever the
antiproton constraint pushes the mass too high, denying a good fit to
the positron data, the overall fit for these
masses is significantly worse.

It is interesting to note that a $~2-3$ TeV DM particles is
sufficient in order to avoid the antiproton bound in the case of
$8$ GeV light gauge field.  This is in contrast to a $10$ TeV DM
required in the case of direct W and Z decays as in the SM
\cite{strumiami}.   A simple way to understand this, is to notice that
the anti-proton spectrum is (almost) identical in both cases when
viewed in the gauge boson restframe.  The spectrum must then be
boosted to the lab frame, which introduces a boost factor of order
$\mdm/m_\phi$.  In both the SM and with light gauge fields, this
factor is of order $10^2$ which is the minimum required in order to
avoid the PAMELA antiproton bound, effectively increasing the QCD scale above $100$ GeV.   Thus we conclude that with light
gauge fields the bound is easy to overcome with, $\mdm/{\rm TeV}
\simeq m_\phi/(10\textrm{ GeV})$.

The photon predictions for the HESS measurements
from the GC and GR are shown in the last two plots. It is apparent that the GR is more constraining
than the GC.   Moreover, we do not include photons from
ICS which may strengthen the bound.  Heavy vector bosons,
at or above $1.2$ GeV are excluded by HESS.  On the other hand,
the $200$ and $500$ MeV vector bosons are marginally within 
$3\sigma$ error bars of the measured data.  Clearly, once a background is added
to the fit the data, the tension becomes larger, as the background is required to
be sufficiently large to allow for a good fit to the data points
above $\mdm$ where a DM signal 
is absent.  We therefore conclude that some further suppression (e.g. local boost factor or
shallower DM profile) is
required to render these models viable.  We return to this issue and quantify the suppression in Section~\ref{sec:imp}.

Finally we note that the Neutrino bound is sufficiently weak to
evade and therefore adds no further constrain on the model.

%%%%%%%%%%%%%%%%%%%%%%%%%%%%%%%%%%%%%%%%%%%%%%%%%%%%%%
\subsubsection{Hidden Sector Shower}\label{sec:shower}
%%%%%%%%%%%%%%%%%%%%%%%%%%%%%%%%%%%%%%%%%%%%%%%%%%%%%%

\begin{figure}[h!]
\begin{center}
$\begin{array}{ll}
\hspace{.25cm}\includegraphics[width=3.2in]{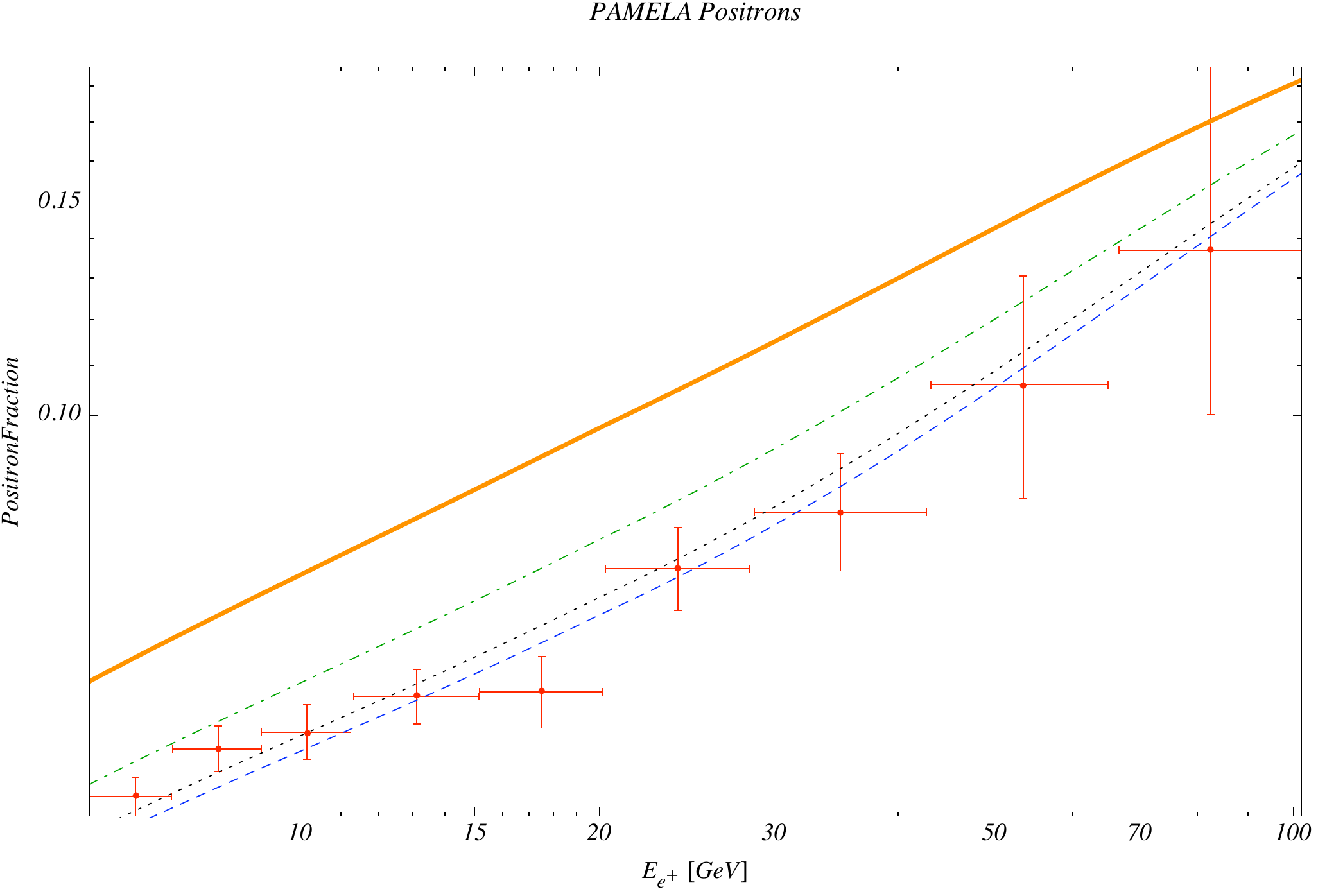} &
\includegraphics[width=3.5in]{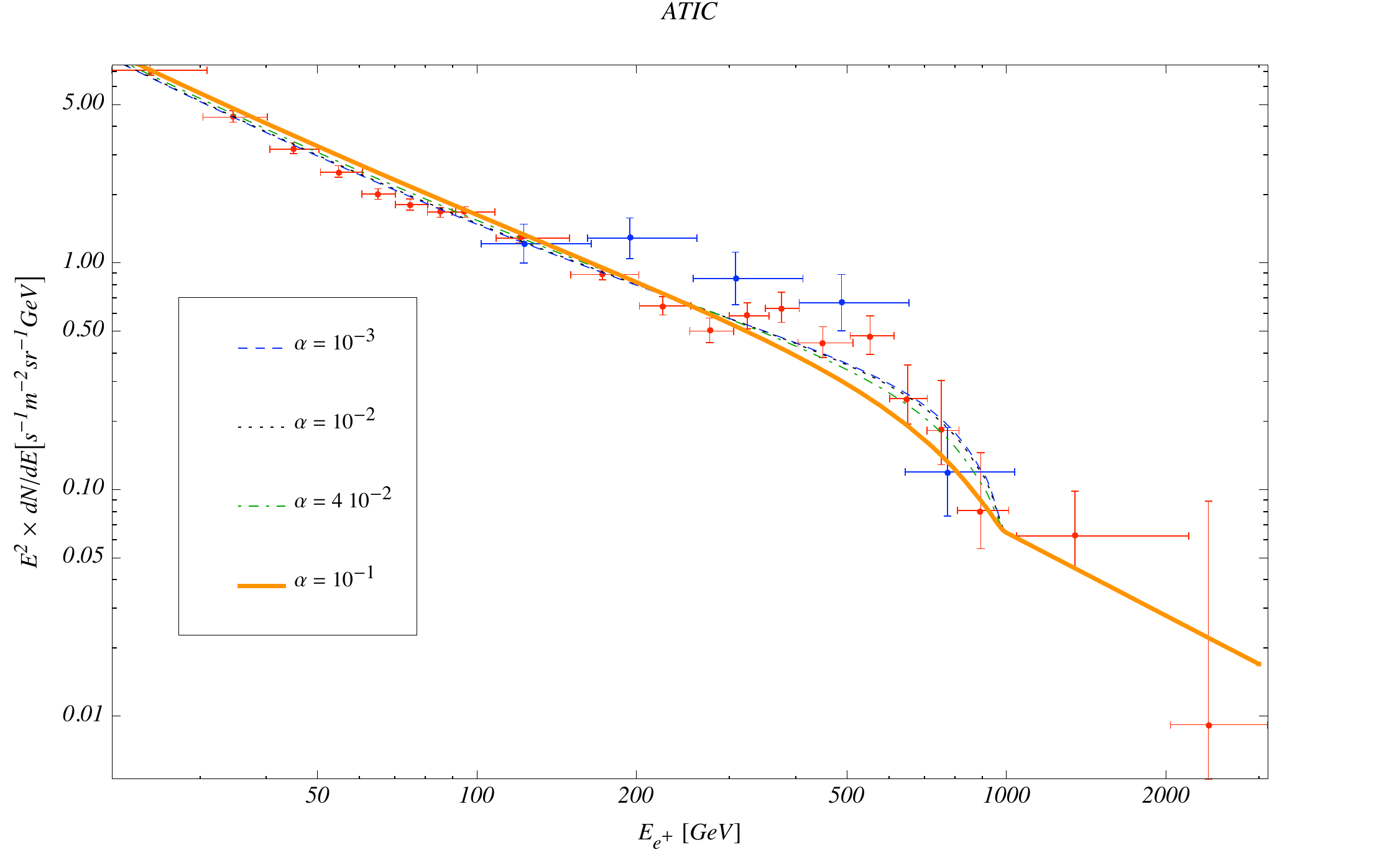}
\\
\includegraphics[width=3.25in]{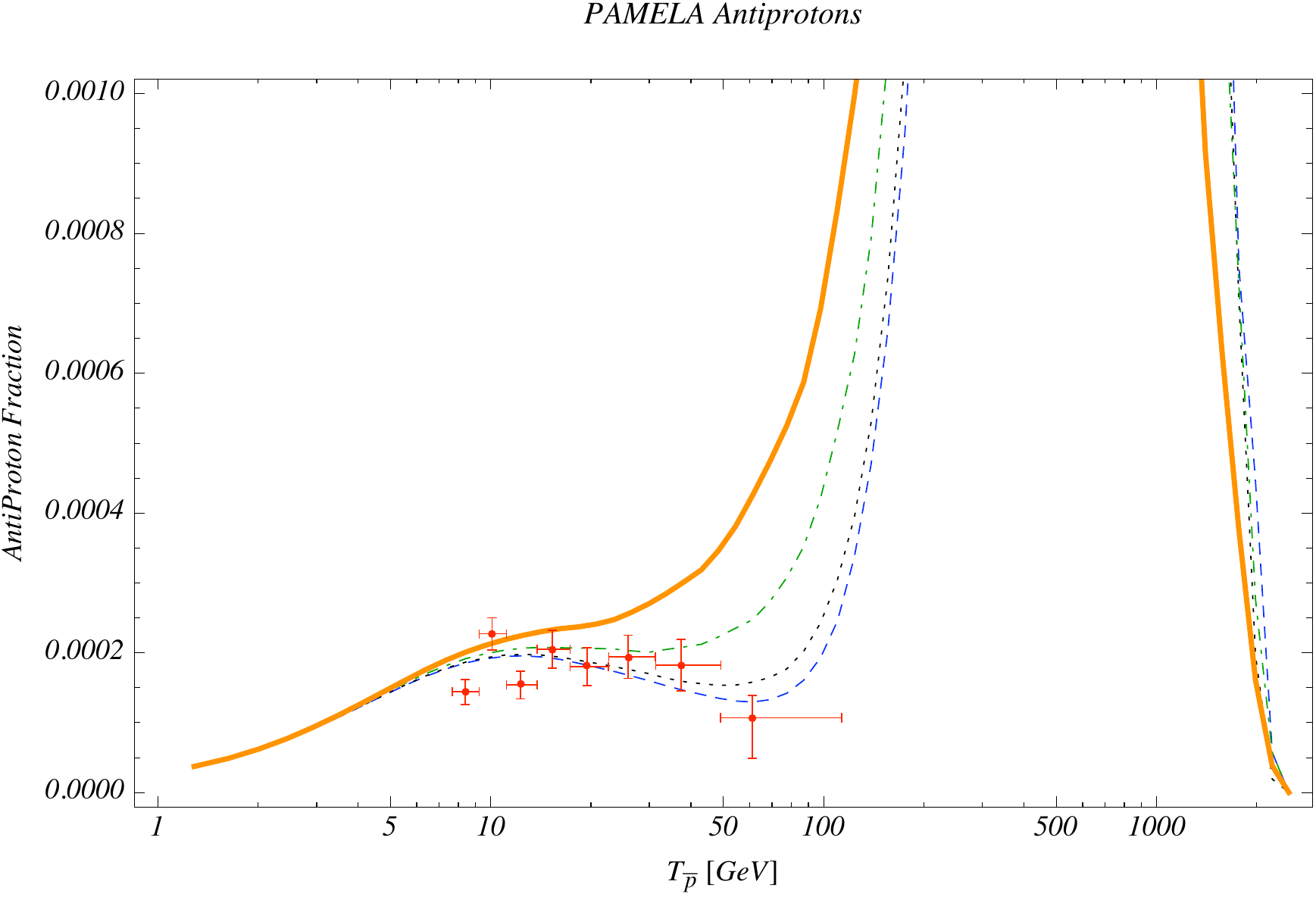}  &
\hspace{-0.3cm}\includegraphics[width=3.35in]{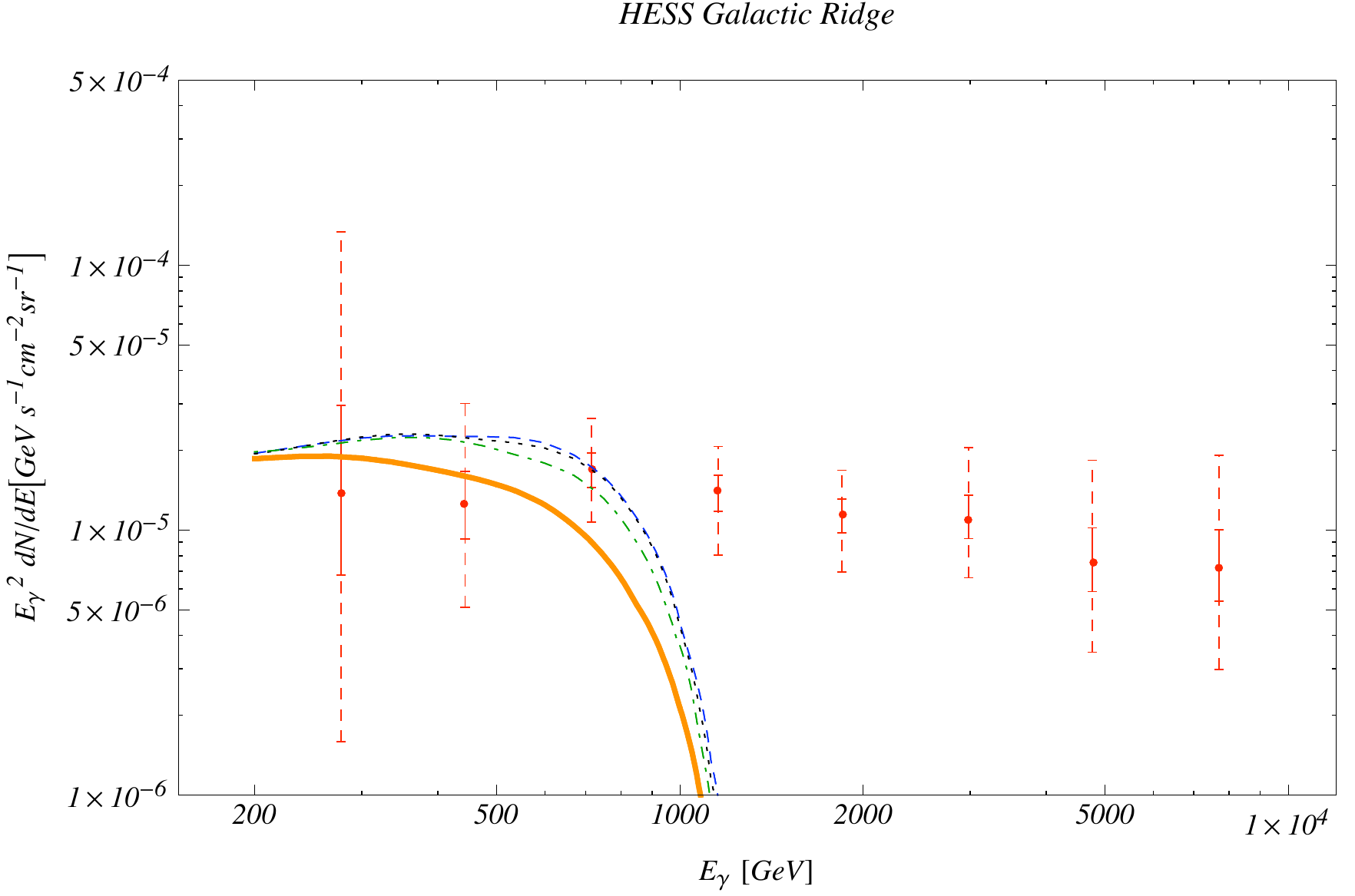}
\end{array}$
\end{center}
\caption{Effects of dark sector showering.  Plotted are the PAMELA,
  ATIC/PPB-BETS and HESS GR data for models with 
  Einasto profile with $\alpha=0.17$, MED propagation model,
  $m_\phi=200$ MeV, $\mdm=1000$ GeV. and $\adm =0.001, 0.01, 0.04,
  0.1$.  These parameters are the best fit for the $\adm=0.001$ case.  In
  this model antiprotons are not produced.  Instead we show the effect
  on antiprotons for the case of $m_\phi=8$ GeV.  Around each bin of the HESS
  data, we indicate the $1\sigma$ (solid) and $3\sigma$ (dasshed)
  error bar. }
\label{fig:iC}
\end{figure}

Let us now study the effects of showering in the dark sector.  As in the case where we isolated the effects of $m_\phi$, we fix an Einasto profile with $\alpha=0.17$ and MED propagation, then  marginalize over the other parameters and plot the fluxes for different values of $\alpha_{DM}$.  Specifically to illustrate the effects of the shower we find the best fit for $\alpha_{DM}=10^{-3}$ and then keeping the same parameters but varying the gauge coupling we plot the resulting fluxes in Fig~\ref{fig:iC}.  To illustrate the
effect of showering on the antiprotons, we repeat the same procedure with $m_\phi=8$ GeV, showing only the antiproton flux ratio.  

There are several effects worth noting.  The first, is that showering
in the DM sector implies a softer electron, positron and antiproton spectra.  As
discussed in the previous section, a softer spectrum fits ATIC less
well.  Moreover, if antiprotons are produced the softening of their spectrum forces $m_\chi$ to be higher. On the other hand, softening of the spectrum implies more
electrons and positrons at the PAMELA energies, which explains the
increase in flux for larger $\adm$ as can be seen in the first plot of
Fig.~\ref{fig:iC}. 

On the other hand, as we will see in Section~\ref{sec:imp}, models
where showering in the DM sector is significant, tend to fit the data
better when taking the HESS measurements into account.  The reason for this
is that the photon spectrum is softened as well due to the showering.  This is
apparent in the last plot of Figure~\ref{fig:iC} where we see
that larger $\adm$ implies less photons.  Since
the strongest constraints come from photons, it is preferred (for
minimizing the $\chi^2$) to allow for softening of both the photons and
electrons, so that the ATIC fit is poorer but the photon fit is
better.

%%%%%%%%%%%%%%%%%%%%%%%%%%%%%%%%%%%%%%%%%%%%%%%%%%%%%%
\subsubsection{SM Ratio and Direct Decays}\label{sec:sm-ratio-direct}
%%%%%%%%%%%%%%%%%%%%%%%%%%%%%%%%%%%%%%%%%%%%%%%%%%%%%%

We now study the implications of DM directly annihilating into SM particles.  A similar study was presented in \cite{strumiami,strumiagamma}.   Here we
confirm part of their results, while emphasizing the difference between
the predictions of direct annihilations versus those through light vector
fields.  

To this end, we concentrate on two possibilities, namely direct
couplings to SM gauge bosons, and direct couplings to $e^\pm$.  From
the point of view of the PAMELA results, the latter is preferable as
it does not incorporate any hardronic decays and therefore no
anomalous antiproton flux.  On the other hand the former is guaranteed to exist at some level if the $\phi$ can mix with the SM vector bosons.

To emphasize the difference between
$2\rightarrow 2$ and $2\rightarrow 4$ decays, we compare the
predictions of direct anihilations to the case of $m_\phi=200$ MeV, for which
only $e^\pm$ production is kinematically accessible.  As was discussed
in subsection~\ref{sec:light-gauge-boson}, this is also the model
which best fits the data if shower is not included.  The best fit parameters are shown in the
first line of Table~\ref{tab:iVfit}.  The $\chi^2$ value for the best
fit of the direct decay is found to be $\chi/{\rm dof} = 2.6$ which is
worse than that of the $2\rightarrow 4$ case. The best fit parameters are,
\begin{eqnarray}
  \label{eq:4}
  &\mdm = 680 \textrm{ GeV}, \qquad \langle\sigma v\rangle = 6.34\times
  10^{-23} \mathrm{\ cm^2}, \qquad \rsm = 2\%,
  \\
  &(N,\gamma)_{e^++e^-} = (0.9,-3.27), \qquad N_{e^+} = 1.
\end{eqnarray}
As in previous cases, $N_i$ denotes the fraction of the
normalization found for the backgrounds, while
$\gamma_{e^++e^-}$ is the spectral index for the correspond
background flux.

\begin{figure}[h!]
\begin{center}
$\begin{array}{ll}
\hspace{.8cm}\includegraphics[width=3.15in]{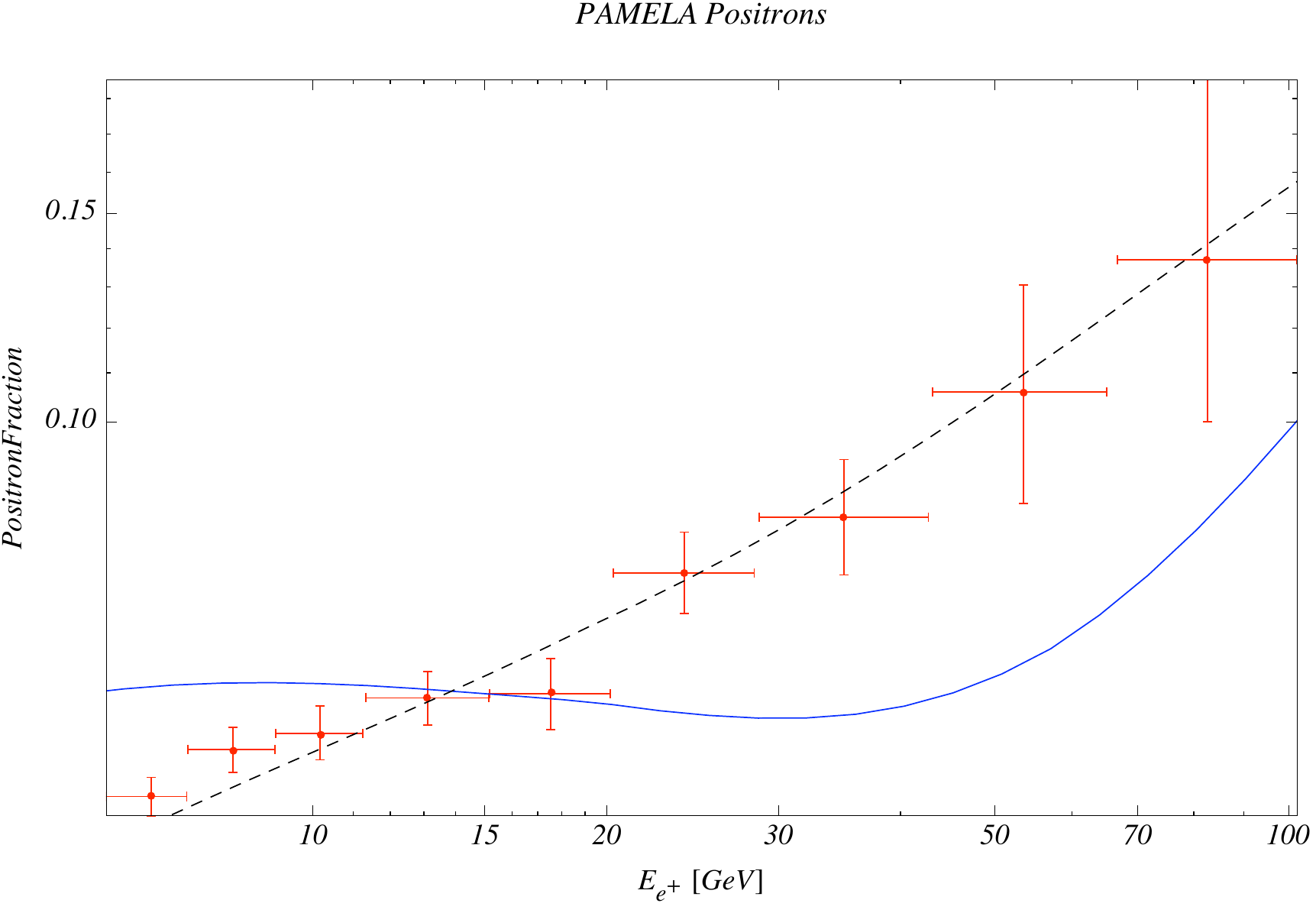}
& \includegraphics[width=3.5in]{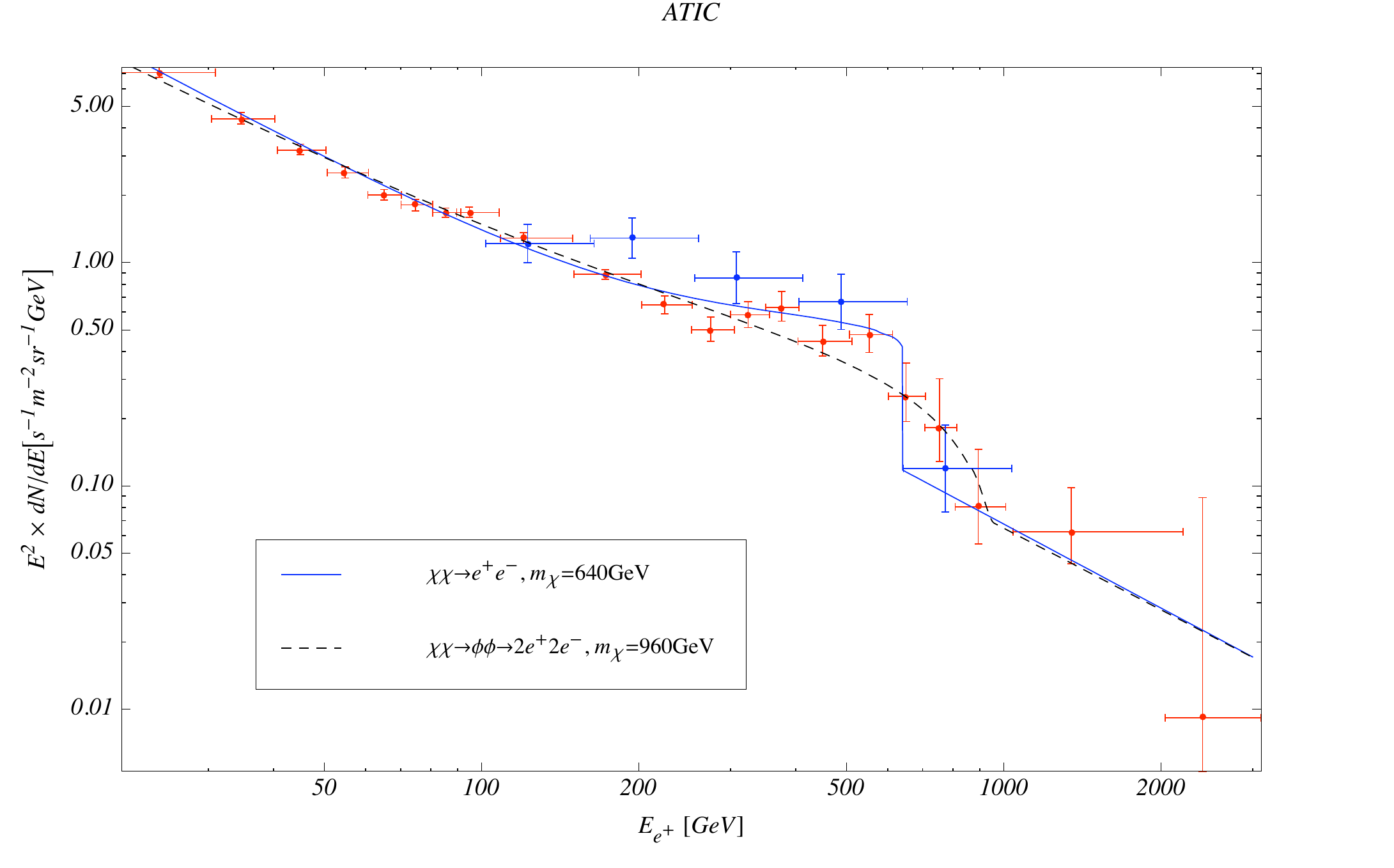}   
\\ 
\includegraphics[width=3.5in]{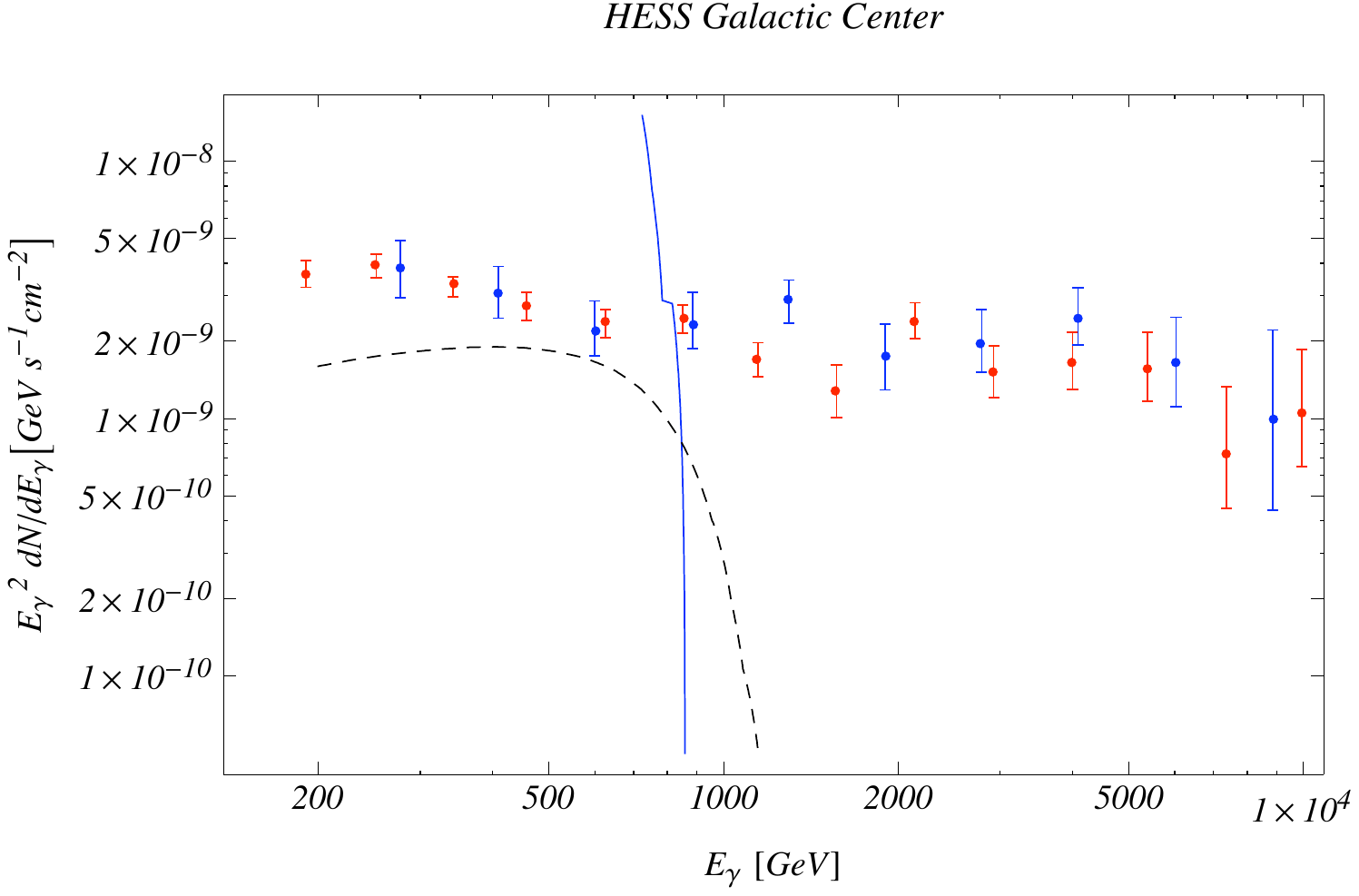}  &  
\hspace{-0.5cm}\includegraphics[width=3.4in]{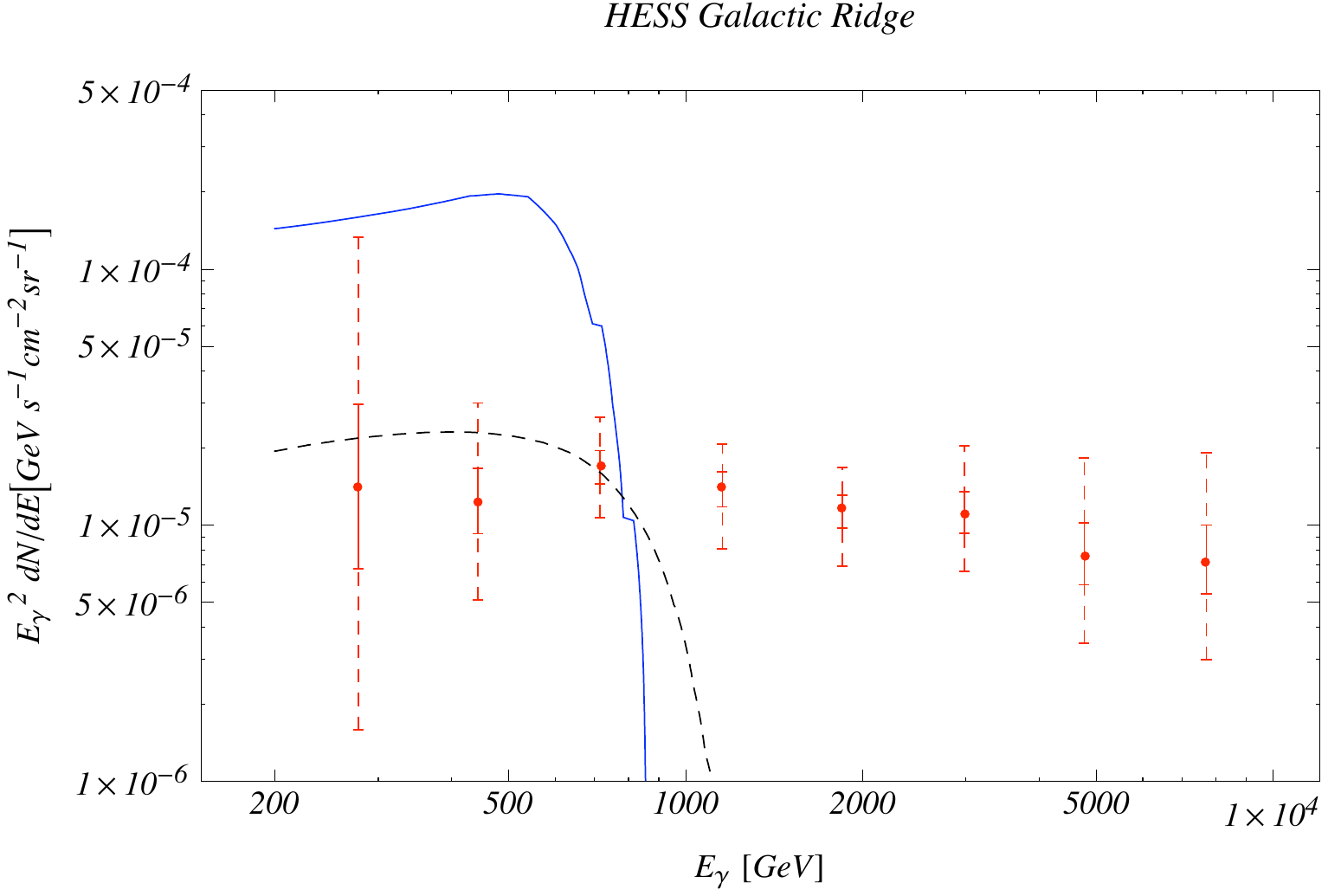}
\end{array}$
\end{center}
\caption{A comparison between an $e^+e^-$ annihilation channel and the
  best fit result for annihilation through two $200$ MeV light state
  which each decays into $e^+e^-$.  The details of the $2\rightarrow4$
  model is shown in Table~\ref{tab:iVfit}.  The fit done with the
  Einasto profile with $\alpha=0.17$ and propagation model MED fixed.
  The antiproton and neutrino spectrum is not shown since those are
  not produced due to kinematics.  The predictions for photons are
  shown in the last two plots overlayed with the HESS measurements.
  Around each bin of the HESS data, we indicate the $1\sigma$ (solid)
  and $3\sigma$ (dasshed) error bar.  The best fit parameters are
  shown in Table~\ref{tab:iVfit}.}
\label{fig:EEflux}
\end{figure}

In Fig~\ref{fig:EEflux} we plot the two models together with the PAMELA
positrons, ATIC/PPB-BETS and HESS measurements.    While the fit to the ATIC/PPB-BETS data is better, it is
clear that direct coupling to $e^\pm$ does not fit 
the PAMELA data well. The reason for this can be traced back to the
injection spectrum for the $e^\pm$ line shown in Fig.~\ref{fig:EE}:  Since the
decay is $2\rightarrow2$, the ATIC
data dictates a low DM mass around the bump.  For such a low mass, the
hard form of the  spectrum is then insufficient 
to explain the PAMELA anomaly. 

To partly compensate for the injection spectrum, a large
cross-section, is required.  This, together with the very hard
spectrum explains the large number of photons
predicted in such a case, as shown in Fig.~\ref{fig:EEflux}.   Clearly,
this model is in contradiction with the HESS data and is therefore
excluded \cite{strumiagamma}.
 
\begin{figure}[t] 
   \centering
   \includegraphics[width=3in]{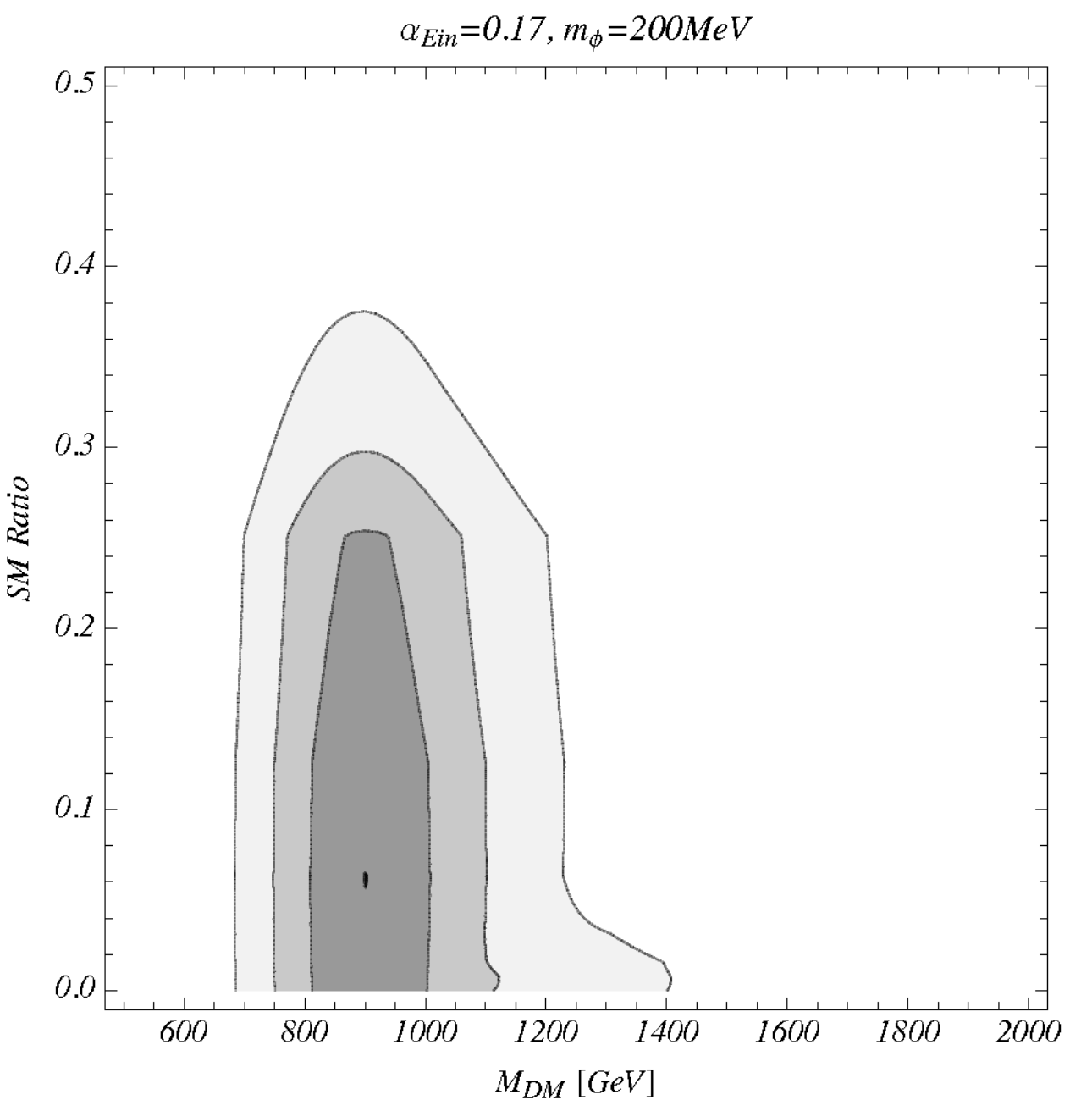}
   \caption{A confidence level contour plot for $m_\phi=200$, and Einasto
     $\alpha=0.2$ as a function of $m_\chi$ and $\rsm$.  The contours
     correspond to the $68\%$, $95\%$ and $99\%$ confidence levels.}
   \label{fig:SMRatio}
\end{figure}

Next we would like to understand the extent to which the DM particle
can couple to the SM gauge vector bosons. A known constraint is that the
massive SM gauge fields decay into hadronic states and may
therefore produce an unacceptable excess of antiprotons.  It is
interesting to quantify this statement.  We do that in
Figure~\ref{fig:SMRatio} where we plot 2D confidence level contours
for the allowed regions of parameter space with  $m_\phi=200$ MeV and 
Einasto $\alpha=0.2$ profile,  as a function of $\mdm$ and
$\rsm$.  We see that while the best fit value prefers $5\%$ of SM
annihilations, at $1\sigma$ one can have as much as $25\%$.  While in
this plot we did not attempt to fit the HESS data, we have checked
that  models which fit the HESS data without difficulty, such as Einasto profile with $\alpha=0.2$, do not change this
result significantly (see Sect.~\ref{sec:imp}).  Conversely, models that are only marginally
consistent with HESS, such as the Einasto profile with $\alpha=0.17$,
reduces the parameter space almost entirely.
For model building this result implies that the DM particle must
couple only weakly to the SM.  This can be implemented either by
coupling $\chi$ to SM singlets, or otherwise making  $\adm$ sufficiently
large.

%%%%%%%%%%%%%%%%%%%%%%%%%%%%%%%%%%%%%%%%%%%%%%%%%%%%%%
%%%%%%%%%%%%%%%%%%%%%%%%%%%%%%%%%%%%%%%%%%%%%%%%%%%%%%
\subsection{Astrophysics Uncertainties}\label{sec:astro}
%%%%%%%%%%%%%%%%%%%%%%%%%%%%%%%%%%%%%%%%%%%%%%%%%%%%%%
%%%%%%%%%%%%%%%%%%%%%%%%%%%%%%%%%%%%%%%%%%%%%%%%%%%%%%

%%%%%%%%%%%%%%%%%%%%%%%%%%%%%%%%%%%%%%%%%%%%%%%%%%%%%%
%%%%%%%%%%%%%%%%%%%%%%%%%%%%%%%%%%%%%%%%%%%%%%%%%%%%%%
\subsubsection{Profile}\label{sec:astroprof}
%%%%%%%%%%%%%%%%%%%%%%%%%%%%%%%%%%%%%%%%%%%%%%%%%%%%%%
%%%%%%%%%%%%%%%%%%%%%%%%%%%%%%%%%%%%%%%%%%%%%%%%%%%%%%

We now study the dependence of the predictions on the DM profile.
As discussed in Section~\ref{sec:dm-halos}, current N-body simulations
do not allow us to pin-point the precise DM profile.  The main
difficulty for these simulations is the resolution, which does not
allow one to probe the DM distribution within $\sim100$ pc from the GC.  Moreover, baryons are not incorporated 
in simulations, while they  may play important roles in the GC.

The effective diffusion scale for electrons is smaller than the
distance of the solar system from the GC and therefore their flux is
not sensitive to the large uncertainties in the inner DM profile.  On the
other hand, photons do not diffuse and therefore most of them come
from the center where the bulk of the DM lies.  This then allows one
to probe and constrain theories of DM in conjunction with the DM profiles
that are extracted from N-body simulations.  

\begin{figure}[hp]
\begin{center}
$\begin{array}{ll}
\multicolumn{2}{c}{\includegraphics[width=3.1in]{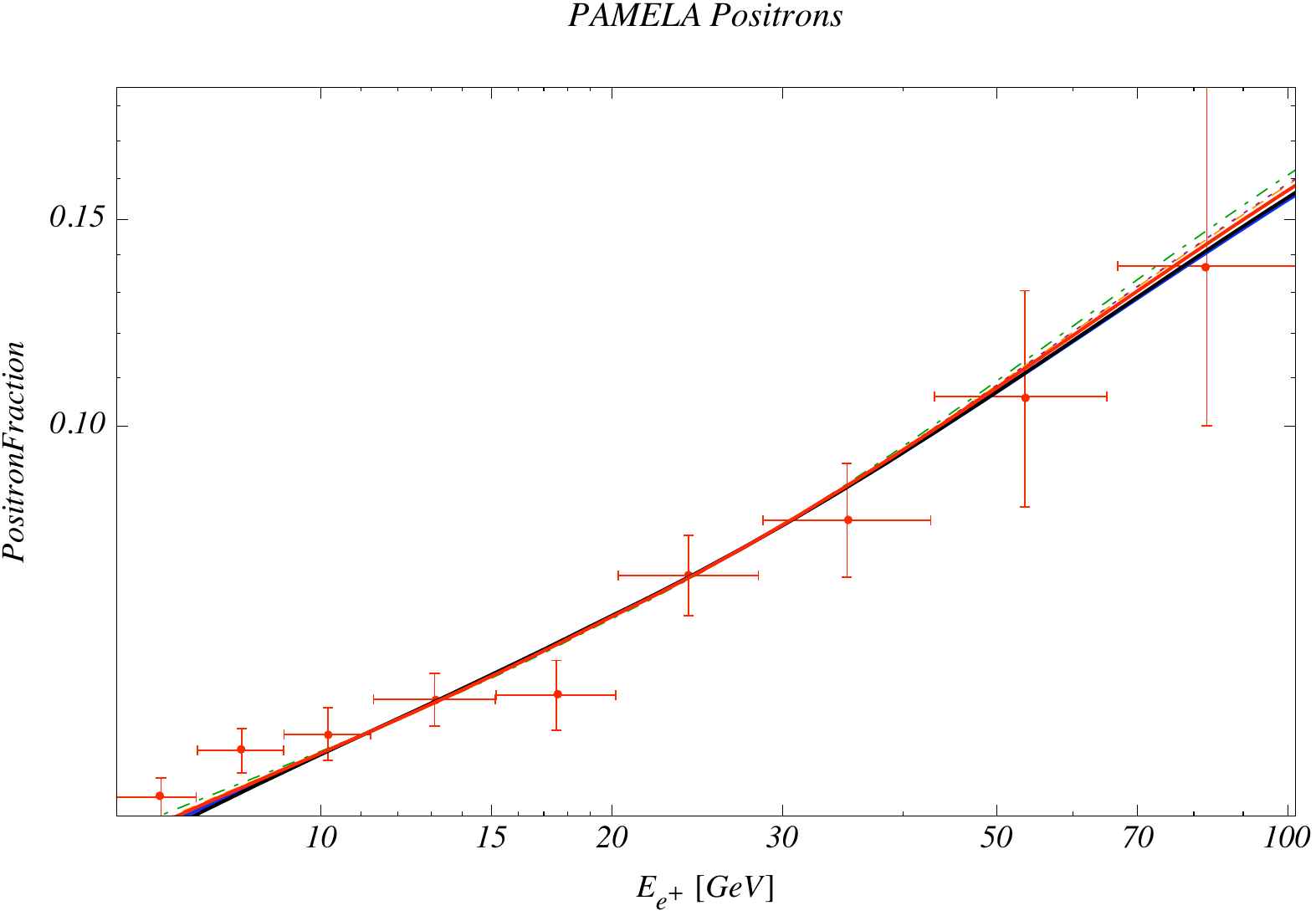}}
\\
\includegraphics[width=3.5in]{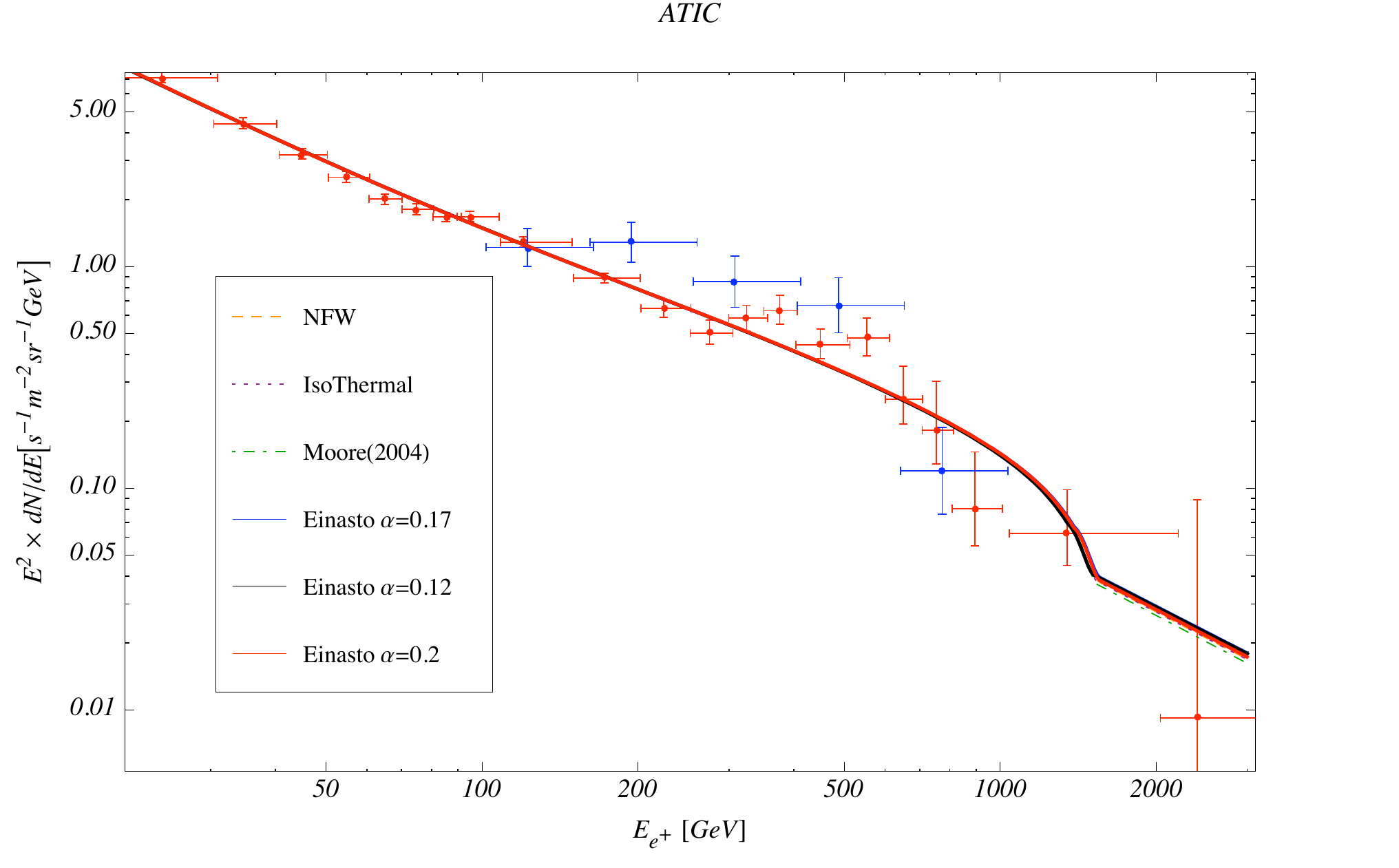}
& \hspace{-0.8cm}\includegraphics[width=3.4in]{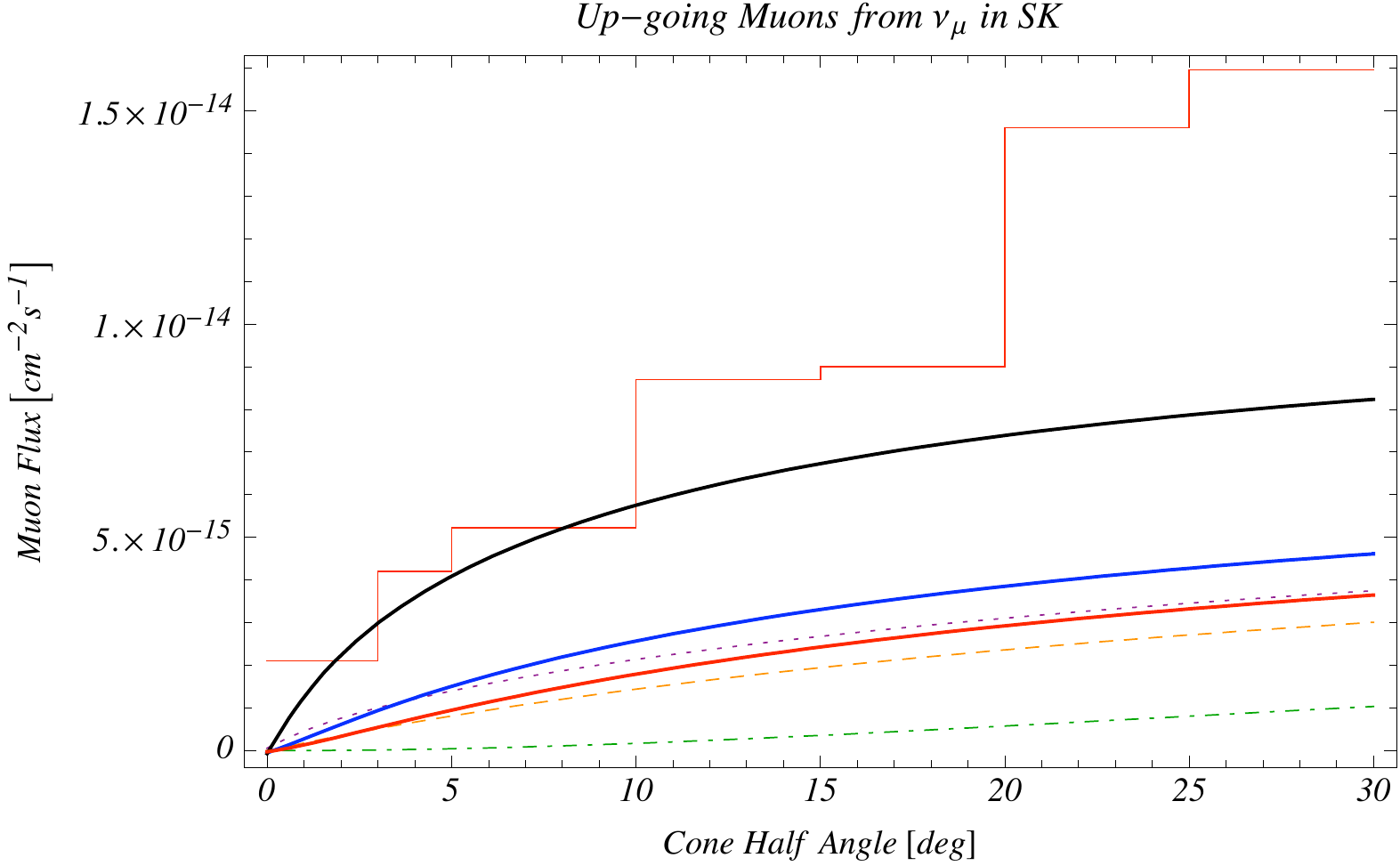}   
\\ 
\hspace{-0.3cm}\includegraphics[width=3.3in]{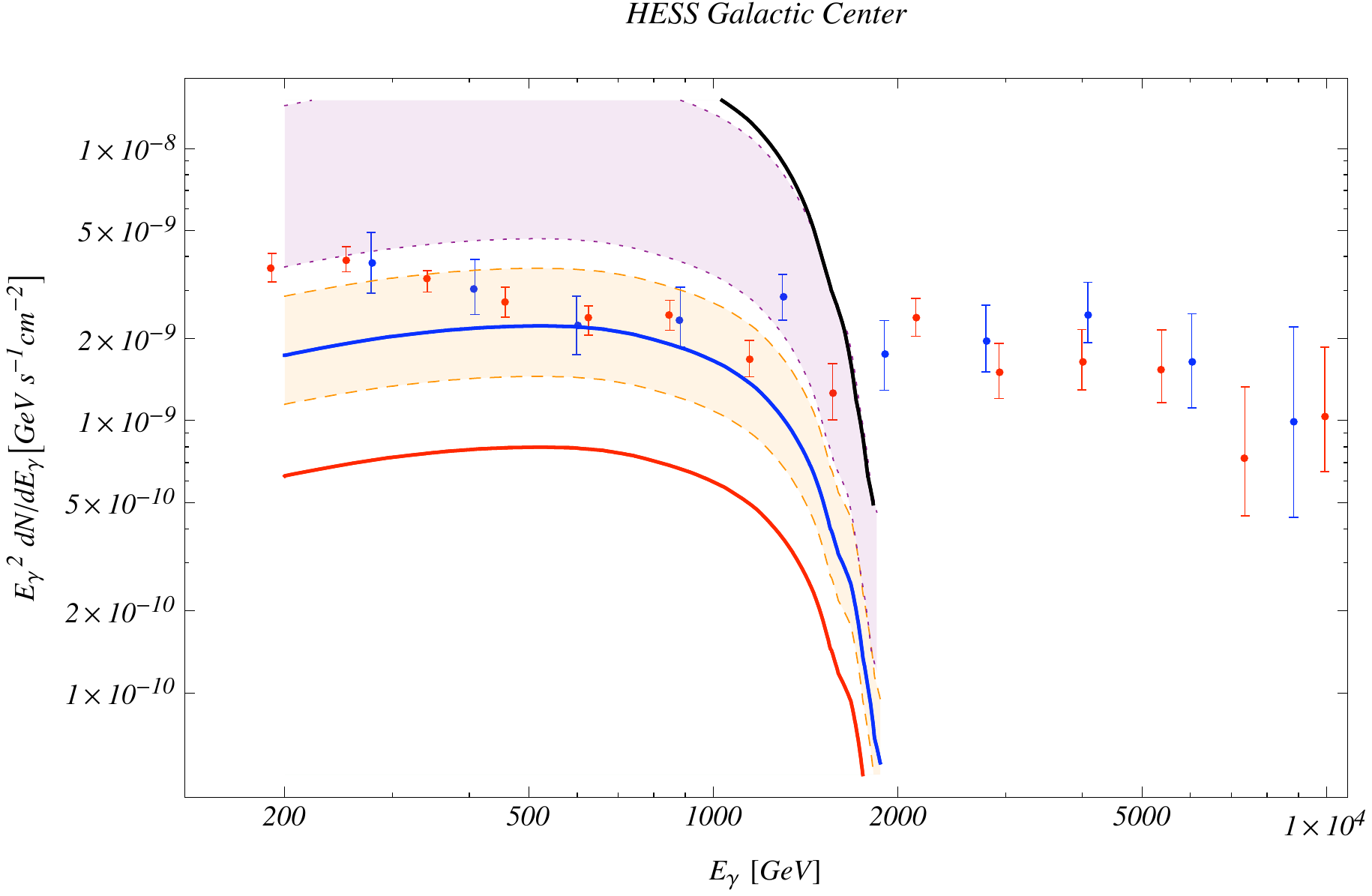}  &  
\hspace{-0.5cm}\includegraphics[width=3.3in]{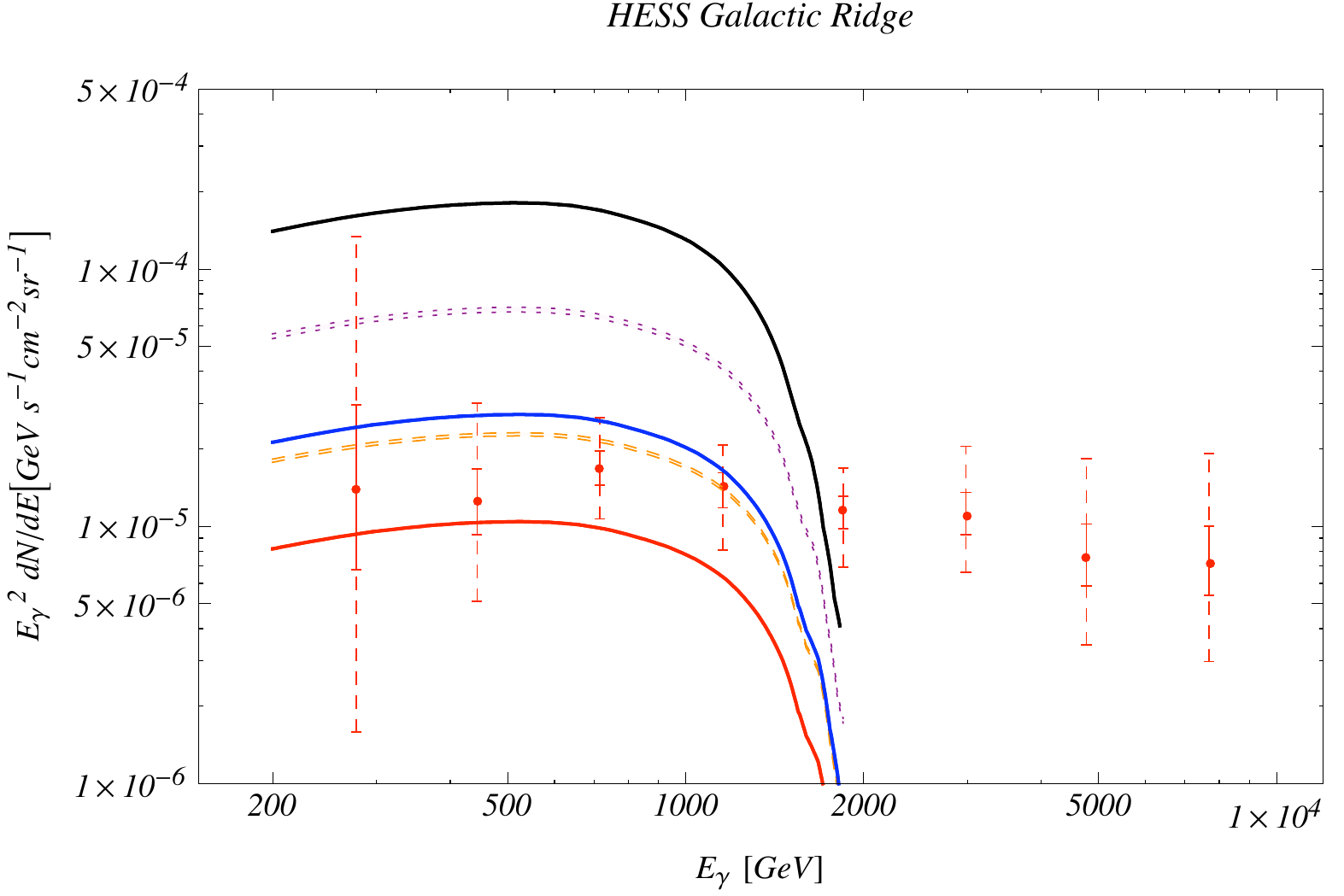}
\end{array}$
\end{center}
\caption{Best fit to the PAMELA, ATIC and PPB-BETS experiments for
  different DM profiles and with $m_\phi=500$ MeV and $\mdm=1.54$ TeV.
   Propagation model is taken to be MED.  No antiproton bound exists
   due to kinematics.  The bands in the photon predictions come from
   regulating the NFW and Moore profiles with $0\leq r_s\leq 100$ pc.  Around
  each bin of the HESS data, we indicate the $1\sigma$ (solid) and
  $3\sigma$ (dasshed)  error bar.  The 
  best fit parameters are shown in Eqs.~\eqref{eq:7},~\eqref{eq:15}.}
\label{fig:iProf}
\end{figure}

To study the profile dependence we take  $m_\phi = 500$ MeV, and scan
over the six profiles: NFW \cite{nfw}, Moore \cite{moore}, Isothermal
Core \cite{isothermal} and Einasto \cite{einasto} with
$\alpha=(0.12,0.17,0.20)$, fitting the rest
of the parameters to PAMELA and ATIC/PPB-BETS.  Because we do not
attempt to fit to the HESS data here, the fits are almost identical in
their resulting parameters.  We find (on average),
\begin{eqnarray}
\label{eq:7}
  &\mdm = 1540 \textrm{ GeV}, \qquad \langle\sigma v\rangle = 2.88\times
  10^{-23} \mathrm{\ cm^2}, \qquad \rsm = 0,
  \\
\label{eq:15}
  &(N,\gamma)_{e^++e^-} = (0.6,-3.21), \qquad N_{e^+} = 0.4-1.
\end{eqnarray}
In Figure~\ref{fig:iProf}, we show the usual plots together with the
predictions for the photons.  The bands in these plots show the
sensitivity of the NFW and Moore profiles to the distance $0\leq
r_s\leq 100$ pc from the center of the Galaxy, below which we
regularize the profiles as discussed in
Section~\ref{sec:dm-halos}.  No such regularization is performed for
the Einasto and Isothermal profiles since they do no diverge at the
GC.  

The HESS plots in the figure demonstrate the strength of the
constraint arising from the HESS data, the strongest coming from the
GR.  These essentially constrain $\bar J$ defined in
Eq.~\eqref{eq:J}.  We stress that, as before, the expected photon
spectrum shown does not include any background.  Adding such a
background is crucial in order to fit the measured data at energies
above that of the DM.  In such a case, the constraints are much
stronger.  We return to this issue in the next section.  

Even without background we learn that both the Moore and Einasto with
$\alpha=0.12$ are excluded by more than $3\sigma$ while NFW and
Einasto with $\alpha=0.17$ are above the data, but within the
$3\sigma$ error bars. As we shall see below, once background is added,
these profiles are excluded unless some kind of suppression is in place, like the hidden sector shower, a local boost factor, etc..

It is important to note that independently of $\bar J$, these photon
predictions provide an conservative estimation of the signal for
two reasons: (i). We have not taken ICS into account.  Such radiation
is certainly important at low energy, but could also be important at
intermediate scales.  (ii).  As discussed in
Section~\ref{sec:gammain}, for the GR, we do not include any photons coming from
the GC.  On the other hand, the GR measurements count photons from the
center, after removing a dominat background source \cite{hessgr}.   This is approximately a $10\%$ effect.

The annihilating DM scenario studied here is somewhat complimentary to
the studies of \cite{strumiagamma}.  We therefore conclude that the
HESS measurements together with PAMELA and ATIC/PPB-BETS, strongly constrain the 
possible DM profiles in the case of annihilating DM scenarios. While
not excluded, future experiments may strengthen the bounds considerably.

%%%%%%%%%%%%%%%%%%%%%%%%%%%%%%%%%%%%%%%%%%%%%%%%%%%%%%
%%%%%%%%%%%%%%%%%%%%%%%%%%%%%%%%%%%%%%%%%%%%%%%%%%%%%%
\subsubsection{Propagation}\label{sec:astroprop}
%%%%%%%%%%%%%%%%%%%%%%%%%%%%%%%%%%%%%%%%%%%%%%%%%%%%%%
%%%%%%%%%%%%%%%%%%%%%%%%%%%%%%%%%%%%%%%%%%%%%%%%%%%%%%

Finally, let us consider the uncertainties arising from the
propagation parameters.  Here we stress again that our choice of
propagation model already entails some uncertainties as
our understanding of Cosmic ray propagation is not complete.  Nevertheless,
we show how our predictions vary as we change the parameters as
discussed in Section~\ref{sec:propagation}, which were shown to span
the possible positron and antiproton spectra.  Such uncertainties have
been studied extensively \cite{mmm,Delahaye:2008ua,Putze:2008ps,Moskalenko:2001ya,Strong:1998pw} so we only concentrate on our predictions.  

\begin{figure}[hp]
\begin{center}
$\begin{array}{ll}
\hspace{.5cm}\includegraphics[width=3.1in]{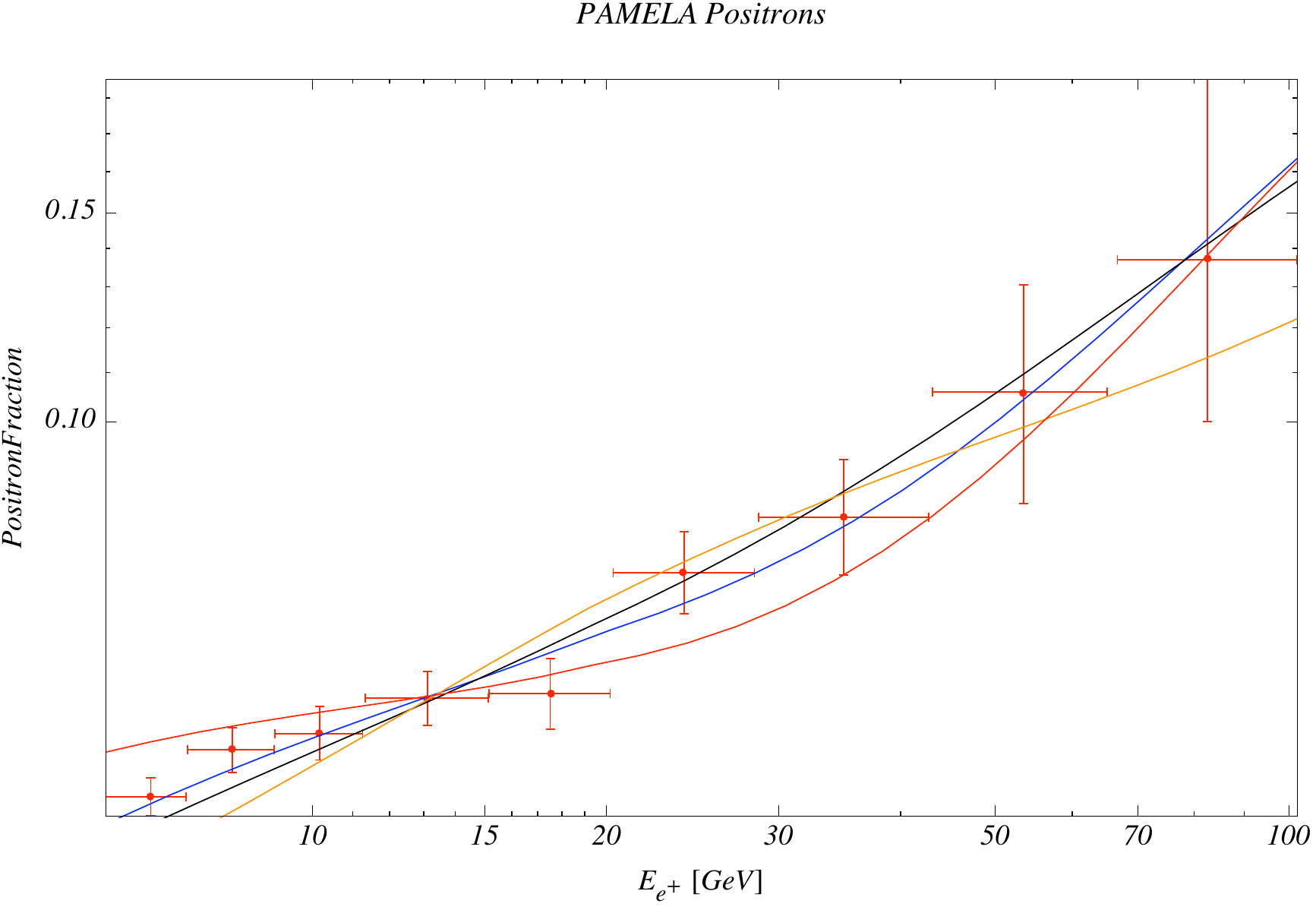} &
\includegraphics[width=3.5in]{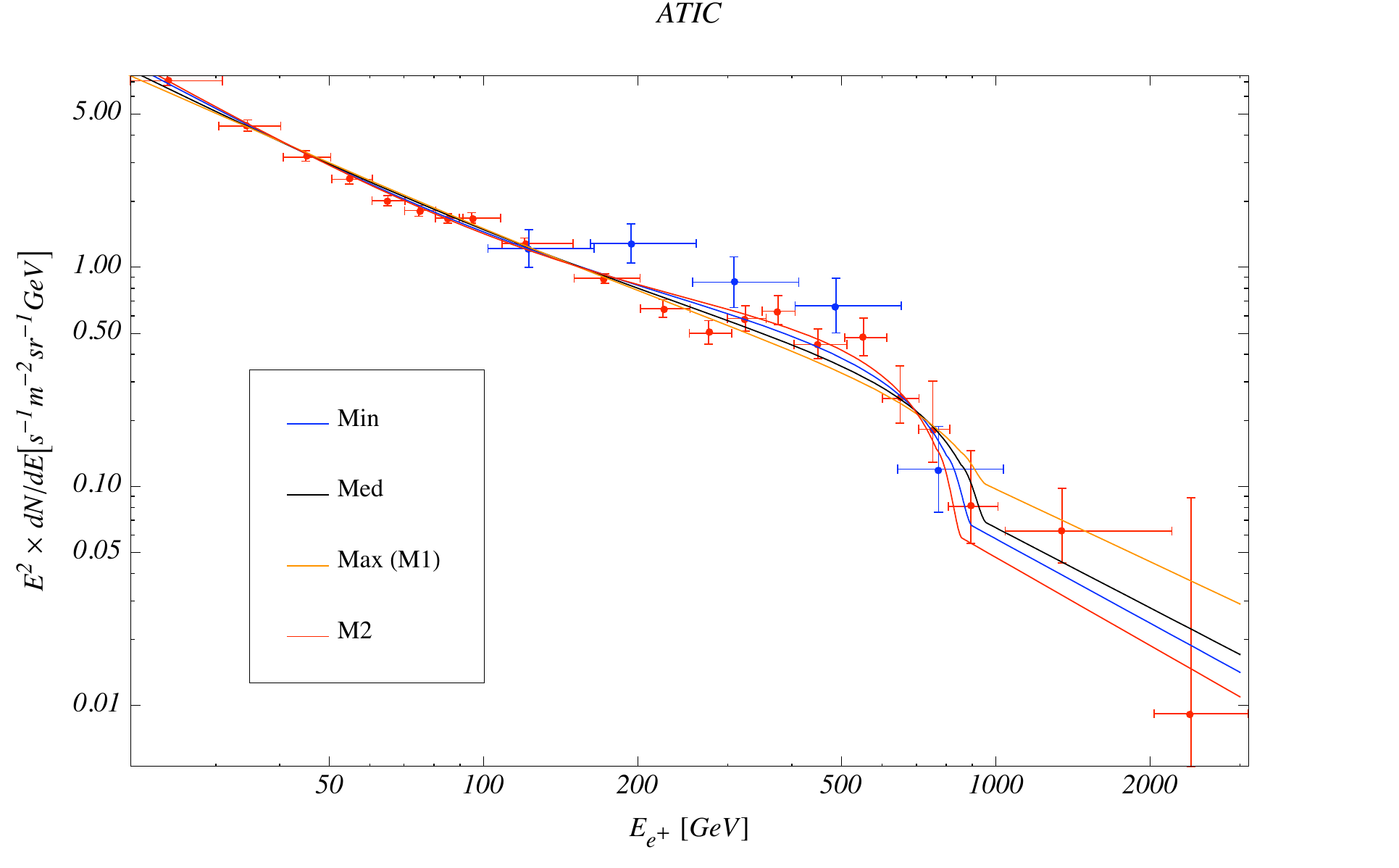}
\\ 
\includegraphics[width=3.3in]{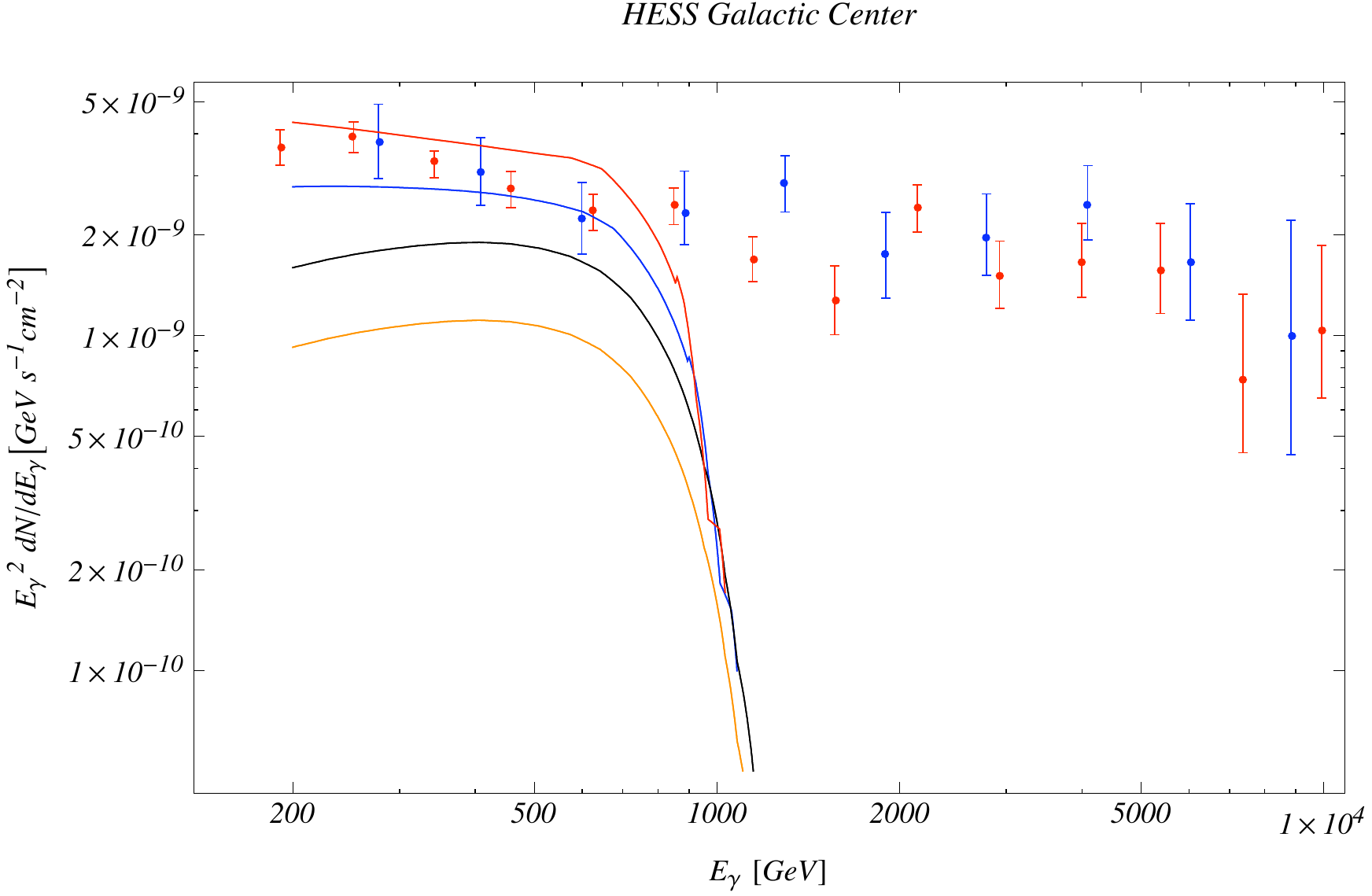}  &  
\hspace{-0.2cm}\includegraphics[width=3.25in]{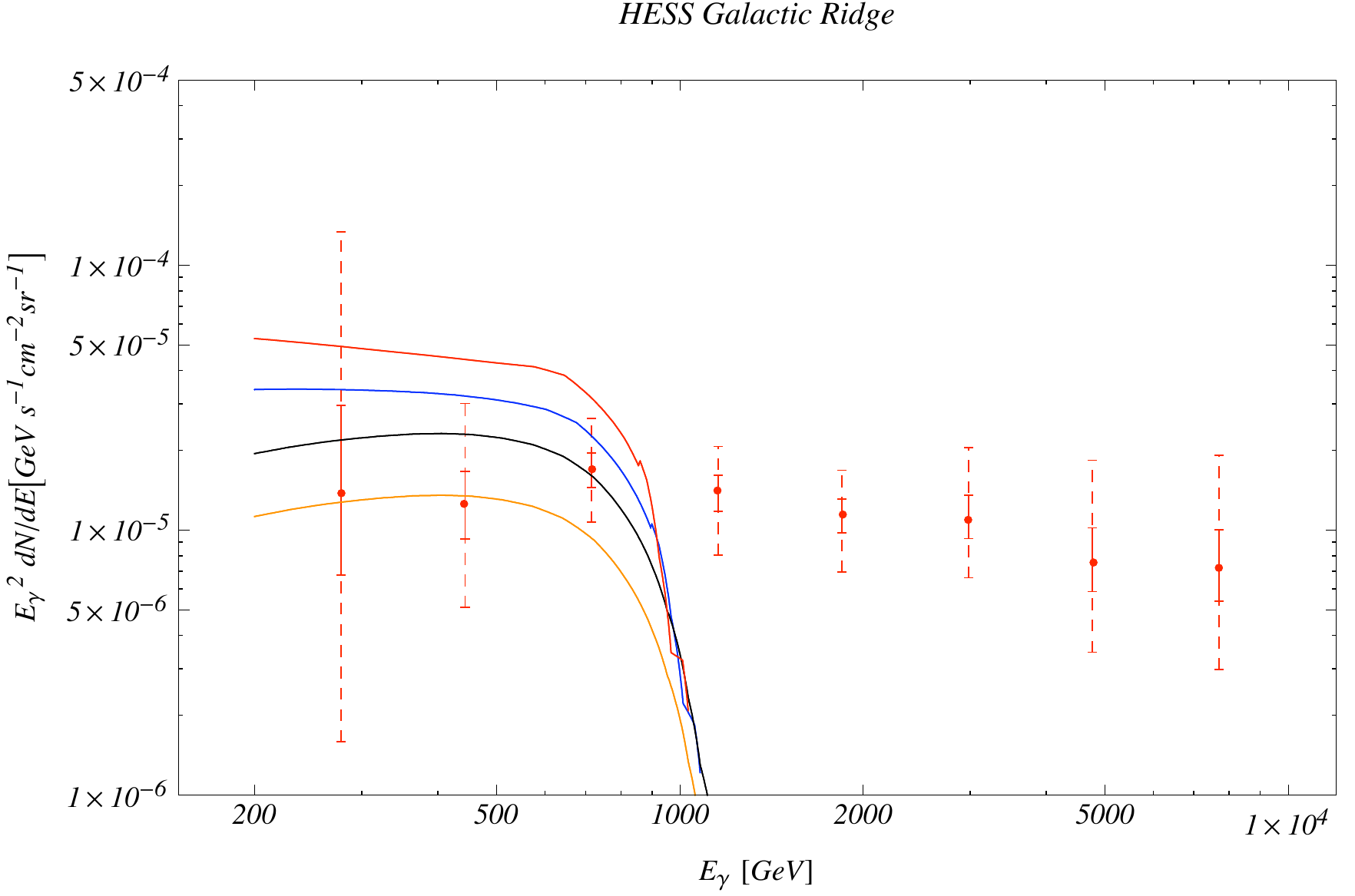}
\end{array}$
\end{center}
\caption{Plots for different propagation models, with $m_\phi=200$
  MeV.  The best fit parameters used are given in
  Table~\ref{tab:ihalofit}.  The neutrino spectrum is not shown since
  non are produced due to kinematics.  The predictions for photons are
  shown in the last two plots overlayed with the HESS
  measurements. Around each bin of the HESS data, we indicate the
  $1\sigma$ (solid) and $3\sigma$ (dasshed) error bar. }
\label{fig:ihalo}
\end{figure}

As in previous sections, we take our benchmark point, $m_\phi=200$ MeV and the Einasto
$\alpha=0.17$ profile, and fit the rest of the parameters for the four
propagation models, MIN/MED/MAX(M1)/M2.  The best fit values of the
parameters are shown in Table~\ref{tab:ihalofit} and the predictions
are shown in Figure~\ref{fig:ihalo}.  Even though antiprotons are not
produced from DM annihilations through the light gauge fields, some of
the best fit models require a 
non-negligible SM couplings and consequently antiprotons and
produced as can be seen in the plots.

The tension anticipated in Section~\ref{sec:propagation} is now
apparent: because of the enhancement of the flux through
$I(\lambda_D)$ for large $\lambda_D$ in the MED and MAX cases, the
electron spectrum in that case is softer (recall that the diffusion
length, $\lambda_D$, is larger for lower detected energies).
Therefore, for these propagation models the feature in the high energy
spectrum is less pronounced and so the fit for ATIC/PPB-BETS is not as
good.  On the other hand, due to the soft spectrum, a smaller
cross-section is needed to fit PAMELA and therefore less photons are
predicted for these models.  This is consistent with
Figure~\ref{fig:ihalo}.  Moreover, in the MED/MAX models the spectral
index for the background $e^++e^-$ flux is harder to compensate for
the bad fit to the ATIC bump.  

We find it is easier to evade the HESS
constraints if propagation of positrons is closer to the MAX model described
above, and in particular if the dependence of the escape time of
Cosmic rays on the energy is weaker (smaller $\delta$).
Interestingly, it is likely that the behavior of the escaping time
flattens somewhere below the knee at $10^{15}$ eV \cite{Hillas:2005cs}.  Moreover,
uncertainties in the spallation cross-sections may point towards a
smaller $\delta$.

\begin{table}[t]
\vskip 0.5cm
\begin{center}
{\begin{tabular}{|c|c|c|c|c|c||c|}
\hline
Model & $\mdm$ (TeV) & $10^{23}\times \langle\sigma v\rangle
(\mathrm{cm^3 s^{-1}})$ & $(N,\gamma)_{e^++e^-}$ & $N_{e+}$ & $\rsm$ &
$\chi^2/ {\rm dof}$\\
\hline \hline
MIN & 0.90 & 1.15 & (0.9,-3.28) & 1.9&$6\%$ &1.8
\\
MED & 0.96 & 0.94 & (0.6,-3.21) & 0.7 &$1\%$ &1.9
\\
MAX(M1) & 1.38 & 0.38 & (0.4,-3.10) & 1.4&$1\%$ &2.3
\\
M2 & 2.3 & 1.44 & (1,-3.33) & 0.9& $11\%$ &2.0
\\
\hline
\end{tabular}}
\end{center}
\caption{The best fit values different propagation models with
  $m_\phi=200$ MeV, shown in the plots of
  Fig.~\ref{fig:ihalo}.  $\gamma_{e^++e^-}$ is the best fit value
  for the spectral index of the background electron plus positron
  flux.  $N_i$ is the fraction of the normalization found for the best
  fit background without a DM signal.  These normalizations are found
  in Section~\ref{sec:ai}.  }
\label{tab:ihalofit}
\end{table}

%%%%%%%%%%%%%%%%%%%%%%%%%%%%%%%%%%%%%%%%%%%%%%%%%%%%%%
%%%%%%%%%%%%%%%%%%%%%%%%%%%%%%%%%%%%%%%%%%%%%%%%%%%%%%
\section{Implications and Future Directions}\label{sec:imp}
\setcounter{equation}{0} \setcounter{footnote}{0}
%%%%%%%%%%%%%%%%%%%%%%%%%%%%%%%%%%%%%%%%%%%%%%%%%%%%%%
%%%%%%%%%%%%%%%%%%%%%%%%%%%%%%%%%%%%%%%%%%%%%%%%%%%%%%

In this section we discuss the implications for models that attempt to
explain the PAMELA and ATIC/PPB-BETS excesses, based on the results
shown in Section~\ref{sec:pb}.  In Section~\ref{sec:pb} we have shown
by varying the particle physics and astrophysics parameters that only
certain regions of parameter space can satisfy all the various
experimental constraints.  Specifically, the most difficult
constraints arose from the HESS's measurement of the GC and GR
regions.  Indeed it was initially believed that hadronic activity from
DM was needed to be suppressed in order to avoid creating an excess in
the antiproton flux.  Conversely, $m_\chi$ could be pushed to scales of
order $10$ TeV, to avoid the bound.  As we have demonstrated in
Section~\ref{sec:pb}, one {\em can} tolerate antiprotons without
having to raise $m_\chi$ significantly above the mass scale of the
purported ATIC/PPB-BETS excesses.  On the other hand, the real
hadronic danger comes from $\pi^0$ decays which produce a significant
amount of photons.  By scanning over $m_\phi$ we have shown that one
does in fact need a model that goes ultimately almost exclusively into
purely leptonic final states.

\begin{figure}[h!]
\begin{center}
$\begin{array}{ll}
\includegraphics[width=3.5in]{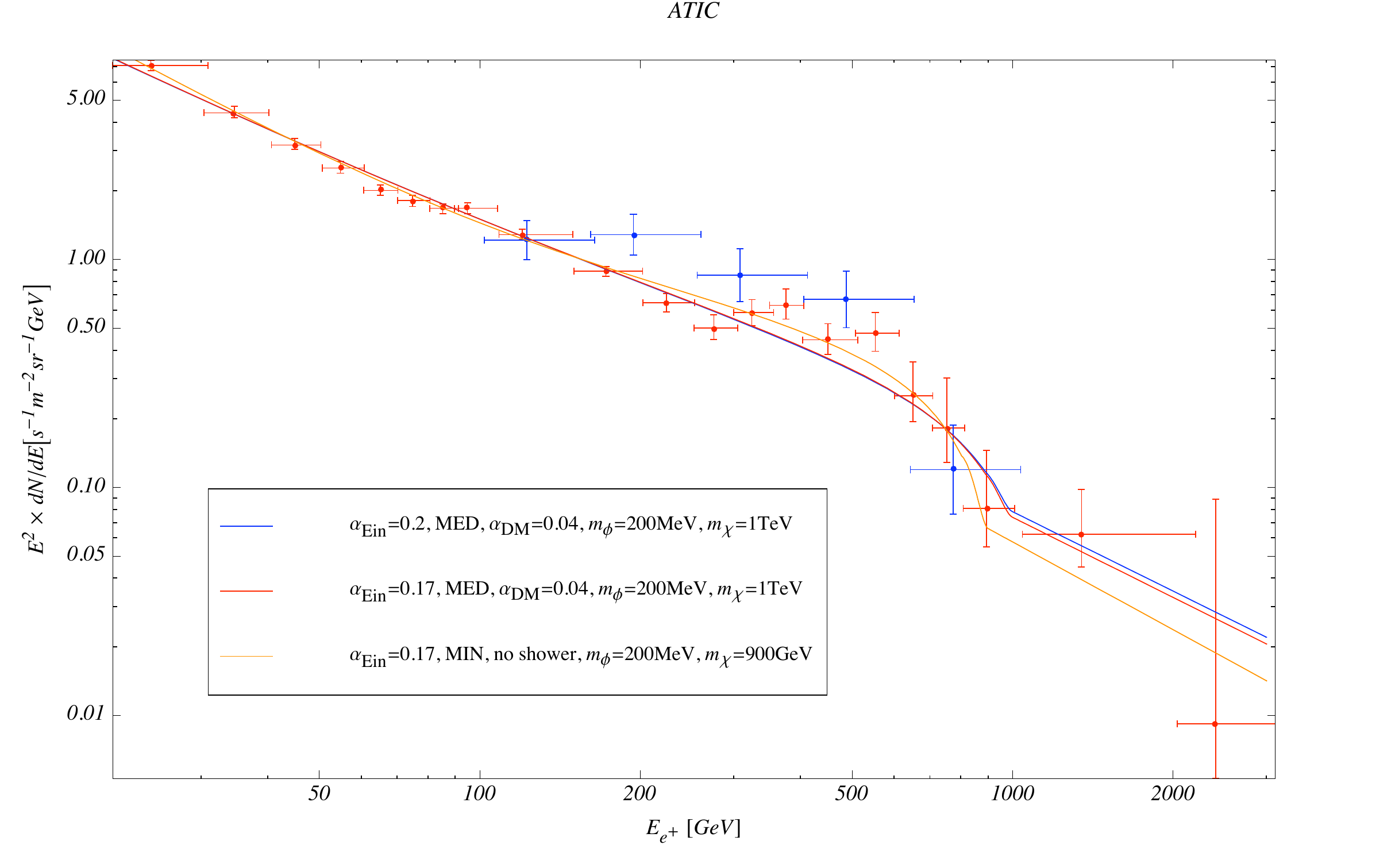}  &  
\hspace{-0.3cm}\includegraphics[width=3.15in]{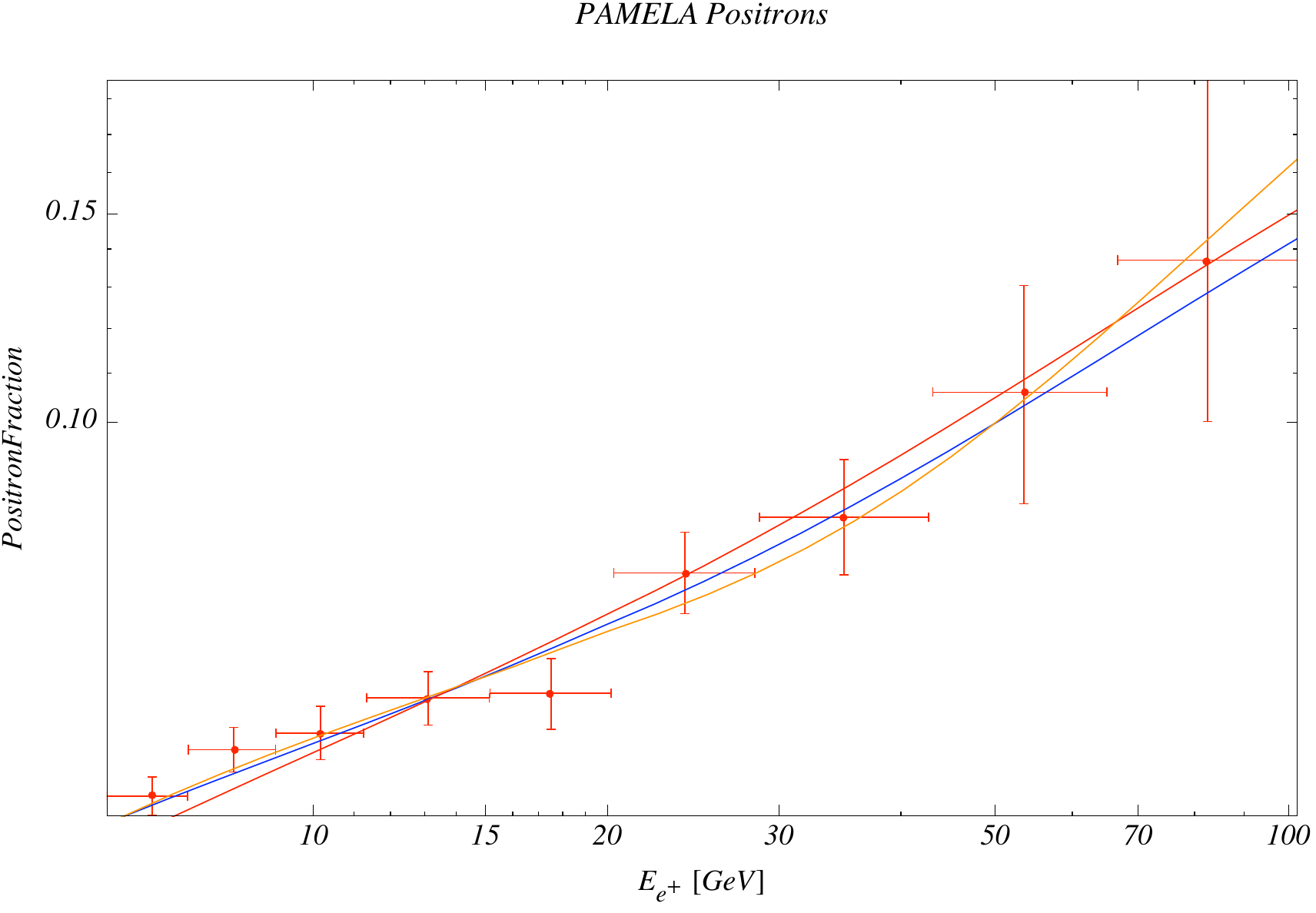}    
\\ 
\hspace{-0.4cm}\includegraphics[width=3.4in]{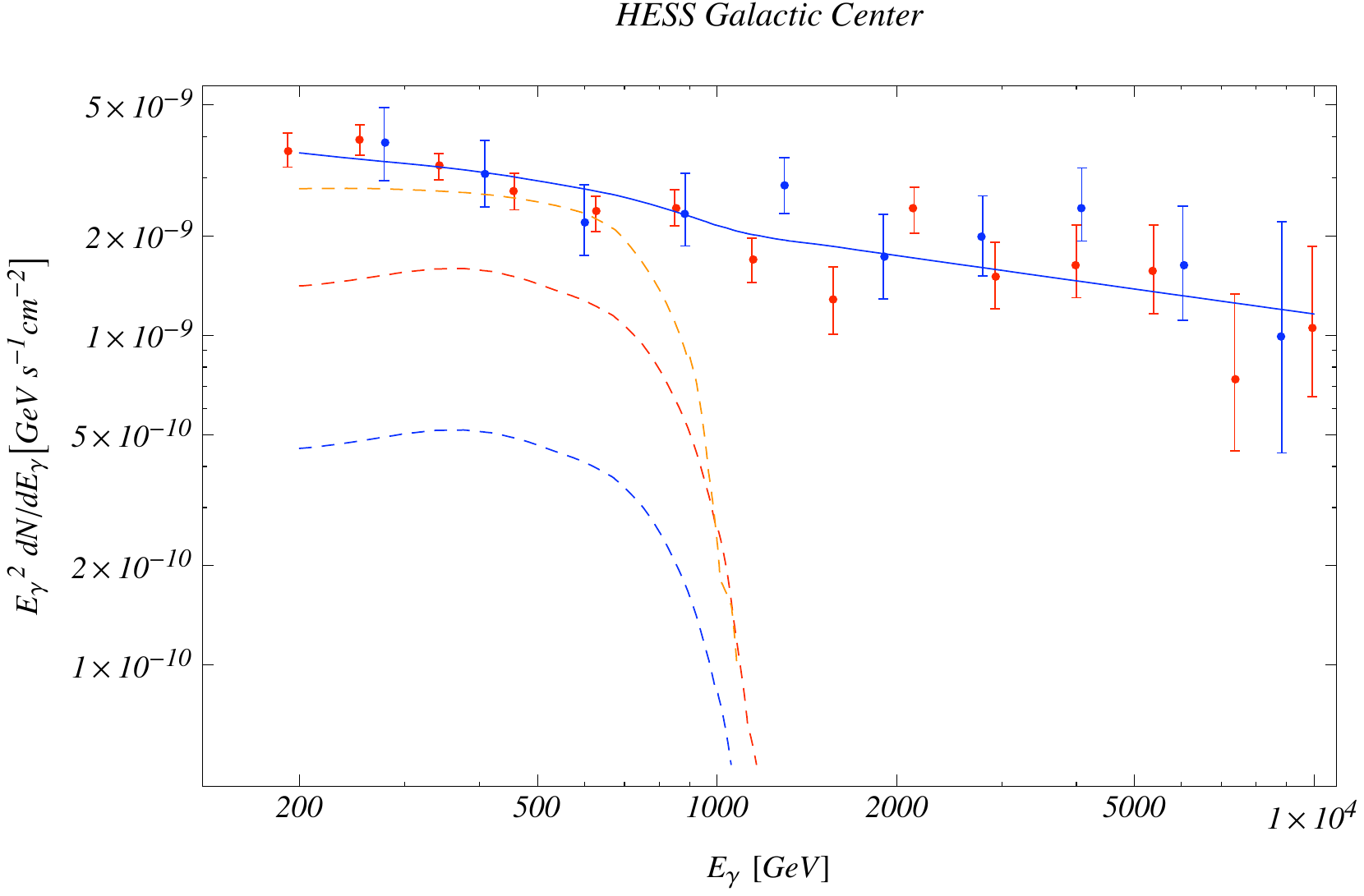}  &  
\hspace{-0.8cm}\includegraphics[width=3.4in]{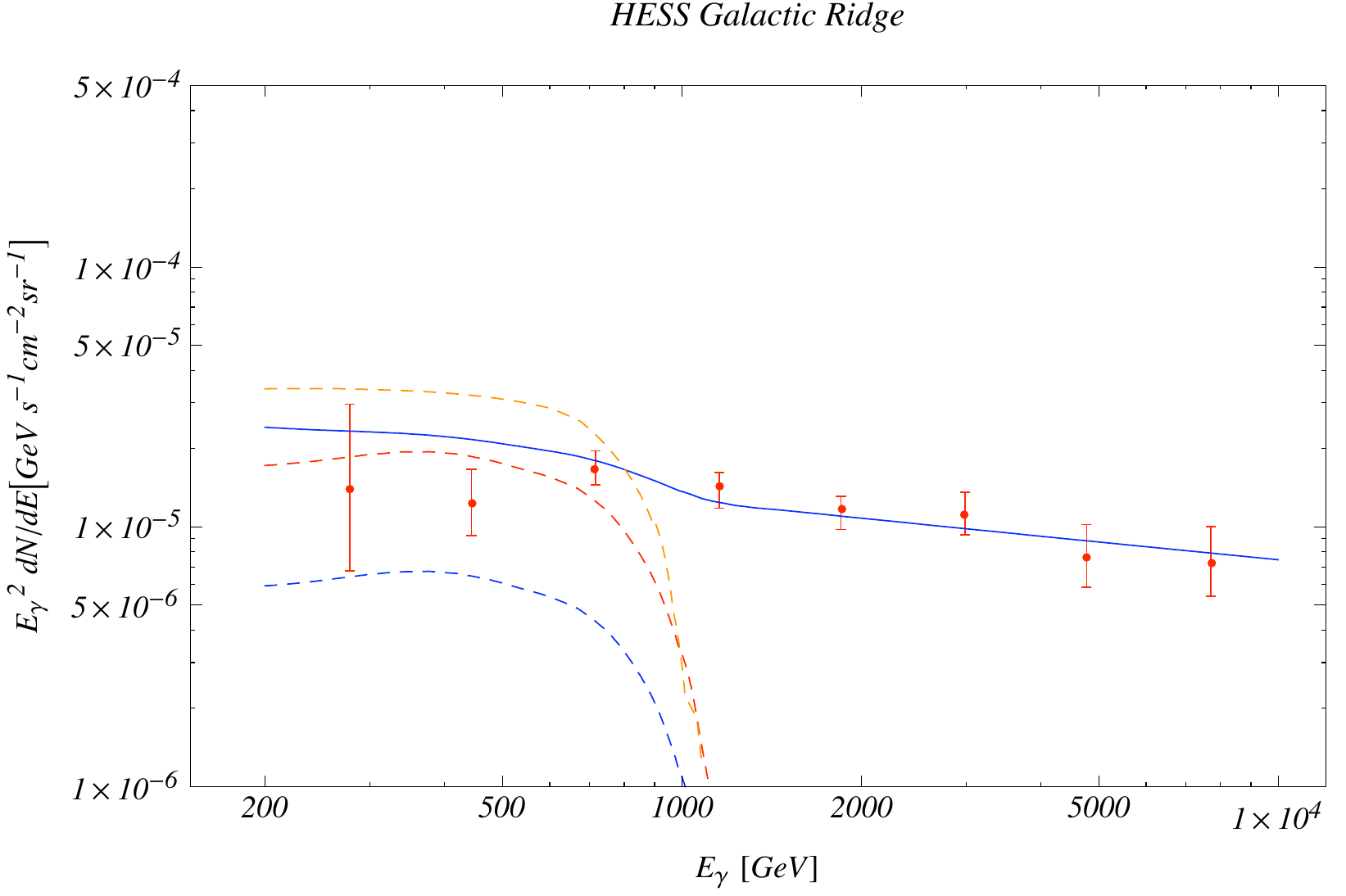}
\end{array}$
\end{center}\
\caption{The best fit including HESS/PAMELA/ATIC/PPB-BETS is plotted
  in blue(lower blue curve is signal only in the HESS plots), which is
  for $m_\chi=1$ TeV, $m_\phi=200$ MeV, $\langle \sigma
  v\rangle=8.2\cdot10^{-24}\,\mathrm{cm}^3/\mathrm{s}$ and
  $\alpha_{DM}=.04$, for an Einasto profile with $\alpha=.2$ and the
  med propagation parameters.  The best fit including
  PAMELA/ATIC/PPB-BETS without HESS is plotted in orange.  The
  parameters are $m_\chi=900$ GeV, $m_\phi=200$ MeV, $\langle \sigma
  v\rangle=1.1\cdot10^{-23}\,\mathrm{cm}^3/\mathrm{s}$, with no parton
  shower, an Einasto density profile with $\alpha=.17$ and MIN
  propagation.}
\label{fig:bestfit}
\end{figure}

In several recent papers~\cite{bergstrom,strumiagamma} it was also argued, or
indirectly demonstrated, that the high energy photons rule out models
that try to explain both the PAMELA and ATIC/PPB-BETS excesses.
This is not a conclusion that we agree with.  These papers
primarily investigated the case where DM annihilated directly into
leptons.  As shown in Section~\ref{sec:udmdndx}, the $dN/dx$ calculated
in these cases is actually quite different, and a simple rescaling of
the overall rate does not interpolate between the case of directly
annihilating into leptons and annihilating through $\phi$.  To
demonstrate that there exist viable models that satisfy the high energy
constraints we show a best fit point from the scan over our
parameters.  In Figure~\ref{fig:bestfit} this point is shown where
$m_\chi=1$ TeV, $m_\phi=200$ MeV, $\langle \sigma
v\rangle=8.2\cdot10^{-24}\,\mathrm{cm}^3/\mathrm{s}$ and
$\alpha_{DM}=0.04$, for an Einasto profile with $\alpha=0.2$ and the MED
propagation parameters.  Additionally, for comparison, Figure~\ref{fig:bestfit} shows the same parameters except we change the DM profile to Einasto $\alpha=0.17$.  Furthermore, we show in the 
figure how HESS alters this fit by giving the
best fit case where only PAMELA and ATIC/PPB-BETS are included in the
$\chi^2$.  This corresponds to a best fit point of $m_\chi=900$ GeV,
$m_\phi=200$ MeV, $\langle \sigma
v\rangle=1.1\cdot10^{-23}\,\mathrm{cm}^3/\mathrm{s}$, with no parton
shower, an Einasto density profile with $\alpha=0.17$ and MIN
propagation parameters. As we can see,
one can find a fit for annihilating background that {\em includes} a
photon background for HESS and satisfies the data.  Still, for a good
fit, the DM profile is required to be less cuspy than the preferred
profile of $\alpha=0.17$.  We can also quantify a suppression factor for an Einasto profile with $\alpha=0.17$, that could come in principle from several sources, so that the effective $\bar{J}$ is as good as our best fit point.  For $m_\phi=200$ MeV one needs a suppression factor of 3-4, for $m_\phi=500$ MeV we need a factor of 4, while for $1.2$ GeV we need a larger suppression of $\mathcal{O}(10)$.

A recent paper~\cite{bergstrom} investigated the case of DM
annihilating through a light $\phi$ and then into leptons.  Their
claim was that models of this type that satisfied PAMELA/ATIC with
standard DM density profiles were ruled out unless an order of
magnitude local boost factor was included. Their most stringent bounds
come from examining Sgr A*, in radio frequencies similar to
~\cite{strumiagamma}.  Unfortunately, such a constraint is highly
sensitive to knowledge of the DM density profile at distances much
below $100$ pc which is beyond the resolution limit of current
simulations.  Nonetheless, if one takes the bound at face value we
find that their band that agrees with the PAMELA measurement is
different than ours, and we can fit both PAMELA and their radio bound.
In the region of high energy gammas this constraint is not more
significant than the HESS GC one.  The authors also consider bounds
from the the Sagitarius dwarf galaxy.  Since the Sagitarrius dwarf
galaxy is being tidally disrupted by the Milky Way it is
dubious to trust bounds coming from this alone.   We have checked for one of the most studied dwarf galaxies, Draco (that is not too close to our Galaxy to be tidally disrupted), that the bounds are easily satisfied by at least an order of magnitude.  As discussed, the most
significant bound we find arises from the study of the
GR region, and as shown in Figure~\ref{fig:bestfit} they can be
satisfied.

\begin{figure}[h!]
\begin{center}
$\begin{array}{c@{\hspace{.1in}}c}
\includegraphics[width=3in]{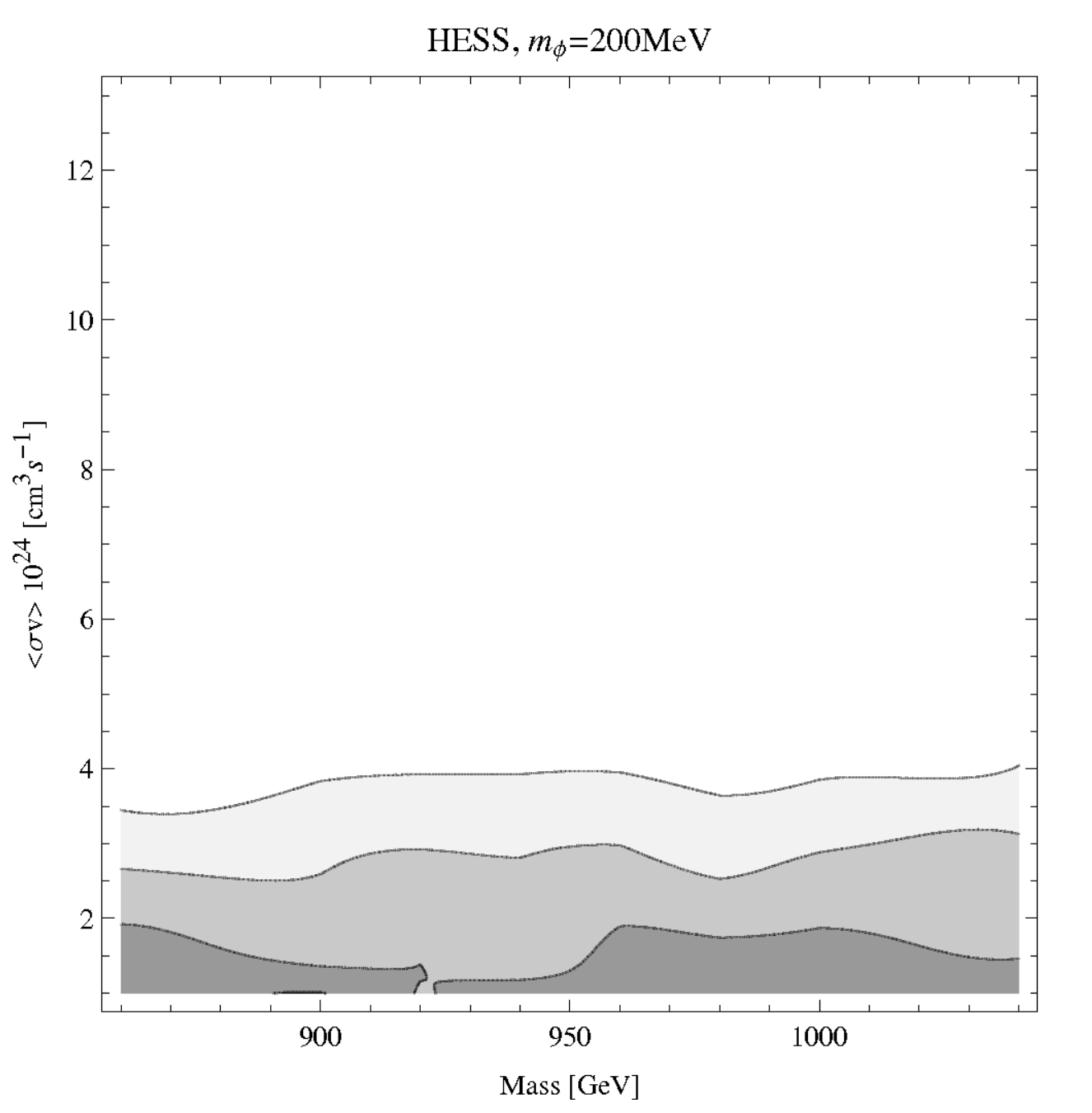}  &  
\includegraphics[width=3in]{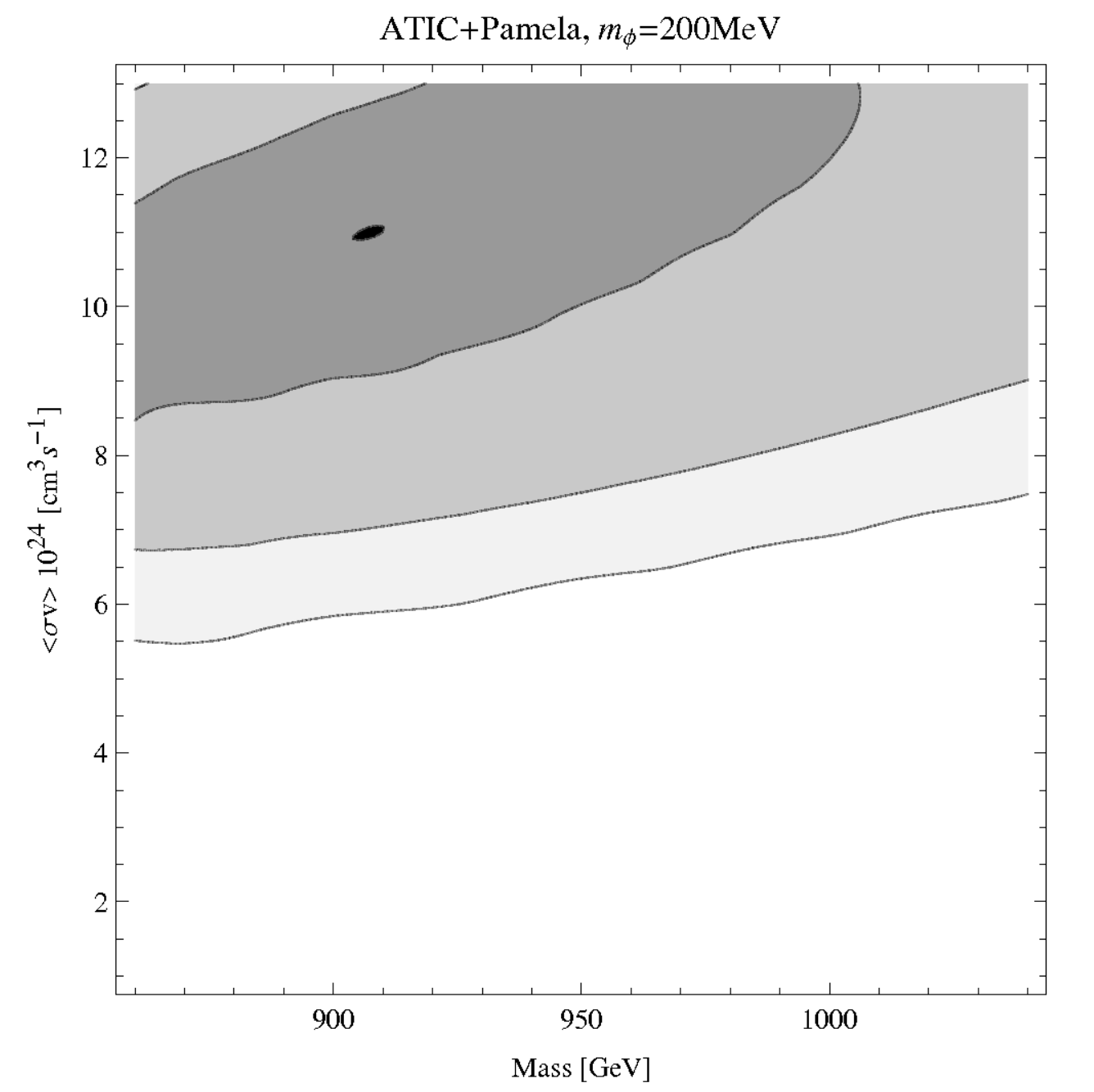}   \\ 
\end{array}$
\end{center}\
\caption{Comparison between the $\chi^2_{\mathrm{HESS}}$ and
$\chi^2_{\mathrm{ATIC/PAMELA}}$ for $m_\phi=200$ MeV as a function of
mass and cross-section.}
\label{fig:tension}
\end{figure}

While there exist points in parameter space that can satisfy all
experimental constraints, this does not mean that generically there is
no tension between DM annihilation models of this type and the
experimental results.  As we can see from Section~\ref{sec:pb}, in
many regions of parameter space the HESS experiment would completely
rule out models from the GR data.  This is due to the fact that the
HESS experiment in both the GC and the GR, records data that has a best
fit to a power law.  Once one includes DM annihilating in the energy
range that HESS studies, it automatically introduces a non-power
law shape on top of the background. Since the PAMELA experiment
currently studies energies less than the photon energies recorded by
HESS, there is no tension between these experiments.  However, the
inclusion of ATIC/PPB-BETS experiments, automatically signal a mass
scale which creates the 
tension.  We demonstrate this in Figure~\ref{fig:tension} where
we separate our $\chi^2$ function into $\chi^2_{\mathrm{HESS}}$ and
$\chi^2_{\mathrm{ATIC/PAMELA}}$ and plot two dimensional contours of
$\chi^2$ for each, in the space of $m_\chi$ and $\langle \sigma
v\rangle$ .
As we can see from Figure~\ref{fig:tension} the best fit for $\chi^2$
taking into account PAMELA/ATIC/PPB-BETS alone prefers a lower mass scale and
higher cross-sections, while HESS prefers a low cross-sections and
potentially higher masses in order to preserve its pure power law background.
 
This generic tension could mean several things.  First, the excesses
in PAMELA and ATIC/PPB-BETS may simply be caused by astrophysics.
Pulsars for instance could explain the leptonic excesses without
necessarily introducing a large component of high energy gamma rays at
the center of our galaxy.  Second, it could be a red herring, and as
we have demonstrated, models of this type do have points in parameter
space that could account for the experimental data. Another way to
avoid the tension would be to investigate models of decaying
DM~\cite{decaying} instead of annihilating DM since the amount of
photons at the center of the galaxy would then scale like $\rho$
instead of $\rho^2$.  While these are all possible interpretations, we
wish to point out another.

The starting point of current DM investigations is to automatically
assume all the recent excesses reported by experiments, and work under
the assumption that there is/will be no conflicting data.  For
instance in the case of the ATIC/PPB-BETS experiments the excesses
that are shown conflict with the existing EC~\cite{EC} data.  It has
been pointed out in the past that EC was based on a much smaller
detector area and thus could be more prone to systematic errors.  This
could be the cause of the apparent discrepancy between the
ATIC/PPB-BETS and EC data.   Still, it is curious nonetheless that
ATIC/PPB-BETS and HESS are the experiments that seem to have a tension if the
picture of annihilating DM is true.  One possibility could be that the anomaly observed by ATIC
 disappears in the future, or that the excess they see is
unrelated to DM but rather to astrophysics (e.g. cooling).
In this case a model would only need to satisfy the PAMELA experiment
and the mass scale could be much lower in principle.

\begin{figure}[h!] 
\begin{center}
$\begin{array}{ll}
\includegraphics[width=3.15in]{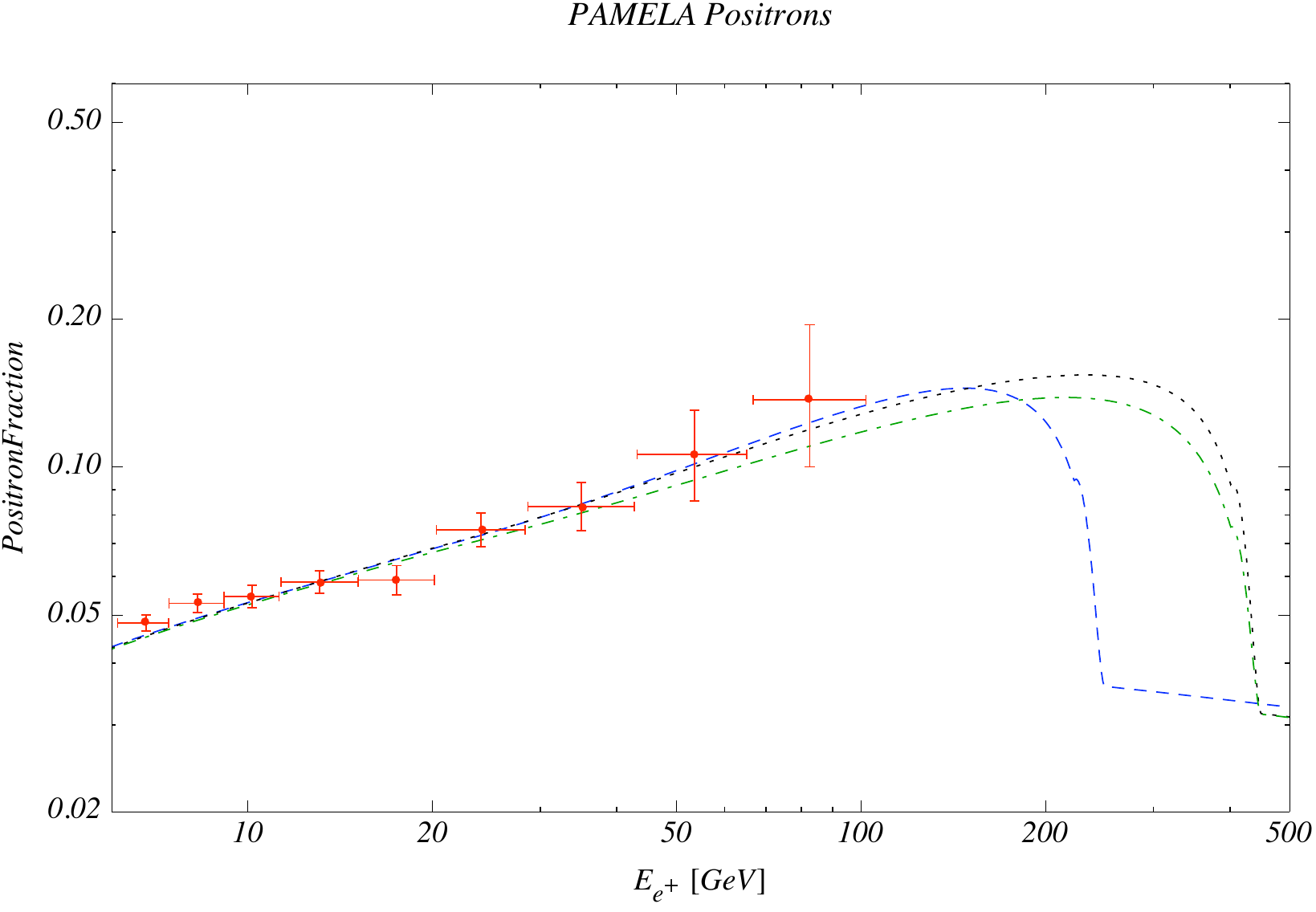}
& \includegraphics[width=3.5in]{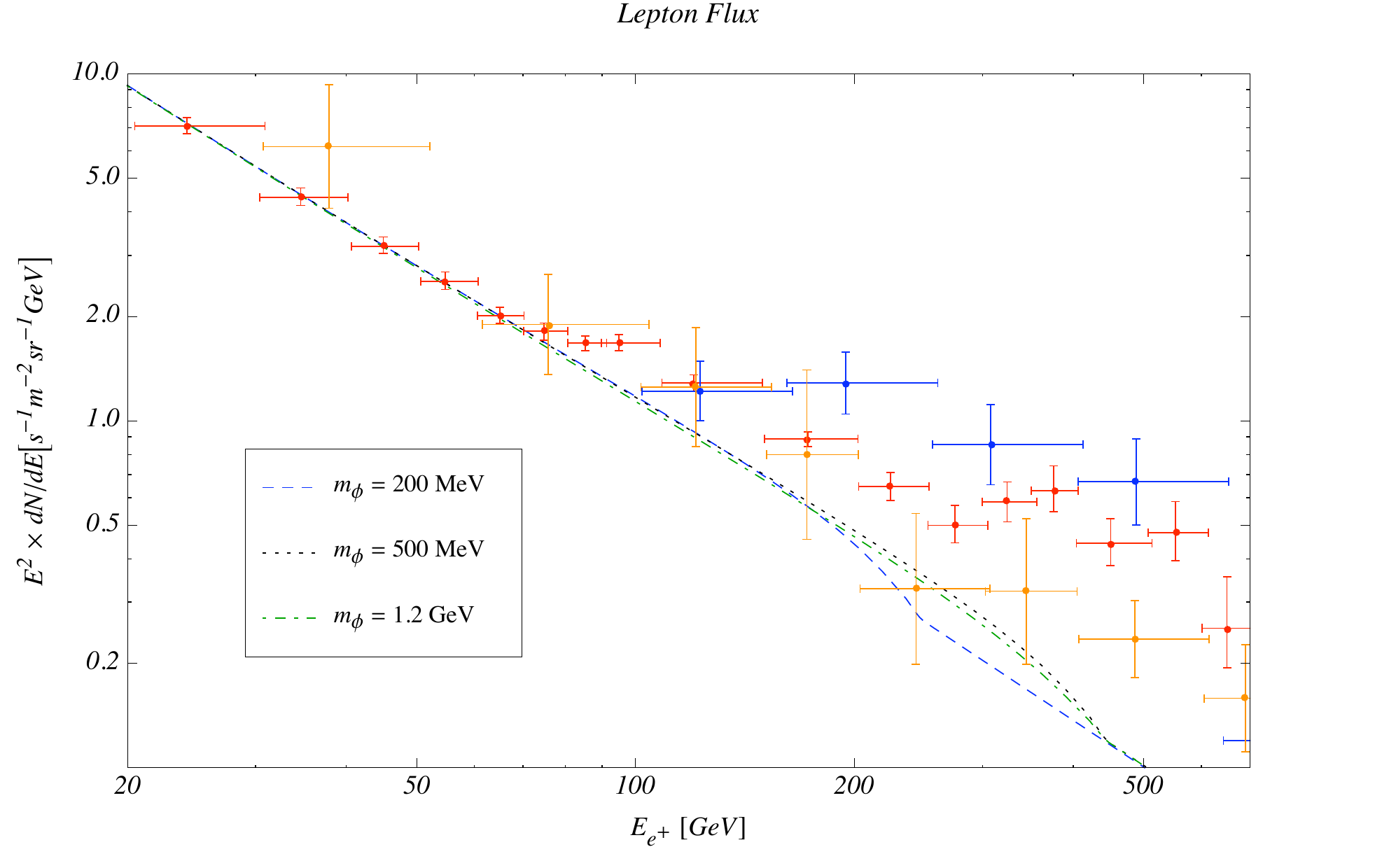} 
\\
\hspace{-0.5cm}\includegraphics[width=3.4in]{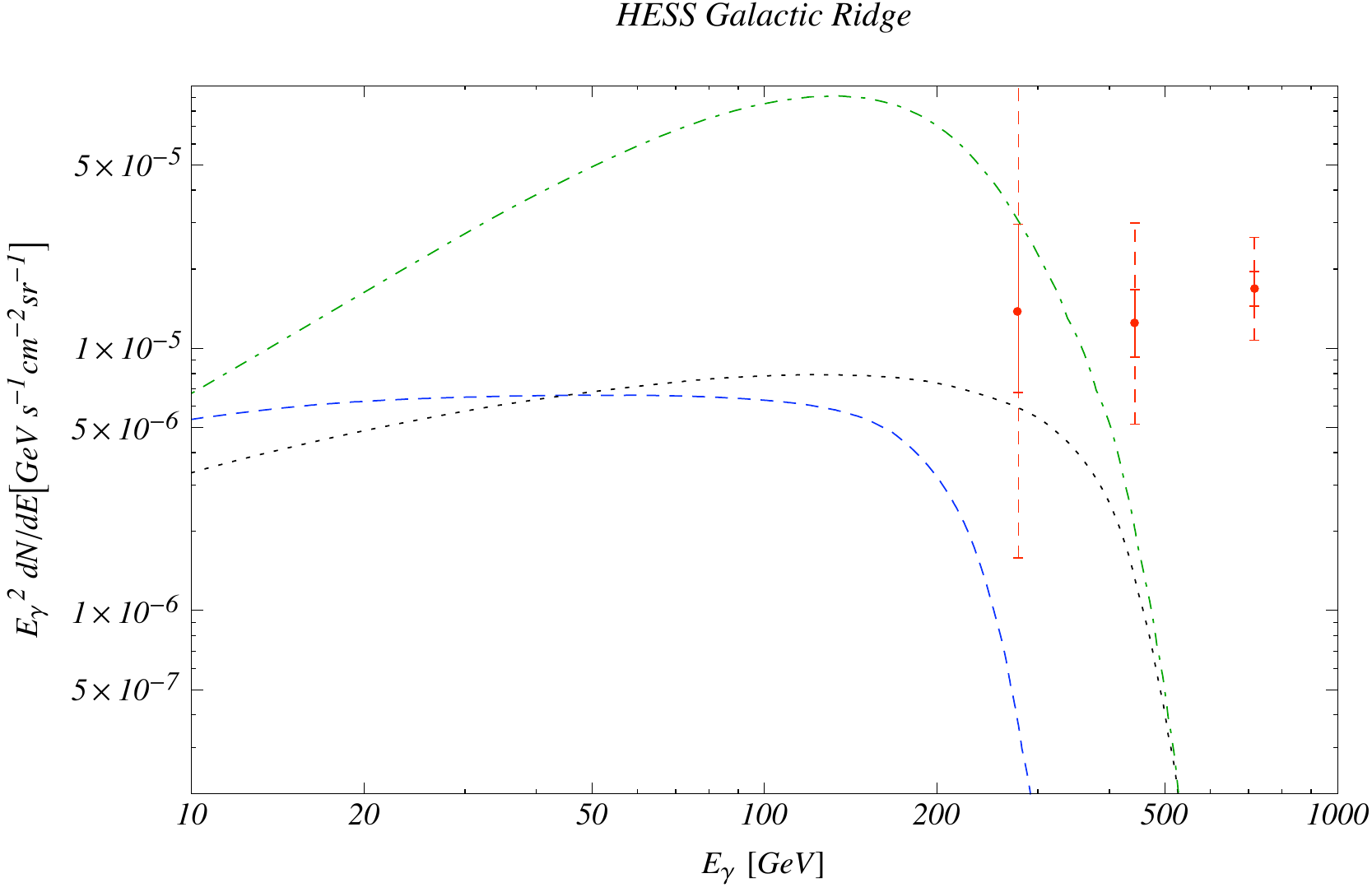}& 
\hspace{-0.5cm}\includegraphics[width=3.35in]{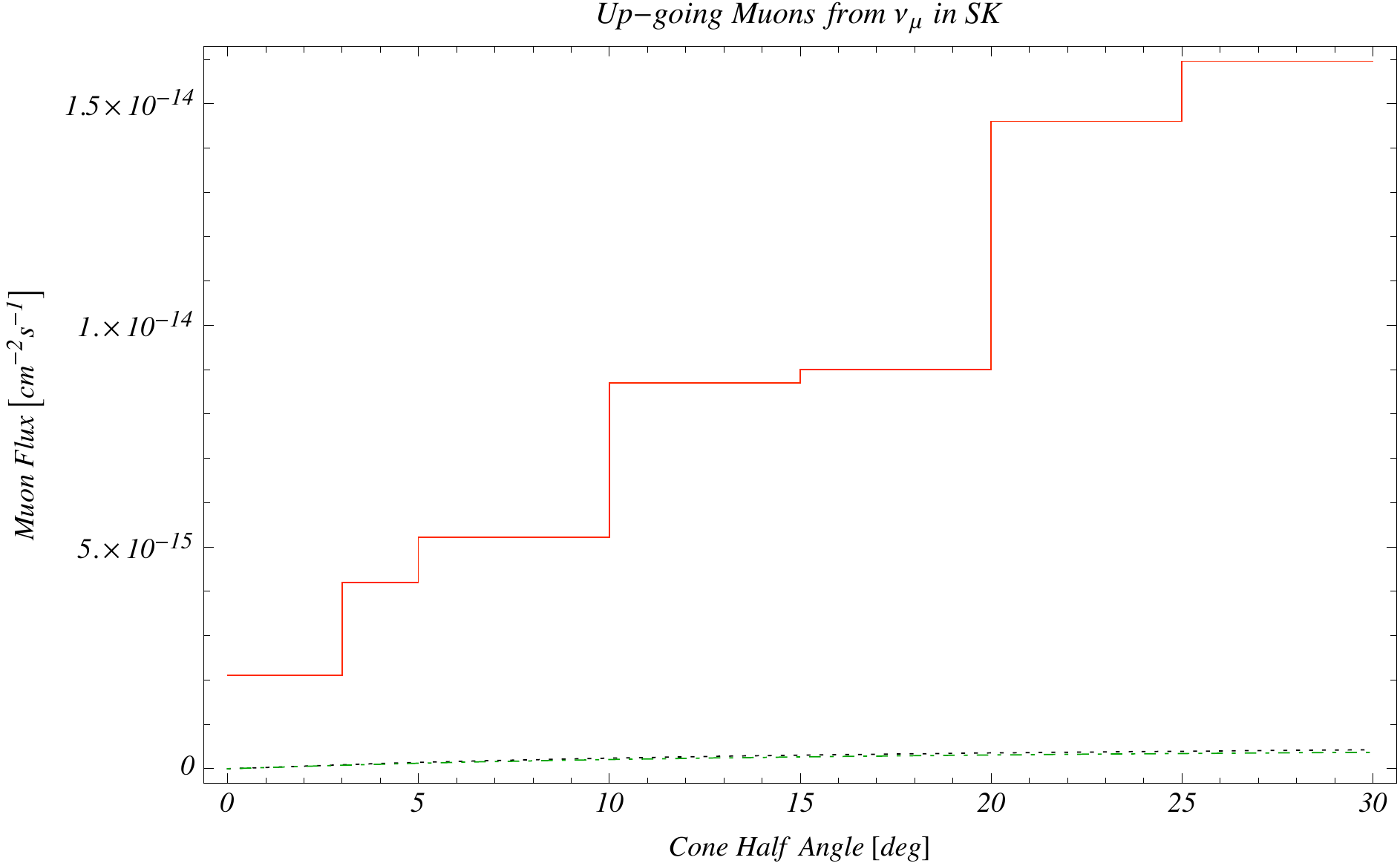}   
\end{array}$
\end{center}\
\caption{We plot the best fit for $m_\phi=0.2,0.5,1.2$ GeV excluding the data from ATIC/PPB-BETS above $100$ GeV.  The best fit point corresponds to $m_\chi=250$ GeV,
$m_\phi=200$ MeV and $\langle \sigma v \rangle = 9.65\cdot
10^{-25}\,\mathrm{cm}^3/\mathrm{s}$, with an Einasto density profile
for $\alpha=0.17$ and MED propagation.}
\label{fig:light}
\end{figure}

In Figure~\ref{fig:light} we demonstrate the consequences of assuming
only the PAMELA experiment and we fit to the electron flux below 100
GeV.  The fact that this is a viable option is not immediately clear
without calculating the results for the other experiments.  Indeed, if
one lowers the mass scale, there are several potential difficulties
that need to be confronted.  One needs to make sure that no feature is
introduced in the electron flux that would have been seen thus far.
At lower energies, even though one can avoid the bounds from the HESS
GR data, in principle one could now be in danger of over-populating
the low energy photons for the EGRET experiment.  As we show in
Figure~\ref{fig:light} for the case of $m_\chi\sim250$ GeV,
$m_\phi=200$ MeV and $\langle \sigma v \rangle = 9.65\cdot
10^{-25}\,\mathrm{cm}^3/\mathrm{s}$, with an Einasto density profile
for $\alpha=0.17$ and MED propagation, one can satisfy all
experimental constraints easily with $\chi^2/\mathrm{dof}\sim1$.  This
is an incredibly interesting prospect as it opens up possible new
avenues for model building and additionally provides even {\em more}
testable predictions.  By lowering the mass scale the high energy
photons are now completely within reach of the FERMI experiment
\cite{fermi} and furthermore a nontrivial turnover in the positron fraction
measured by PAMELA could be observed.  In
Figure~\ref{fig:light} we plot the photon flux expected from the high energy
photons observable by FERMI when looking at the GC.  This is well
within the sensitivity range of the experiment, and it is an open
question whether or not the high energy photons
are observable compared to the ICS contribution.  We postpone this for
future work~\cite{usinthefuture}.

As we have stated, there are several possibilities for explaining the
excesses:  annihilating DM for low or high $m_\chi$, decaying DM,
Astrophysics/pulsars, some unknown idea or combination thereof.
Whatever it turns out to be, we are in a golden age of experiments and
it's useful to review what are the most important experimental
results/possible measurements that could be done to sort out potential
DM candidates.
\begin{itemize}
\item{\bf PAMELA:} Future measurements of the positron flux
  can tell us whether or not the fraction continues to increase
  throughout its mass range and has bearing on whether or not DM is
  heavy or not.
\item {\bf FERMI:} Has the ability to confirm or reject the
  ATIC/PPB-BETS excesses by measuring the electron flux.  Additionally
  if the DM is light FERMI will be in the exact range needed to study
  it's properties.
\item {\bf ATIC:} The release of ATIC-4 data will offer better
  statistics and will allow for a better comparison to HESS's recent
  release of the lepton flux at the high end of ATIC's reach
  \cite{HESSelectrons}.
\item {\bf HESS:} Has already bounded high scale annihilating DM
  by studying the GR with only a relatively short amount of data
  taking.  In principle by collecting more data, it could rule out
  beyond a shadow of the doubt high scale annihilating DM.
\end{itemize}
These possible results combined with other experiments should allow us
within the next few years to confirm or rule out many possibilities
for DM.  As it stands now, we have demonstrated that the particle
physics module we have implemented can account for existing data.  In
the near future we will hopefully be able to further pin down or rule
out the properties of models that have the features we examined.

\vspace{1cm}
{\bf Note Added: } While this paper was in preparation we learned of a similar work in progress by Jeremy Mardon, Yasunori Nomura, Daniel Stolarski,  and Jesse Thaler~\cite{mardon}.

%%%%%%%%%%%%%%%%%%%%%%%%%%
\section*{Acknowledgments}
We would like to thank Nima Arkani-Hamed, Kfir Blum, Marco Cirelli,
Boaz Katz, Michael Kuhlen, Gilad Perez, Alessandro Strumia, Eli Waxman and  Neal Weiner
for useful discussions. PM and TV are supported in part by DOE grant
DE-FG02-90ER40542.  MP is supported in part by NSF grant PH0503584.
%%%%%%%%%%%%%%%%%%%%%%%%%%

\appendix

%%%%%%%%%%%%%%%%%%%%%%%%%%%%%%%%%%%%%%%%%%%%%%%%%%%%%%
%%%%%%%%%%%%%%%%%%%%%%%%%%%%%%%%%%%%%%%%%%%%%%%%%%%%%%
\section{The Leaky Box Approximation}\label{app:box}
\setcounter{equation}{0} \setcounter{footnote}{0}
%%%%%%%%%%%%%%%%%%%%%%%%%%%%%%%%%%%%%%%%%%%%%%%%%%%%%%
%%%%%%%%%%%%%%%%%%%%%%%%%%%%%%%%%%%%%%%%%%%%%%%%%%%%%%

In this Appendix we describe the leaky box approximation which is used
to estimate the background for the positron flux for energies
$\lesssim 100$ GeV.
This model
is essentially a simplified version of that described  in section
\ref{sec:propagation}.  Its value lies in
its minimal set of assumptions, allowing one to examine the data with as few extra theoretical inputs as possible.  

The confinement time of charged particles in the Galactic disk is
enhanced due to galactic magnetic fields and is of order $t_{\rm esc}
\sim 10^7 (E/\mathrm{GeV})^{-0.6}$ yr.
On the other hand, the cooling time due to ICS and synchrotron
radiation is of order $t_{cool} \sim 2\times 10^8 (E/{\rm GeV})^{-1}$ yr.
Thus for energies below $\sim100$ GeV, the particles either reach
the Earth or escape the Galaxy before loosing energy and therefore
cooling can be neglected.  For the background computation below, we
assume this is the case.  

The leaky box approximation is a simplified diffusion model which takes into
account the confinement of charged particles.  The model assumes a
free homogeneous diffusion of charged particles within the galactic
disk.  At the galactic boundaries, particles are either reflected or
escape with finite energy-dependent probability.  In its simplest
form, the only independent parameter is the
mean density of matter,  $\lambda_{\rm esc}$, traversed by the changed particle before
escaping.  One has $\lambda_{\rm esc} = \rho_{\rm
  ISM} \beta c t_{\rm esc}$ where $\rho_{\rm ISM}$ is the average
interstellar matter (ISM) density in the galaxy (not to be confused
with the density in the galactic disk) and $t_{\rm esc}$ is the
escaping time.  Under the assumption that $\lambda_{\rm esc}$ depends only on the
rigidity, $\R = pc/Ze$ of the charged particle, $\lambda_{\rm esc}$ is
extracted by measuring secondary to primary ratios.  The B/C
measurements from the HEAO-3 experiment \cite{Engelmann:1990} and other
balloon experiments (the most recent being ATIC-2
\cite{Panov:2007fe}), provide the most stringent constraint.  For
particles with rigidity $\R > 4.5$ GV one finds
\cite{Gupta:1989,Blasi:2008ch},
\begin{eqnarray}
  \label{eq:9}
  \lambda_{\rm esc} = 23.8\ \beta
  \left(\frac{\R}{\mathrm{GV}}\right)^{-\delta} \mathrm{ g\ cm^{-2}}.
\end{eqnarray}
with $\delta \sim 0.6$.  As discussed in section
\ref{sec:propagation}, other
propagation models allow variations in $\delta$
between $\delta \sim .45 - .85$.  To be conservative we consider these values here.

In the absence of  cooling effects and losses due to collisions, the master transport equation 
at equilibrium for a stable nuclei takes a simple form, 
\begin{eqnarray}
  \label{eq:10b}
  \frac{n_i(E_i)}{\tau_i(E_i)} = Q_i^{\rm prim}(E_i) + Q_i^{\rm
    sec}(E_i).
\end{eqnarray}
Here $Q_i^{\rm prim (sec)}$ is the primary (secondary) source term for a
particle of type i with, 
\begin{eqnarray}
  \label{eq:12a}
 Q_i^{\rm
    sec}(E_i) = 
\frac{\beta c\ \rho_{\rm ISM}}{m}\ \sum_j
  \int dE_j \ n_j(E_j)\ \frac{d \sigma(E_j\rightarrow E_i)}{dE_i} .
\end{eqnarray}
$n_i$ is
the number density per unit energy which is related to the flux
through $\Phi_i =(v_{i}/4\pi)n_i$.    For positrons, where 
primary sources are absent, the first term on the RHS of
(\ref{eq:10b}) vanishes and one finds,
\begin{eqnarray}
  \label{eq:11a}
  \fpos(E) = Q_{e^+}^{\rm sec}(E) t_{\rm esc}(E)  =
  \frac{\lambda_{\rm esc}}{m} \langle n_p \sigma_p\rangle,
\end{eqnarray}
were $Q_{e^+}^{\rm sec}$ is the secondary source of positrons
generated at the point of interaction, and $\langle n_p\sigma_p\rangle
\equiv \sum_j \int dE_j \ n_j(E_j)\
\frac{d\sigma}{dE_i}(E_j\rightarrow E_i)$.  For positrons spallation
occurs through interactions of protons and $\alpha$ particles with
ISM.  Since to a good approximation $\sigma_p$ does not depend on
energy (other than a sharp kinematical cutoff \cite{Kamae:2006bf}),  a
prediction of this theory is that the ratio of positron-to-proton flux
is proportional to the escape time,
\begin{eqnarray}
  \label{eq:13}
  \frac{\fluxpos}{\fluxp} \propto \lambda_{\rm esc} \sim \left(\frac{{\cal
      R}_{e^+}}{{\rm GV}}\right)^{-\delta}.
\end{eqnarray}

The positron injection spectrum, $Q_{e^+}^{\rm sec}$, can be computed
using the measurements of the protons flux which leads to the fit
given in Eq. (\ref{eq:15}) and the table below it.  The differential
cross-section for the spallation was derived in \cite{Kamae:2006bf}
and the theoretical uncertainties were analyzed in
\cite{Delahaye:2008ua}.  In light of these uncertainties, we give a
rough estimate for the positron source (taking for positrons $e{\cal R}\simeq
E$), 
\begin{eqnarray}
  \label{eq:5a}
  Q_{e^+} \simeq 4 \times 10^{-27}\ \left(\frac{E_{e^+}}{{\rm GeV}}\right)^{-2.84\pm 0.02} \ {\rm \ cm^{-3}\ s^{-1}\ sr^{-1}\
  GeV^{-1}},
\end{eqnarray}
were we took $\rho_{\rm ISM}/m = 0.9 {\rm \ cm^{-3}}$.  Using
eq. \eqref{eq:9}, the results, Eq. (\ref{eq:8a}) follows.
We stress that the coefficient for the positron flux may suffer from 
corrections of order $100\%$.

\end{document}